%% file: dvojvit.tex
\documentclass{aa}
\usepackage{amsmath,graphicx,amssymb}
\usepackage[varg]{txfonts}
\usepackage{natbib}
\usepackage{pst-eps}
\usepackage{fixltx2e}
\bibpunct{(}{)}{,}{a}{}{,}

\newcommand{\zav}[1]{\left(#1\right)}
\newcommand{\hzav}[1]{\left[#1\right]}
\newcommand{\szav}[1]{\left\{#1\right\}}
\newlength\staretab
\newcommand{\Teff}{\mbox{$T_\mathrm{eff}$}}
\newcommand\de{\text{d}}
\newcommand\x[1]{\ensuremath{#1_\text{X}}}
\newcommand\lx{\ensuremath{\x L}}

\makeatletter
\def\sgn{\mathop{\operator@font sgn}\nolimits}
\makeatother

\begin{document}

\title{X-ray irradiation of the winds in binaries with massive components}

\author{J.~Krti\v{c}ka\inst{1} \and J. Kub\'at\inst{2} \and
I.~Krti\v{c}kov\'a\inst{1}}


\institute{\'Ustav teoretick\'e fyziky a astrofyziky, Masarykova univerzita,
           Kotl\'a\v rsk\' a 2, CZ-611\,37 Brno, Czech
           Republic
           \and
           Astronomick\'y \'ustav, Akademie v\v{e}d \v{C}esk\'e
           republiky, Fri\v{c}ova 298, CZ-251 65 Ond\v{r}ejov, Czech Republic}

\date{Received}

\abstract{Binaries with hot massive components are strong X-ray sources. Besides
the intrinsic X-ray emission of individual binary members originating in their
winds, X-ray emission stems from the accretion on the compact companion or
from wind collision. Since hot star winds are driven by the light absorption
in the lines of heavier elements, wind acceleration is sensitive to the
ionization state. Therefore, the over-ionization induced by external X-ray source
strongly influences the winds of individual components.} {We studied the effect
of external X-ray irradiation on hot star winds.} {We used our kinetic
equilibrium (NLTE) wind
models to estimate the influence of external X-ray ionization for different
X-ray luminosities and source distances. The models are calculated for
parameters typical of O stars.} {The influence of X-rays is given by the X-ray
luminosity, by the optical depth between a given point and the X-ray source, and
by a distance to the X-ray source. Therefore, the results can be interpreted in 
the diagrams of X-ray luminosity vs.~the optical depth parameter.
X-rays are negligible
in binaries with low X-ray luminosities or at large distances from the X-ray
source. The influence of X-rays is stronger for higher X-ray luminosities and in
closer proximity of the X-ray source. There is a forbidden area with high
X-ray luminosities and low optical depth parameters, where the X-ray ionization
leads to wind inhibition. There is excellent agreement between the positions of
observed stars in these diagrams and our predictions. All wind-powered
high-mass X-ray binary (HMXB)
primaries lie outside the forbidden area. Many of them lie close to the border
of the forbidden area, indicating that their X-ray luminosities are
self-regulated. We discuss the implications of our work for other binary types.}
{X-rays have a strong effect on the winds in binaries with hot components. The
magnitude of the influence of X-rays can be estimated from the position of a
star in the diagram of X-ray luminosity vs.~the optical depth parameter.}


\titlerunning{The effect of the X-ray irradiation on the wind in binaries with
hot component}

\authorrunning{J.~Krti\v{c}ka et al.}
\maketitle

\section{Introduction}

Binary stars with hot massive components are well-known X-ray sources. A high-mass
X-ray binary (HMXB) is a system consisting of a massive luminous hot star and a
compact object (either a neutron star or a black hole). In these systems a
fraction of the primary star matter, coming from either a hot star wind or a
Roche-lobe overflow, is trapped in the gravitational well of the compact object
\citep{davos,laheupet}. These objects belong to the most powerful stellar X-ray
sources with highest X-ray luminosities $\x
L\approx10^{37}-10^{38}\,\text{erg}\,\text{s}^{-1}$.

Be/X-ray binaries also constitute a hot star donor and a compact companion,
however, the matter is accreted from the circumstellar disk of the Be-star
primary in these objects \citep[see][for a review]{reig}. The X-ray luminosities
of Be/X-ray binaries may reach up to $\x
L\approx10^{37}\,\text{erg}\,\text{s}^{-1}$. Finally, the X-ray emission may
originate in the wind collision in binaries with non-degenerate hot components
\citep{usaci,kapitan,skoda}. This kind of interaction results in relatively weaker X-ray
sources with $\x L\approx10^{32}-10^{33}\,\text{erg}\,\text{s}^{-1}$
\citep[e.g.,][]{sane,igorkar}.

The hydrodynamics of the circumstellar matter in binary systems is very complex
and, in fact, constitutes a time-dependent problem of radiative hydrodynamics
\citep{frc,blok,felan}. Therefore, most studies concentrate on the dynamics
of the matter leading to the X-ray generation. However, there is also a feedback
effect of the produced X-rays on the stellar wind of hot component(s).

The feedback effect is connected with X-ray photoionization of the stellar wind
\citep{mav,vanlok,viteal}. Because the stellar wind of hot stars is driven by
light absorption in lines of heavier elements (e.g., carbon, nitrogen, and
iron), photoionization may affect wind driving \citep{viteal}. This may even result
in wind stagnation and possibly fall back on the source star
\citep{velax1}.

The problem of the X-ray photoionization should be treated using 3D
hydrodynamical simulations coupled with radiative transfer and statistical
equilibrium equations. These kinds of calculations are likely beyond the possibilities of
current computers. To make the problem more tractable, one either uses
hydrodynamical simulations with a simplified form of the radiative force
\citep[e.g.][]{frc,blok,felan,hadrvitr,pars} or 1D stationary wind models that
account for the influence of X-ray photoionization on the level populations
\citep[e.g.,][]{velax1}. Time dependent hydrodynamical simulations can explain
the wind structure in binaries in detail, but they cannot predict for which
binary parameters the effect of X-ray irradiation becomes important.

We concentrate on a detailed solution of statistical equilibrium equations and
neglect the time-dependent phenomena. We provide a grid of stationary 1D wind
models with external X-ray irradiation for a wide grid of O star parameters. The
X-ray irradiation is parameterized by the corresponding X-ray luminosity and the
distance of the X-ray source from the star with wind. The wind is only calculated
in the direction towards the X-ray source. This approach is sufficient to infer
the effect of X-ray irradiation on the radiative force and yields a reliable
estimate of the wind velocity in the direction of the X-ray source.

\section{Influence of X-rays on the wind ionization}

The influence of the X-rays on the ionization state of the wind is stronger
the lower the X-ray optical depth between the X-ray source and a given point
in the wind. The radial optical depth between the X-ray source at radius $D$ and
point at radius $r$ is given by 
\begin{equation}
\label{zamlhou}
\tau_\nu(r)=\left|\int_r^D\kappa_\nu(r')\rho(r')\,\de r'\right|,
\end{equation}
with distances measured with respect to the centre of wind-losing star.
The absolute value in this equation ensures that the optical depth is positive
even for $r>D$. The X-ray optical depth $\tau_\nu(r)$ depends on density, which
follows from the continuity equation as $\rho(r)=\dot M/(4\pi r^2 \varv_r)$,
where $\dot M$ and $\varv_r$ are the wind mass-loss rate and the radial
velocity, respectively. Assuming that the X-ray mass-absorption coefficient
$\kappa_\nu$ is spatially constant and that the wind already reached its
terminal velocity $\varv_\infty$, Eq.~\eqref{zamlhou} can be written as
\begin{equation}
\label{zamlhoup}
\tau_\nu(r)\approx
\frac{\kappa_\nu\dot M}{4\pi \varv_\infty}
\left| {\frac{1}{r}-\frac{1}{D}} \right| .
\end{equation}
The wind opacity in the X-ray domain mostly stems from the direct and Auger
ionization. Eq.~\eqref{zamlhoup} may motivate one to plot stars in the $\x
L-\tau_\nu$ diagram to estimate the influence of X-rays on the wind ionization
state \citep[][Fig.\,4]{velax1}.

Inspired by Eq.~\eqref{zamlhoup},
we introduce a non-dimensional X-ray optical depth parameter
\begin{equation}
\label{tx}
\x t=\frac{\dot M}{\varv_\infty}\zav{\frac{1}{R_*}-\frac{1}{D}}
\zav{\frac{10^3\,\text{km}\,\text{s}^{-1}\,1\,R_\odot}
{10^{-8}\,{M}_\odot\,\text{year}^{-1}}}
\end{equation}
to characterize the radial optical depth between the stellar surface and the
X-ray source. This parameter estimates the influence of X-rays on the ionization
state in the region of the critical point where the wind mass-loss rate is
determined. The critical point is defined as a point where the speed of the
Abbott waves, the fastest waves in the wind, is equal to the wind velocity
\citep{abbvln,owor}. If the matter is highly overionized close to the critical
point, then the wind is not accelerated radiatively since higher ions are not
efficient wind drivers \citep[e.g.][]{velax1}. This leads to wind inhibition.
Therefore, we expect a strong influence of X-rays in stars with large X-ray luminosity {\lx} and low X-ray optical
depth parameter $\x t$, whereas we assume weak X-ray influence in stars
with low $\x L$ and high $\x t$. For medium values
of these parameters, the X-rays may just affect the wind velocity and lead
possibly to wind stagnation.

A description of the influence of X-ray irradiation on the wind state via the optical
depth parameter $\x t$ introduced in Eq.~\eqref{tx} can only be used for stars
with a similar ionization and density structure. For a more complete description
of this problem, one also has to account for the $n^2$ dependence of the
recombination that follows the X-ray ionization ($n$ is the electron number
density). Therefore, with the mean intensity of the radiation $J_\nu^\text{X}$
at the distance $d=|D-r|$ from the X-ray source \citep[see also][Eq.~4]{velax1},
\begin{equation}
\label{xneutron}
J_\nu^\text{X}=\frac{L_\nu^\text{X}}{16\pi^2d^2}\text{e}^{-\tau_\nu(r)},
\end{equation}
one can introduce the ionization
parameter 
\begin{equation}
\label{xi}
\xi(r)=\frac{1}{n d^2}\int L_\nu^\text{X}\text{e}^{-\tau_\nu(r)}\,\de\nu,
\end{equation}
which estimates the influence of X-ray radiation on the ionization equilibrium of
the wind. Here $L_\nu^\text{X}$ is the X-ray luminosity per unit of frequency
and the integration goes over the X-ray domain. In the optically thin limit this
gives the ionization parameter $\xi\sim\lx/(n d^2)$ introduced by \citet[][see
also \citealt{sekerka}]{tenci}, where $\x L=\int L_\nu^\text{X}\,\de\nu$.

\section{Description of the X-ray irradiated wind models}

For our calculations, we used spherically symmetric stationary wind models of
\citet{cmf1}. The radiative force was calculated using the solution of the
comoving frame (CMF) radiative transfer equation with occupation numbers derived
from the kinetic equilibrium (NLTE) equations. The model enables us to predict
the wind structure, including the wind mass-loss rate and terminal velocity,
consistently from global stellar parameters only.

The wind ionization and excitation state was calculated from the NLTE equations.
Ionic models for the NLTE calculations are based on the Opacity and Iron Project
results \citep{topt,zel0} and on the data described by \citet{pahole}. Part of
the ionic models was adopted from TLUSTY model atmosphere input files
\citep{ostar2003,bstar2006}. The line radiative force was derived using the
radiative flux calculated from the CMF radiative transfer equation \citep{mikuh}
with the actual occupation numbers calculated from the NLTE equations. The line
opacity data used in the line force calculation were extracted from the VALD
database (Piskunov et al. \citeyear{vald1}, Kupka et al. \citeyear{vald2}). The
occupation numbers derived from the NLTE equations were also used to calculate
the radiative cooling and heating terms \citep{kpp}. The emergent surface flux
(inner boundary condition) was taken from the H-He spherically symmetric NLTE
model stellar atmospheres of \citet[and references therein]{kub}. The resulting
hydrodynamical equations, that is, the continuity equation, equation of motion
(with CMF radiative force), and the energy equation (with radiative cooling and
heating), were solved iteratively together with NLTE and radiative transfer
equations to obtain the wind density, velocity, and temperature structure. 

Because the velocity field may
become non-monotonic in the presence of the external irradiation, we do not use the CMF procedure of the radiative force
calculation directly. Instead, as in \cite{velax1}, we calculate the ratio of
the CMF and Sobolev line force \citep[see][]{cmf1} for a model without external
X-ray irradiation and use this ratio to correct the Sobolev line force in the
models with external X-ray irradiation. The numerical test showed that this
approach introduces very minor difference (below 1\%) between models with CMF
line force and models with scaled Sobolev line force. Moreover, we use the model
atmosphere flux for the calculation of the Sobolev line force. Therefore, we
neglect nonlocal radiative coupling between absorption zones, which occurs in
the non-monotonic flows \citep{rybashumrem,felnik}.

The influence of the secondary component is only taken into account by inclusion
of external X-ray irradiation. This is supposed to originate in the wind
accretion on the compact companion or in the wind-wind collision. The X-ray
irradiation is modelled by an additional term in the mean intensity $J_\nu$ in
the form of Eq.~\eqref{xneutron}. The optical depth $\tau_\nu(r)$ along a radial
ray in the direction of the X-ray source is given by Eq.~\eqref{zamlhou}. The
frequency distribution of emergent X-rays $L_\nu^\text{X}$ is for simplicity
approximated by the power law $L_\nu^\text{X}\sim\nu^{-1}$ from 0.5 to 20~keV
\citep [cf.] []{viteal}. The total X-ray luminosity $\x L=\int
L_\nu^\text{X}\,\de\nu$ and the distance of the X-ray source $D$ are free
parameters of the model grid.

The absorption coefficient and the density in Eq.~\eqref{zamlhou} can be derived
directly from the actual model. However, to simplify the calculation of the
external irradiation term Eq.~\eqref{xneutron} in the presence of the kink in
the velocity law (see below), for the calculation of $J_\nu^\text{X}$ we use
density and absorption coefficient in the form of
\begin{equation}
\begin{split}
\rho(r)&=\frac{\dot M}{4\pi r^2 v(r)},\\
\varv(r)&=\min(\tilde \varv(r),\varv_\text{kink}),\\
\kappa_\nu(r)&=\tilde \kappa_\nu^\text{X},
\end{split}
\end{equation}
where $\tilde \kappa_\nu^\text{X}$ is the depth-independent approximation of
X-ray opacity, $\varv_\text{kink}$ is the velocity of the kink (see below) if it
is present in the models, and otherwise $\varv_\text{kink}=\infty$. The fits to
the wind velocity $\tilde \varv(r)$ and absorption coefficient
$\tilde\kappa_\nu^\text{X}$ are derived from the model with no external
irradiation. The wind velocity is fitted as
\citep{betyna}
\begin{multline}
\label{vrfit}
\tilde 
\varv
(r)=\hzav{\varv_1\zav{1-\frac{R_*}{r}}+\varv_2\zav{1-\frac{R_*}{r}}^2+
     \varv_3\zav{1-\frac{R_*}{r}}^3}\\\times
     \szav{1-\exp\hzav{\gamma\zav{1-\frac{r}{R_*}}^2}},
\end{multline}
where $\varv_1$, $\varv_2$, $\varv_3$, and $\gamma$ are free parameters of the
fit. The X-ray opacity per unit of mass averaged for radii $1.5\,R_*-5\,R_*$ is
approximated as
\begin{equation}
\label{kapafit}
\log\zav{\frac{\tilde \kappa_\nu^\text{X}}{1\,\text{cm}^2\,\text{g}^{-1}}}=
  \left\{\begin{array}{l}
    \min(a_1\log\lambda+b_1,\log a_0)\quad \lambda<\lambda_1,\\
    a_2\log\lambda+b_2,\quad \lambda>\lambda_1,\\
  \end{array}\right.\\
\end{equation}
where $\lambda_1=20.18$. The parameter $\lambda$ is non-dimensional, and
has the same value as the wavelength in units of \AA. Here $a_0$, $a_1$, $b_1$,
$a_2$, and $b_2$ are parameters of the fit.

The models were calculated in the direction towards the X-ray source. The
influence of the X-ray irradiation is strongest in this direction and it can
be expected to decrease for the rays with increasing angular distance from the
source.

\section{Calculated wind models}

\begin{table}
\caption{Stellar parameters of the model grid.}
\centering
\label{ohvezpar}
\begin{tabular}{rccrc}
\hline
\hline
& Model &$\Teff$ & $R_{*}$ & $M$  \\
& & $[\text{K}]$ & $[{R}_{\odot}]$ & $[{M}_{\odot}]$ \\
\hline supergiants (I)
& 300-1 & 30\,000 &22.4 & 28.8 \\
& 375-1 & 37\,500 &19.8 & 48.3 \\
& 425-1 & 42\,500 &18.5 & 70.3 \\
\hline main
& 300-5 & 30\,000 & 6.6 & 12.9 \\ sequence (V)
& 375-5 & 37\,500 & 9.4 & 26.8 \\
& 425-5 & 42\,500 &12.2 & 45.0 \\
\hline
\end{tabular}
\end{table}

\begin{table}
\caption{Parameters of the velocity law fits.}
\centering
\label{rychpar}
\begin{tabular}{rrrrr}
\hline
\hline
Model & $\varv_1$ [$\text{km}\,\text{s}^{-1}$] & $\varv_2$ [$\text{km}\,\text{s}^{-1}$]&
$\varv_3$ [$\text{km}\,\text{s}^{-1}$] & $\gamma$\\
\hline
300-1 & 2280 & -680& 0 & -610 \\
375-1 & 2700 & -980& 0 & -6300 \\
425-1 & 2720 & 0 & -380& -18600\\
\hline
300-5 & 2680 & -130& 0 & -11700\\
375-5 & 3380 & -1320& 0 & -6900\\
425-5 & 2770 & 0 & -510& -13700 \\
\hline
\end{tabular}
\end{table}

\begin{table}
\caption{Parameters of the opacity fits.}
\centering
\label{opapar}
\begin{tabular}{rrrrrr}
\hline
\hline
Model & $a_0$ & $a_1$ & $b_1$ & $a_2$ & $b_2$ \\
\hline
300-1 & 215 & 2.619 & -0.917 & 2.655 & -1.478\\
375-1 & 206 & 2.558 & -0.857 & 2.665 & -1.516\\
425-1 & 191 & 2.482 & -0.794 & 2.620 & -1.530\\
\hline
300-5 & 210 & 2.618 & -0.917 & 2.649 & -1.466\\
375-5 & 200 & 2.516 & -0.819 & 2.670 & -1.534\\
425-5 & 197 & 2.497 & -0.804 & 2.647 & -1.537\\
\hline
\end{tabular}
\end{table}

The adopted grid of stellar parameters corresponds to O stars in the effective
temperature range $30\,000-42\,5 00\,\text{K}$. The stellar masses and radii in
Table~\ref{ohvezpar} were calculated from the effective temperature using
empirical relations of \citet{okali}. These are, together with the X-ray
luminosity and the distance from the X-ray source, the parameters of the model
grid. The X-ray luminosities and the X-ray source distances were selected to
correspond to typical hot star X-ray sources. Wind parameters for the stars in
Table~\ref{ohvezpar} are given in \citet{fosfor}. The parameters of the velocity
law fit Eq.~\eqref{vrfit} and opacity fits Eq.~\eqref{kapafit} are given in
Tables \ref{rychpar} and \ref{opapar}, respectively.

\newrgbcolor{sarlatova}{0.79 0.37 0.37}
\newrgbcolor{plosar}{0.86 0.55 0.67}
\newrgbcolor{nejakazelena}{0.20 0.64 0.65}

\begin{figure*}[t]
\centering
\resizebox{0.49\hsize}{!}{\includegraphics{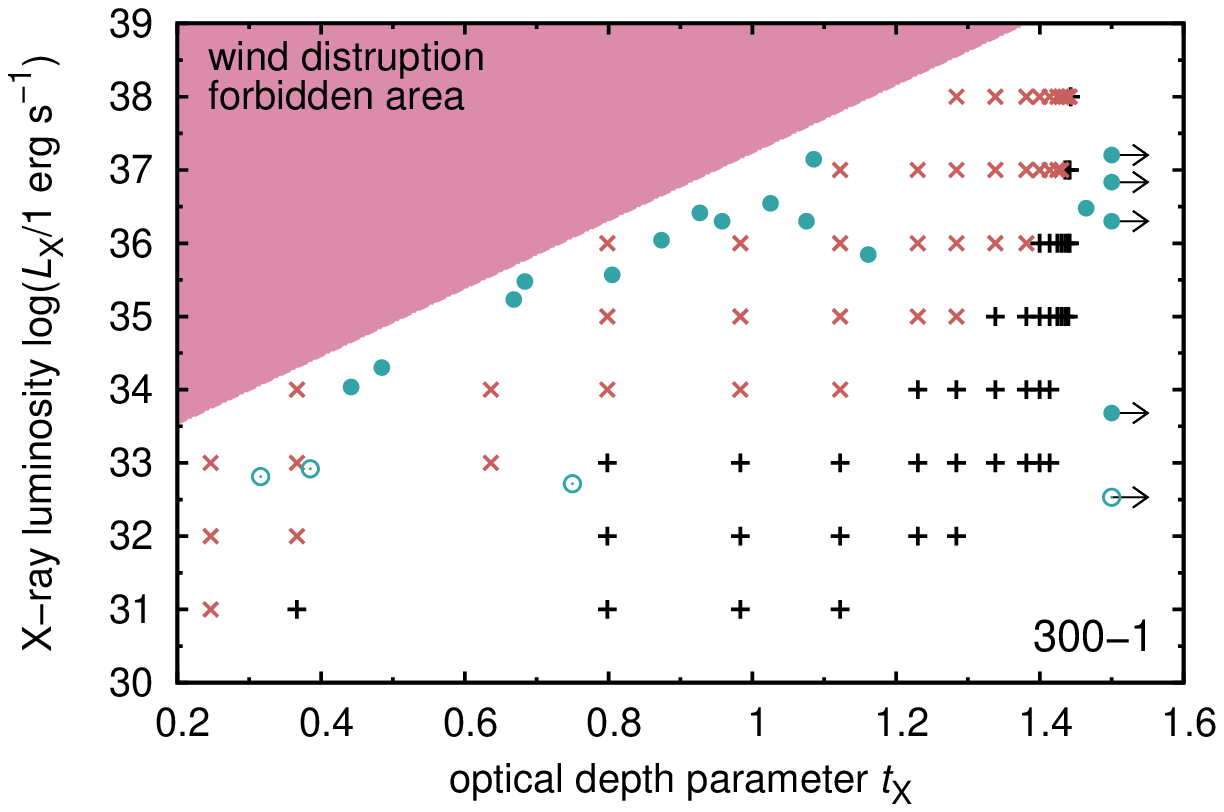}}
\resizebox{0.49\hsize}{!}{\includegraphics{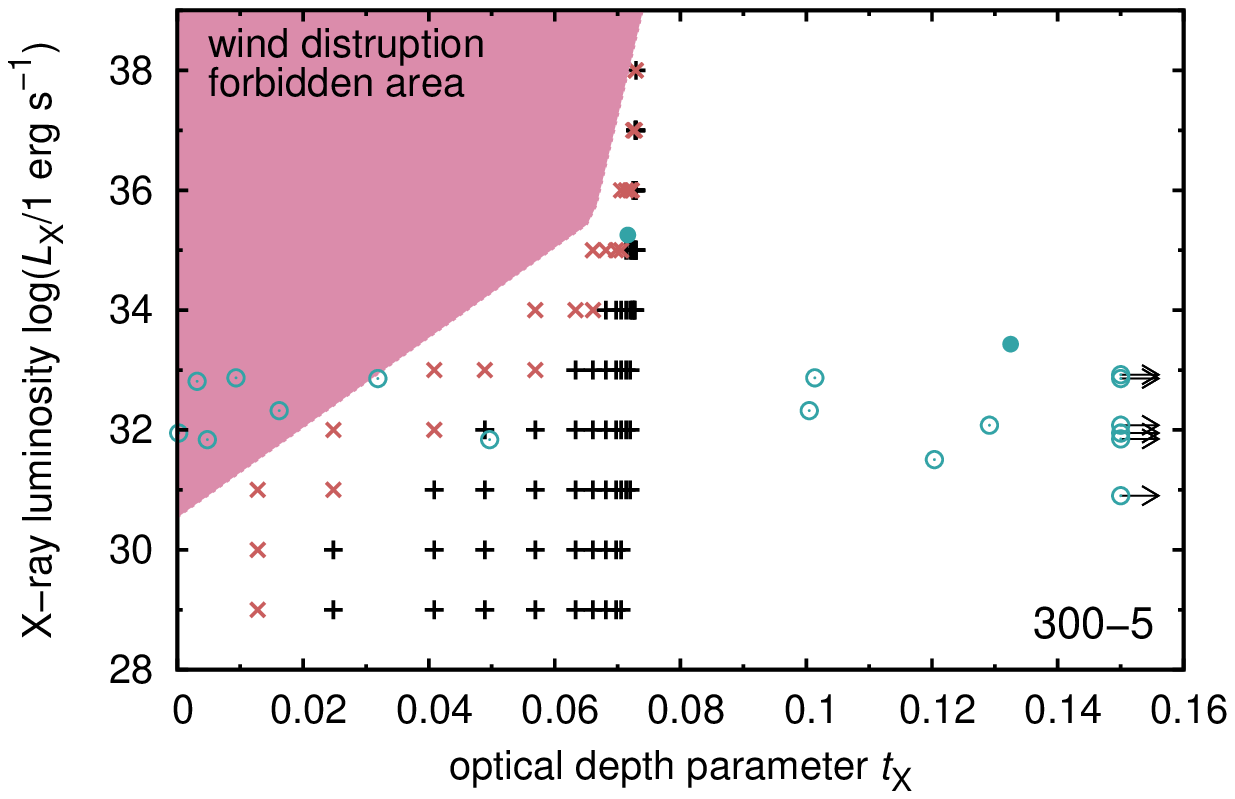}}
\resizebox{0.49\hsize}{!}{\includegraphics{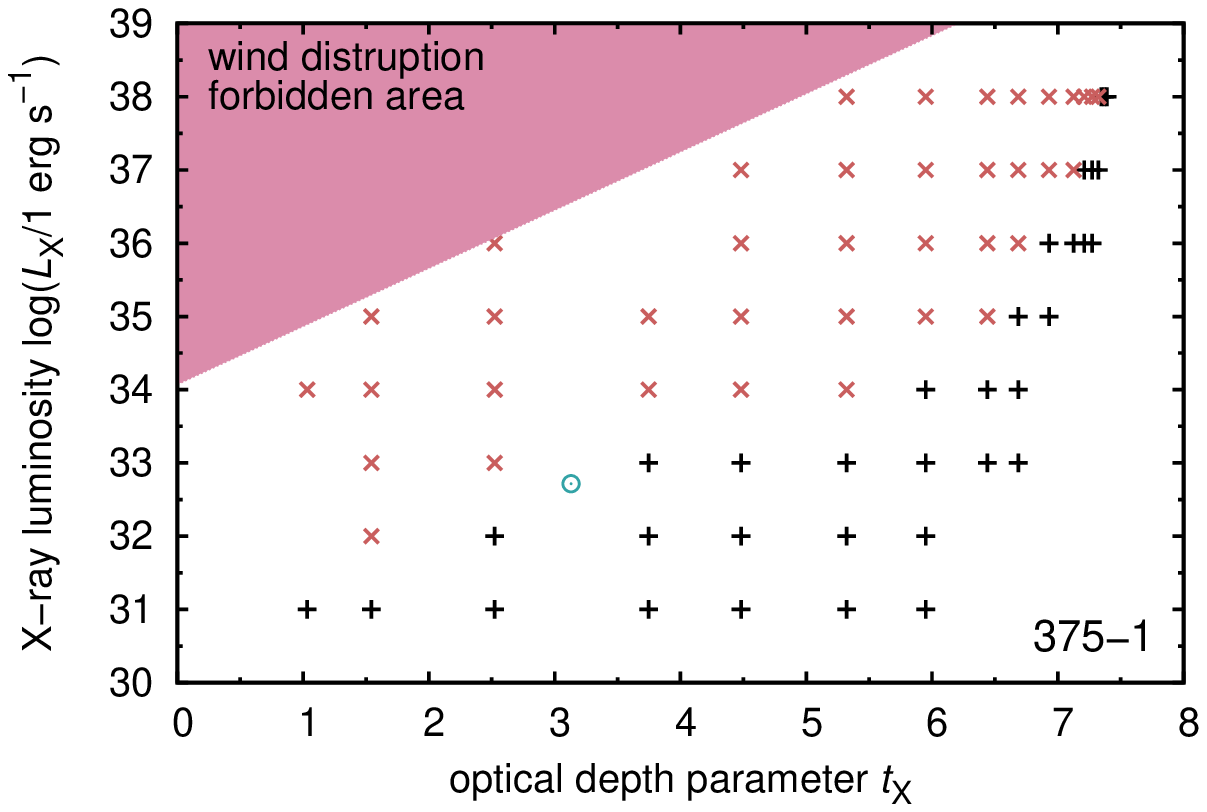}}
\resizebox{0.49\hsize}{!}{\includegraphics{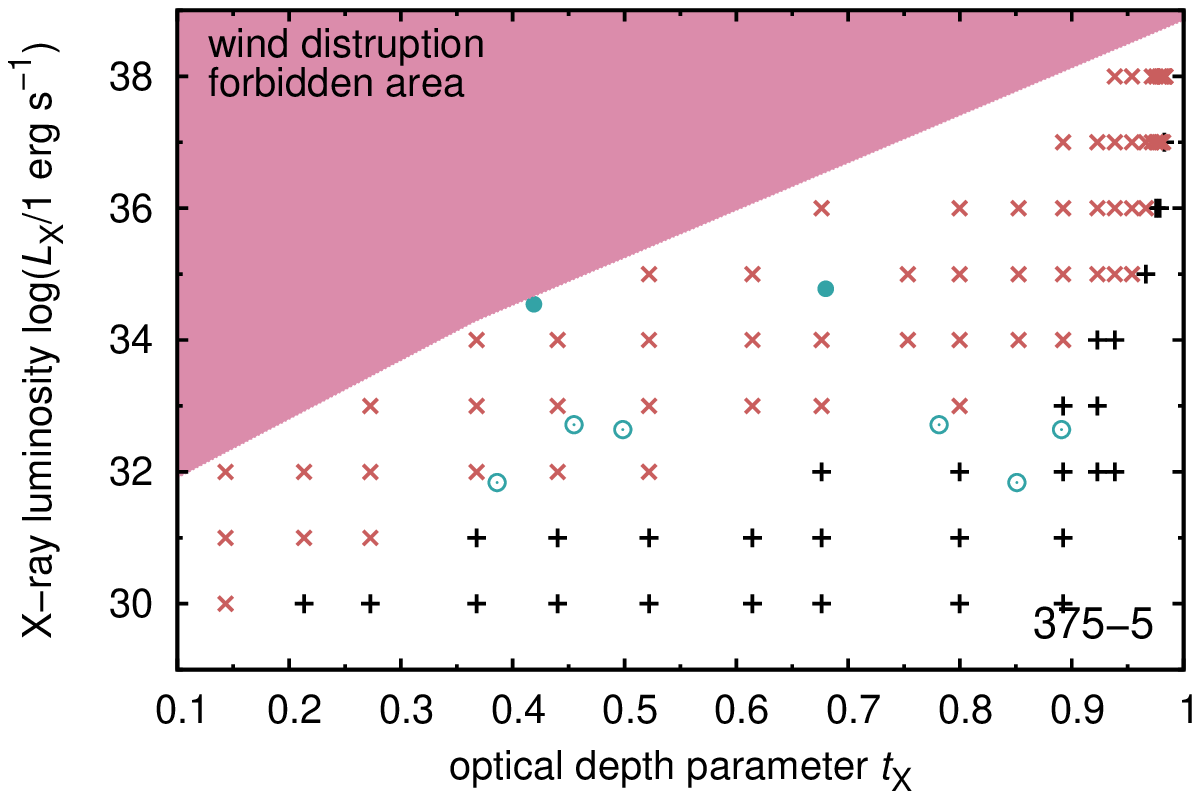}}
\resizebox{0.49\hsize}{!}{\includegraphics{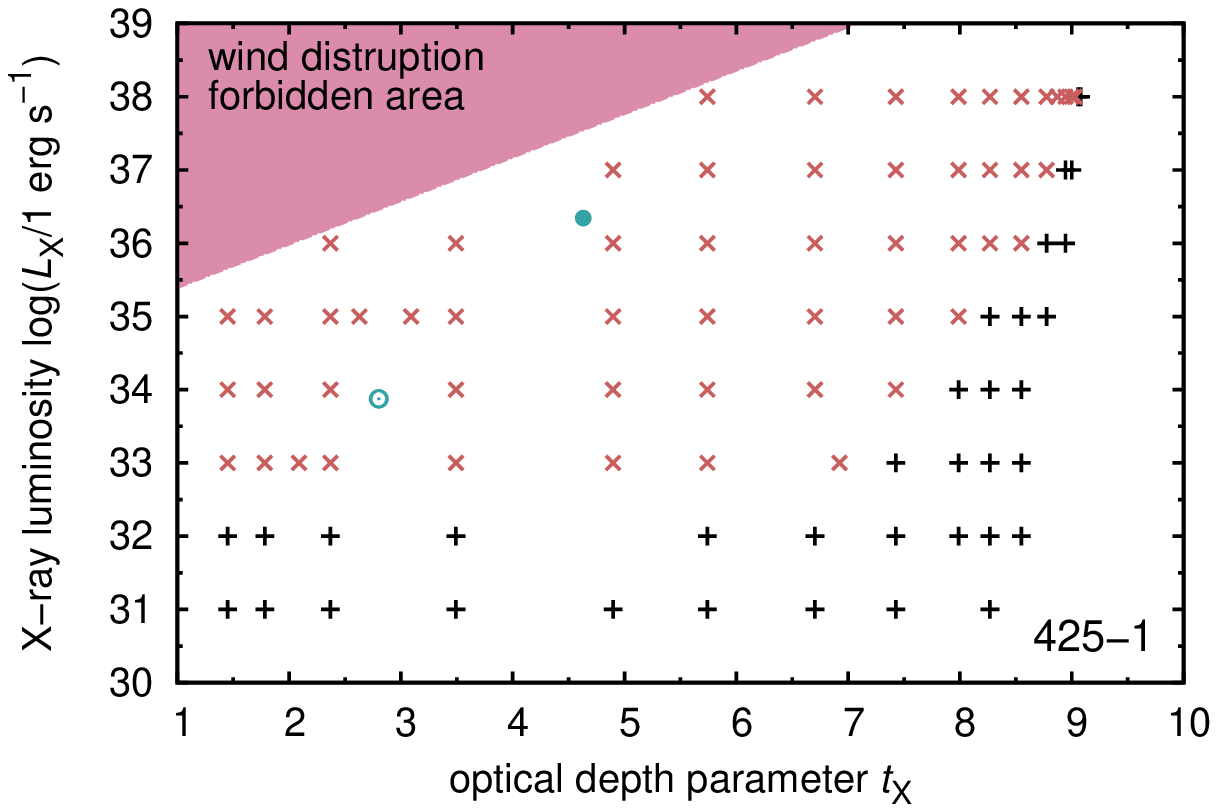}}
\resizebox{0.49\hsize}{!}{\includegraphics{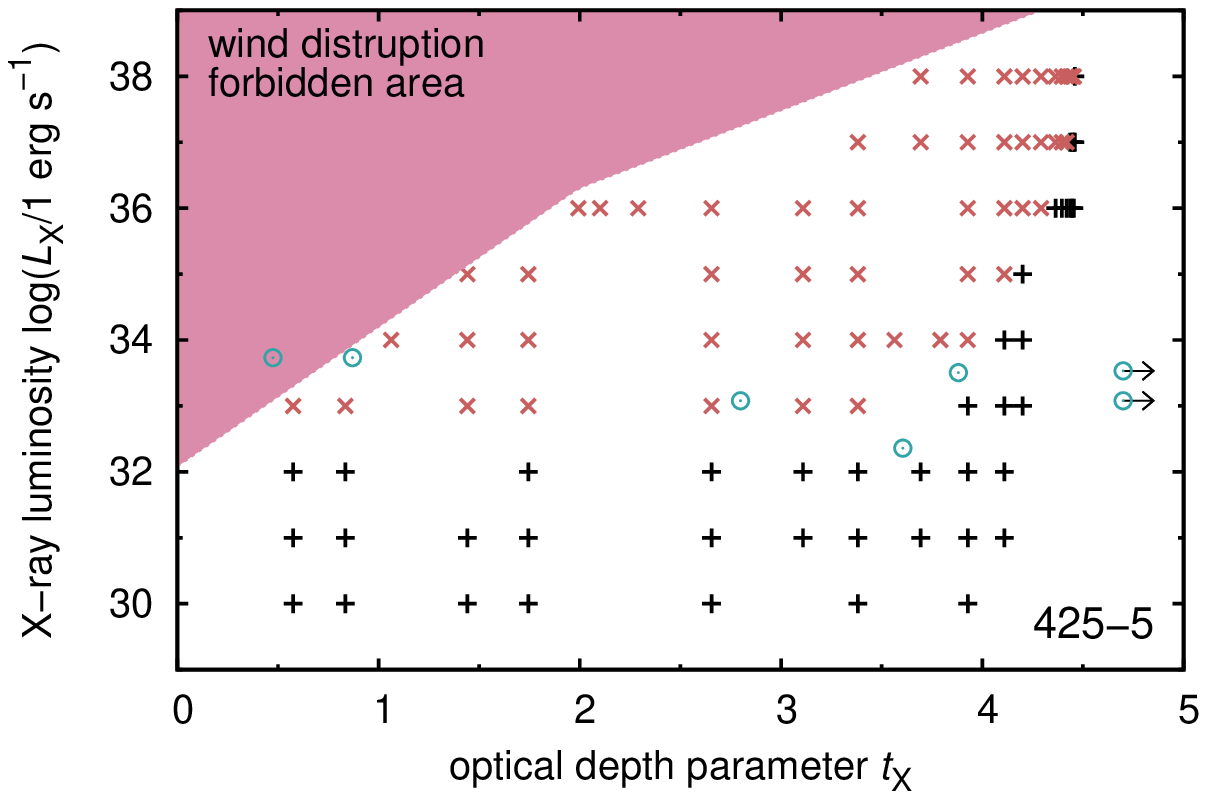}}
\caption{Regions with different effect of the X-ray irradiation in
the diagrams of X-ray luminosity (\lx) vs.~the optical depth parameter ($\x t$).
Graphs are
plotted for individual model stars in Table~\ref{ohvezpar}. Individual symbols
denote positions of: models with negligible influence of X-ray irradiation
(black plus, $\boldsymbol+$), models where X-ray irradiation leads to the
decrease of the wind terminal velocity (red cross,
{\sarlatova$\boldsymbol\times$}), non-degenerate components of HMXBs from
Table~\ref{neutron} (filled circles, {\nejakazelena\Large\raise
-2pt\hbox{\textbullet}}), and individual components of X-ray binaries from
Table~\ref{dvojhve} (empty circles, {\nejakazelena$\boldsymbol\odot$}). The
regions of the $\lx-\x t$ parameters that lead to the wind inhibition are
denoted using the shaded area ({\plosar\protect\rule{8pt}{8pt}}).}
\label{txobr}
\end{figure*}

\begin{figure}[ht]
\centering
\resizebox{0.33\hsize}{!}{\includegraphics{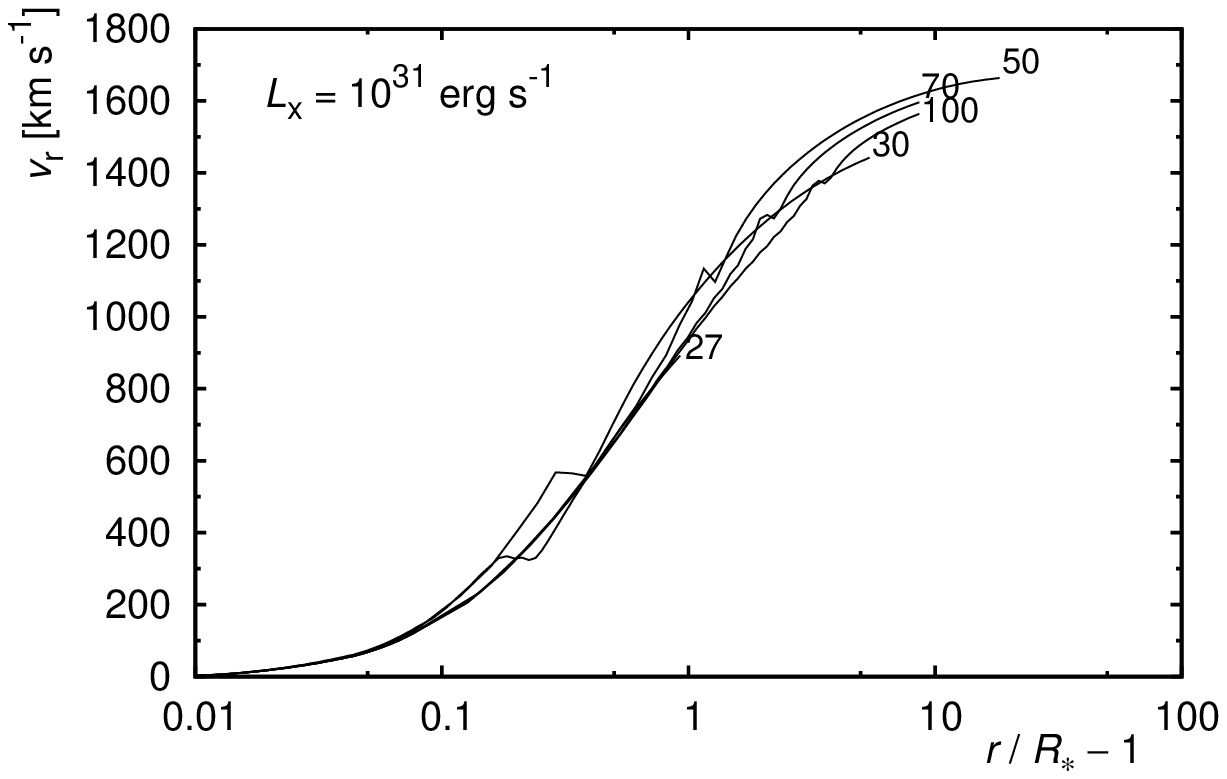}}
\resizebox{0.33\hsize}{!}{\includegraphics{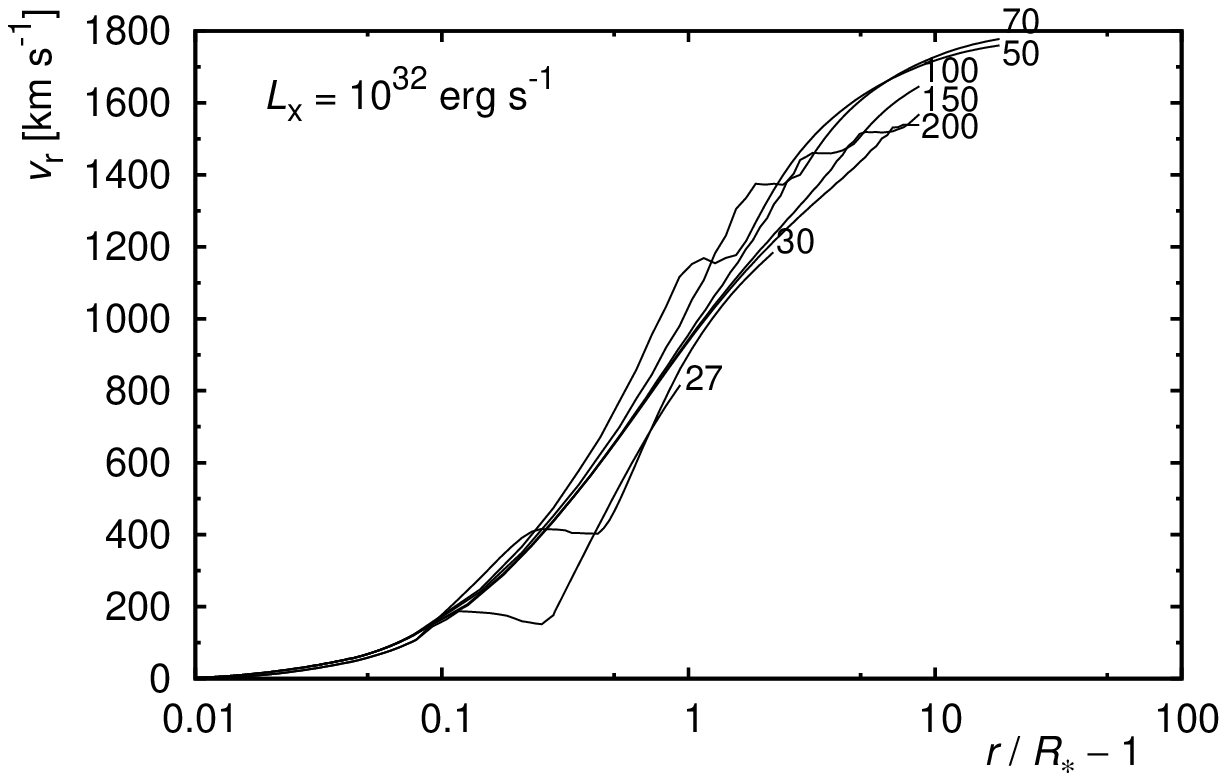}}
\resizebox{0.33\hsize}{!}{\includegraphics{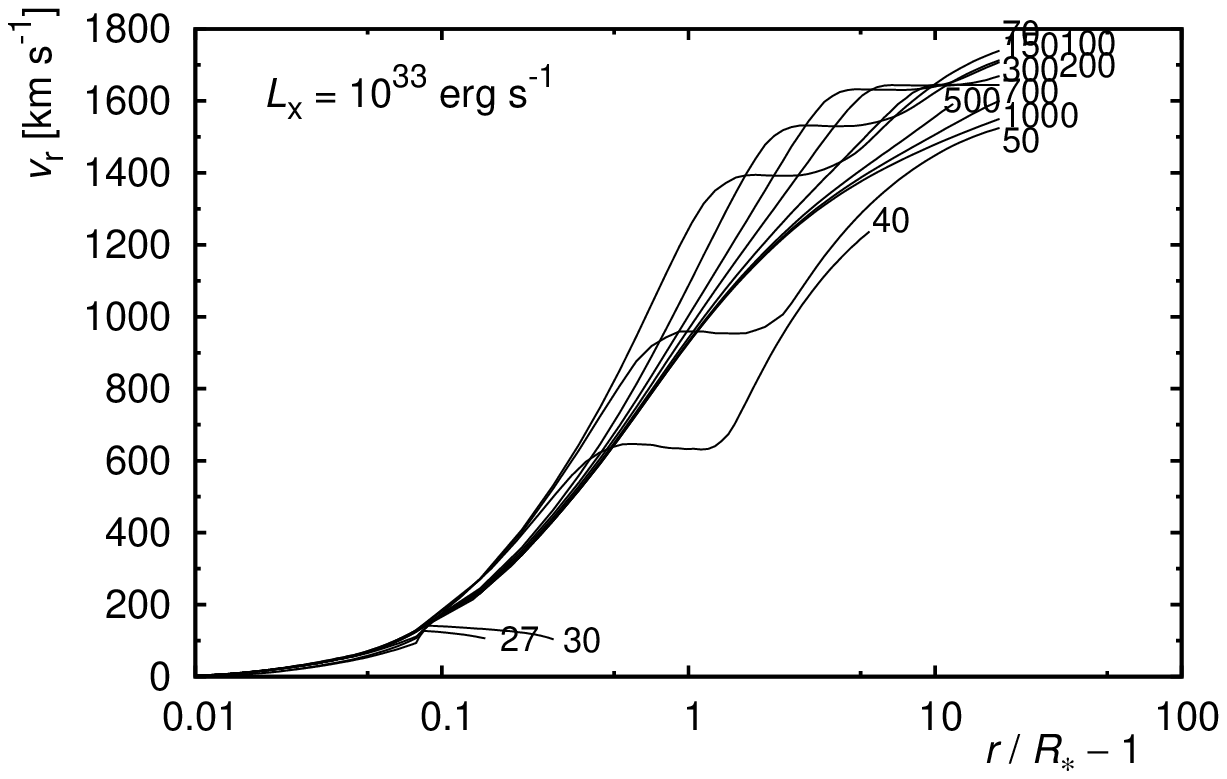}}
\resizebox{0.33\hsize}{!}{\includegraphics{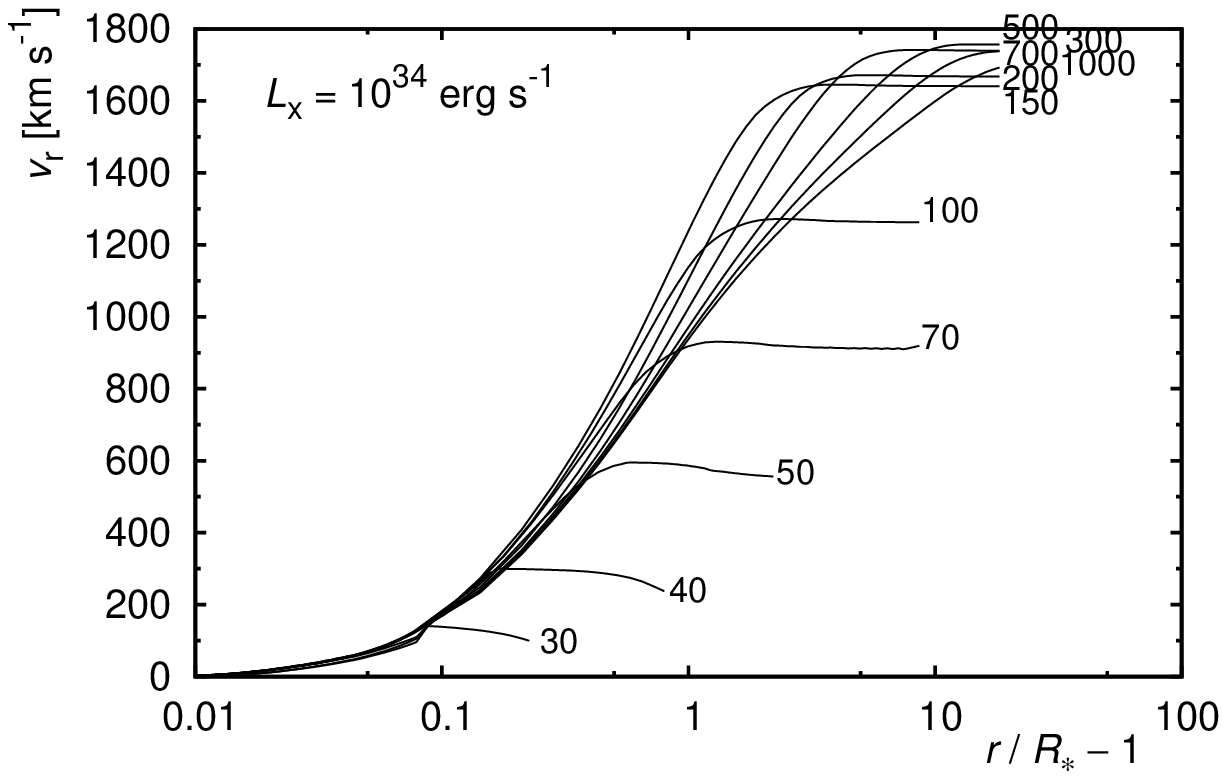}}
\resizebox{0.33\hsize}{!}{\includegraphics{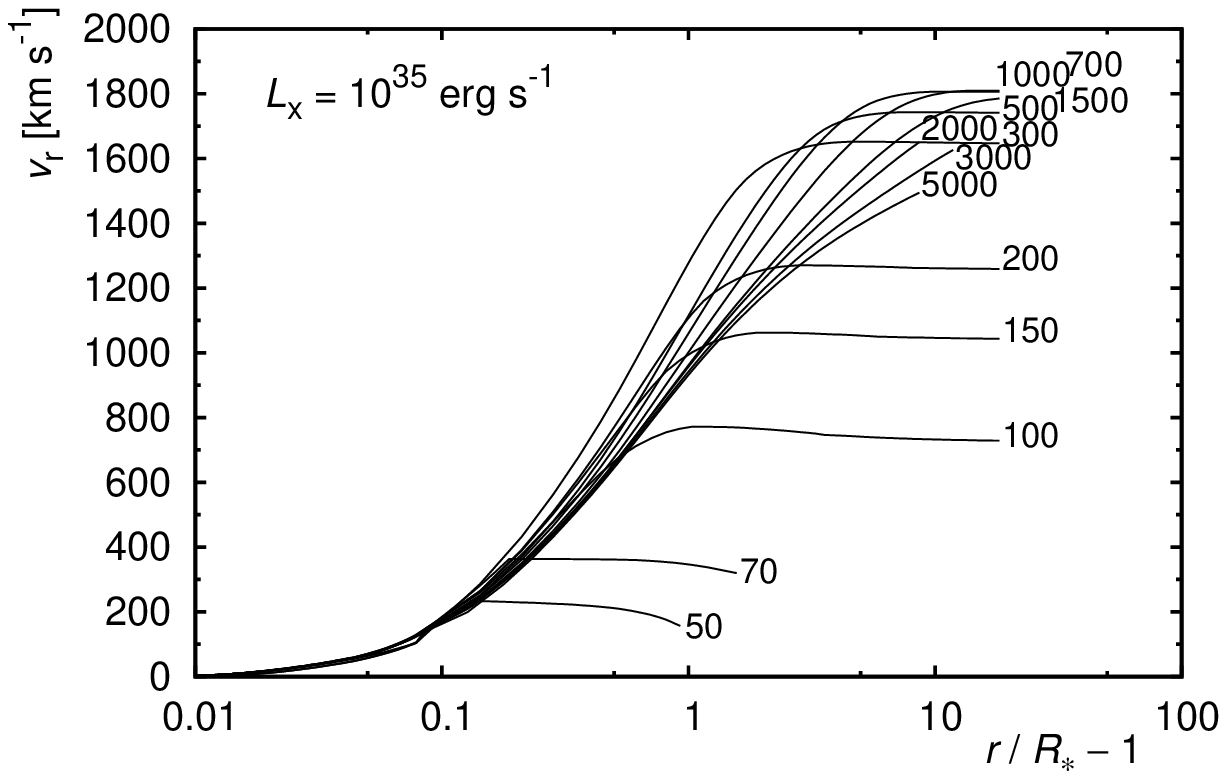}}
\resizebox{0.33\hsize}{!}{\includegraphics{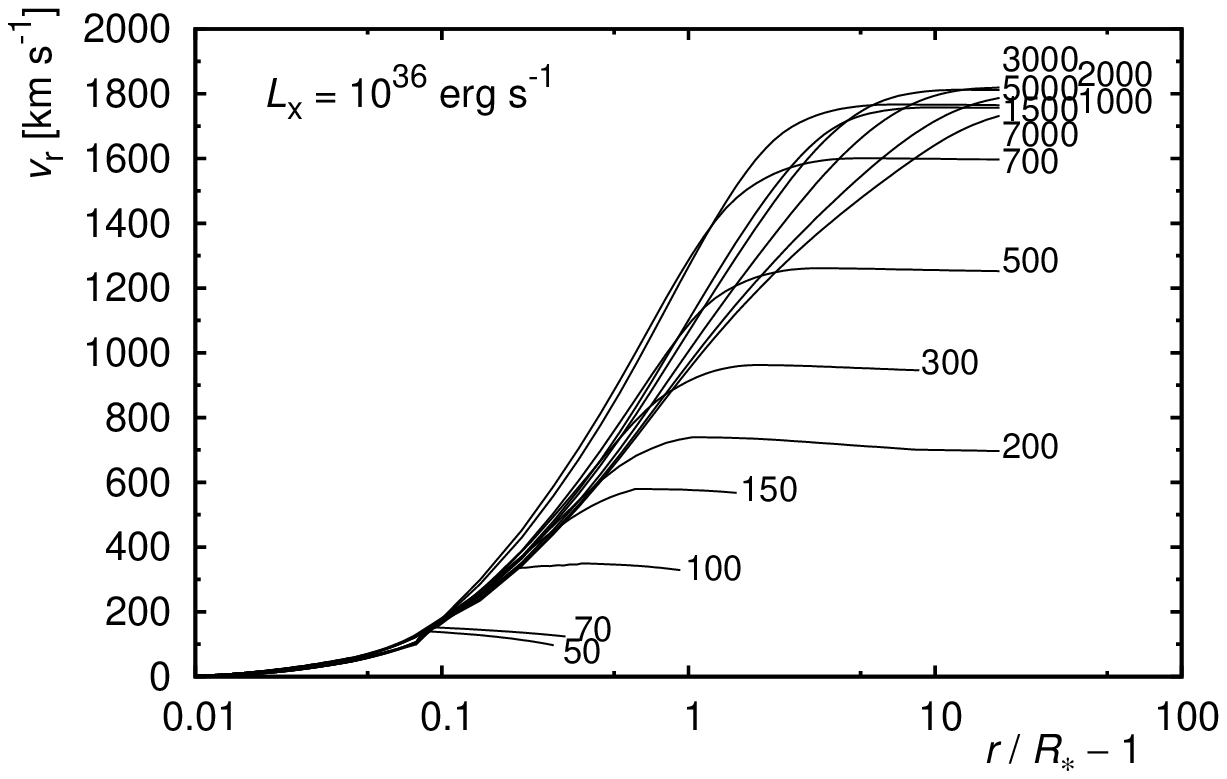}}
\resizebox{0.33\hsize}{!}{\includegraphics{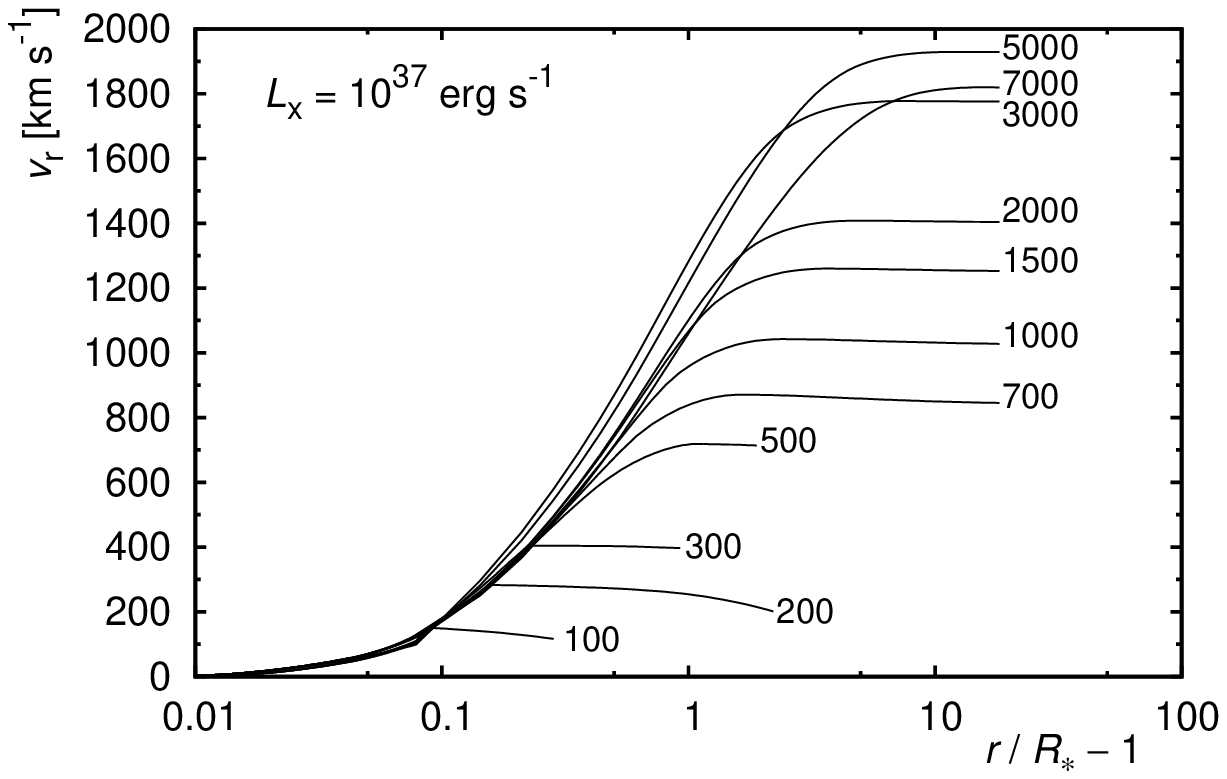}}
\resizebox{0.33\hsize}{!}{\includegraphics{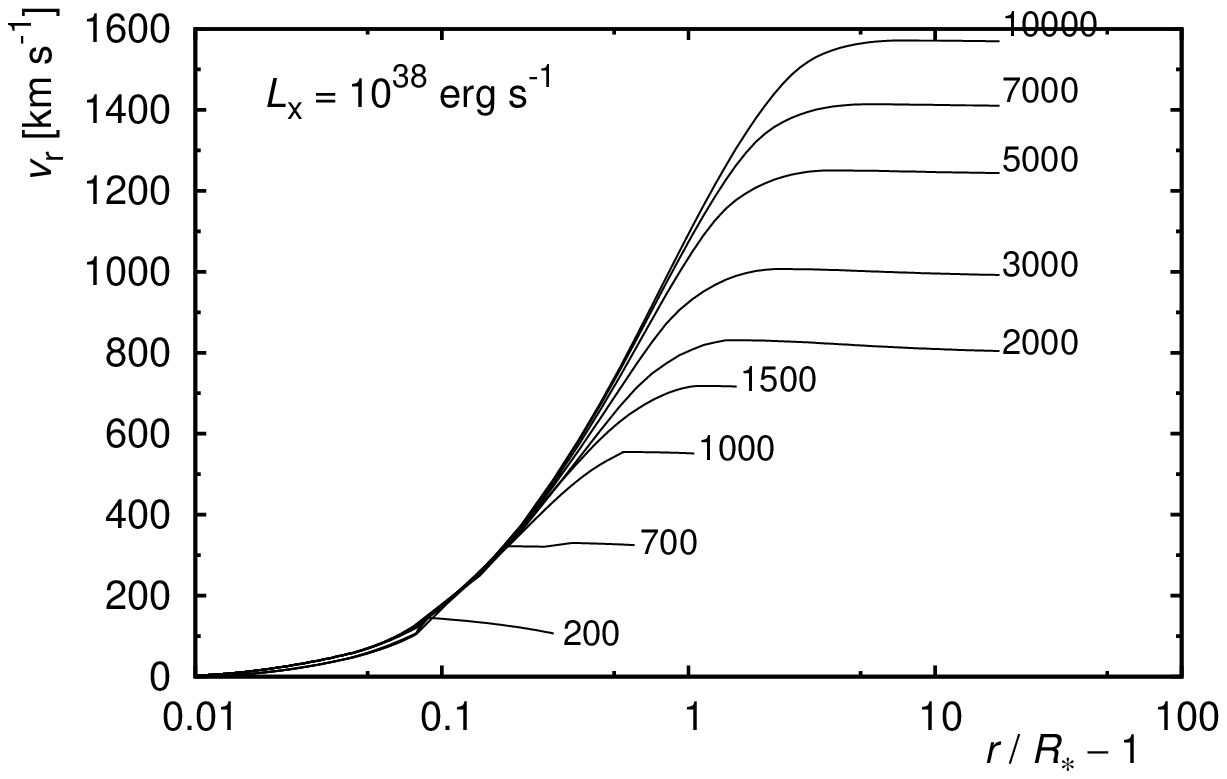}}
\caption{Dependence of the radial velocity on the radius in the wind of
300-1 star model. Each graph corresponds to different luminosity of additional
X-ray source \lx\ labelled in the graph. Each radial velocity plot is labelled
with a corresponding X-ray source distance $D$ in $R_\odot$.}
\label{300-1vr}
\end{figure}

\begin{figure}[ht]
\centering
\resizebox{0.33\hsize}{!}{\includegraphics{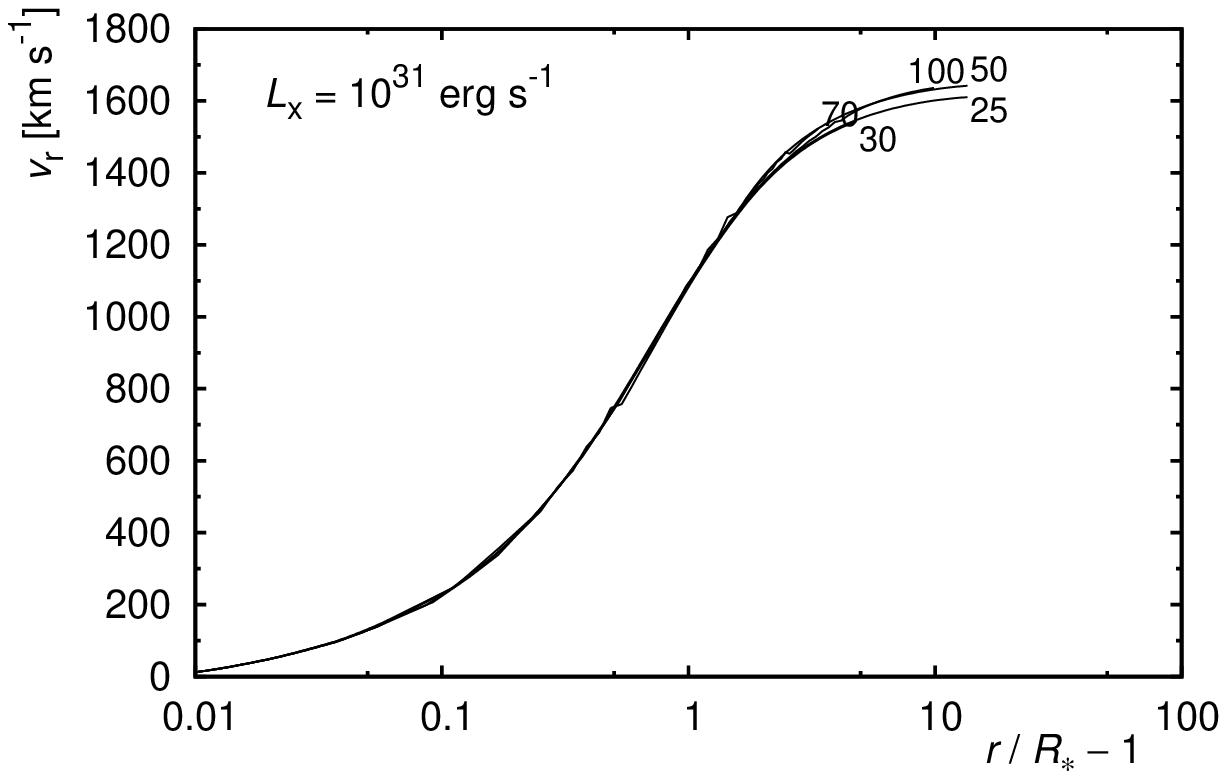}}
\resizebox{0.33\hsize}{!}{\includegraphics{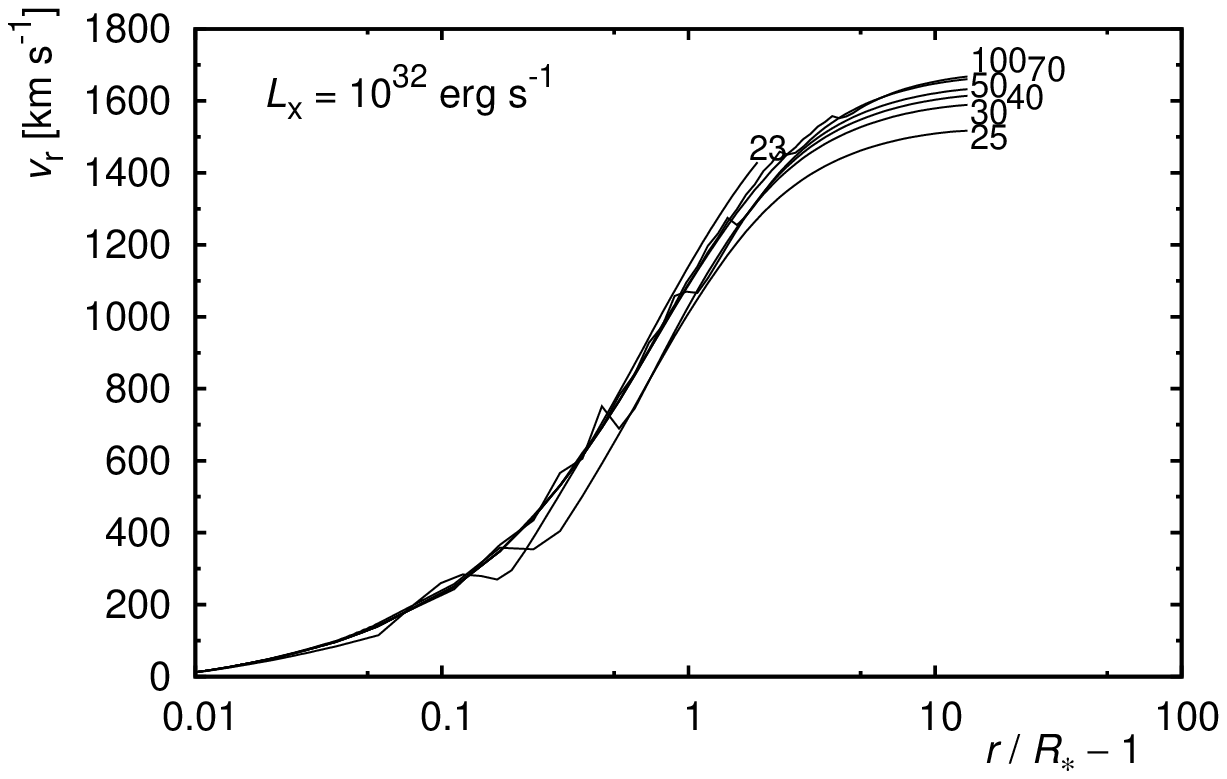}}
\resizebox{0.33\hsize}{!}{\includegraphics{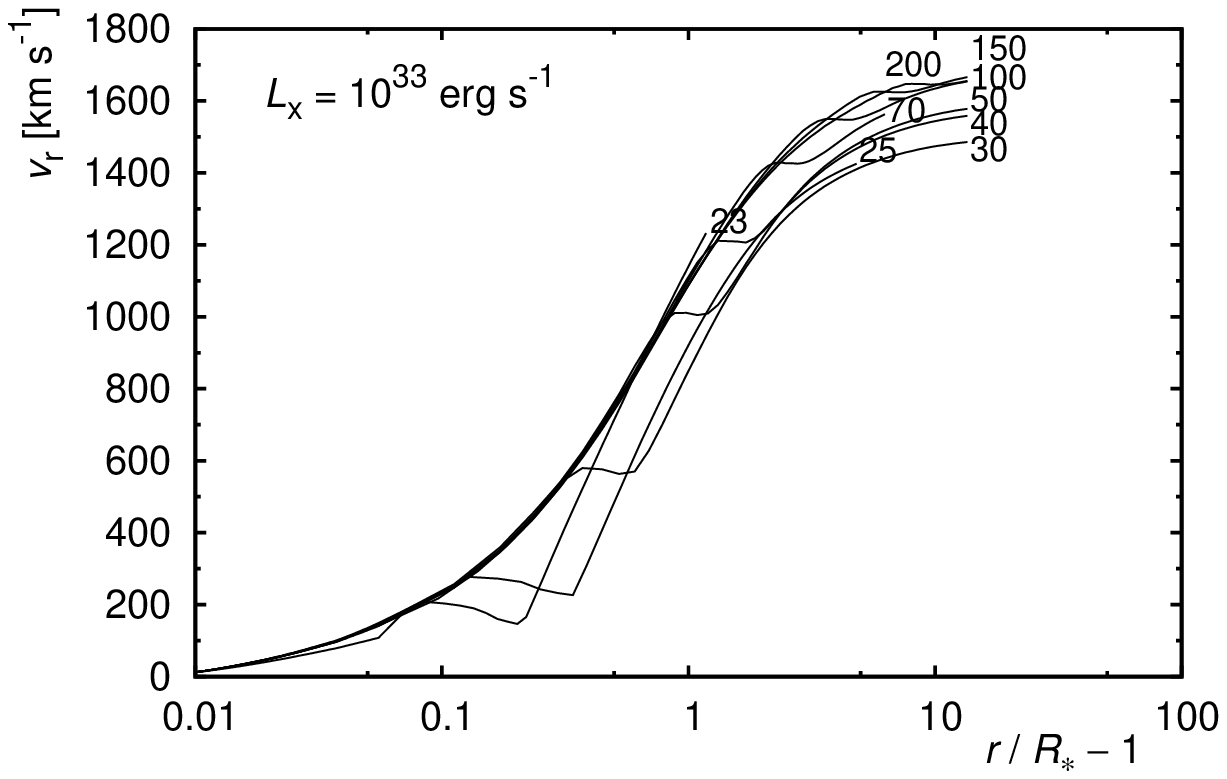}}
\resizebox{0.33\hsize}{!}{\includegraphics{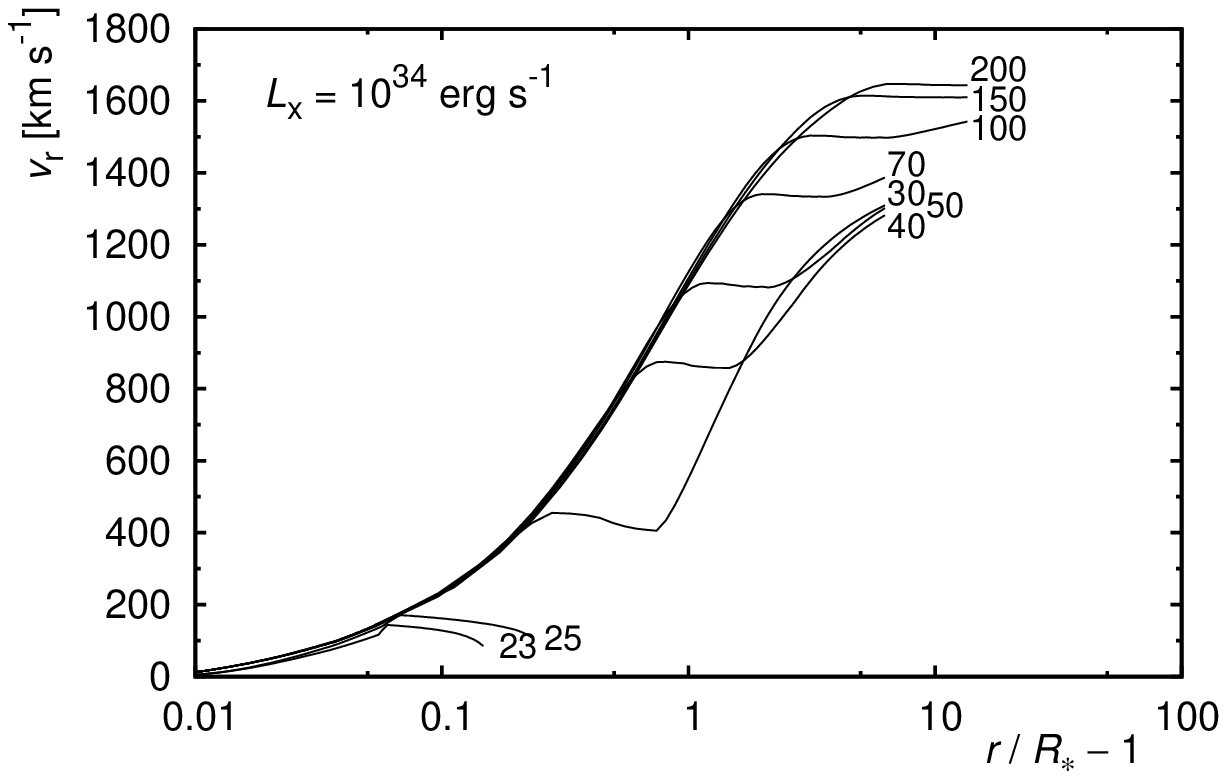}}
\resizebox{0.33\hsize}{!}{\includegraphics{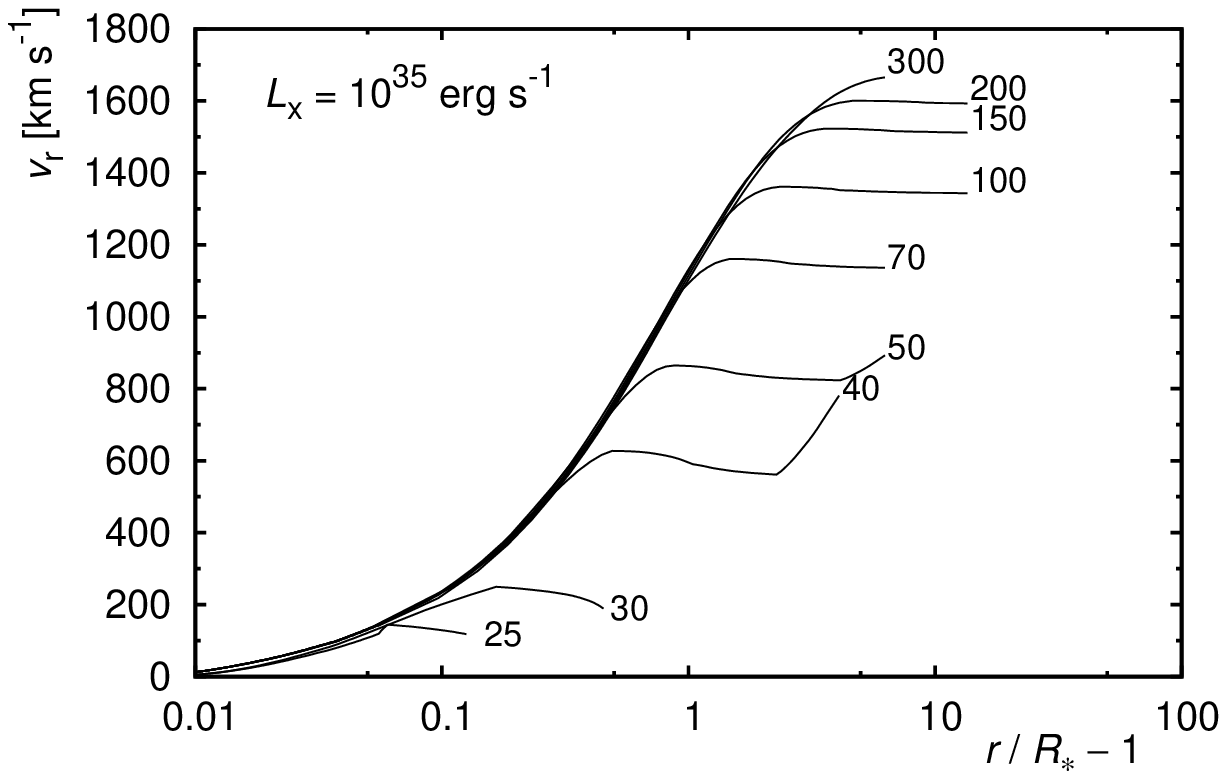}}
\resizebox{0.33\hsize}{!}{\includegraphics{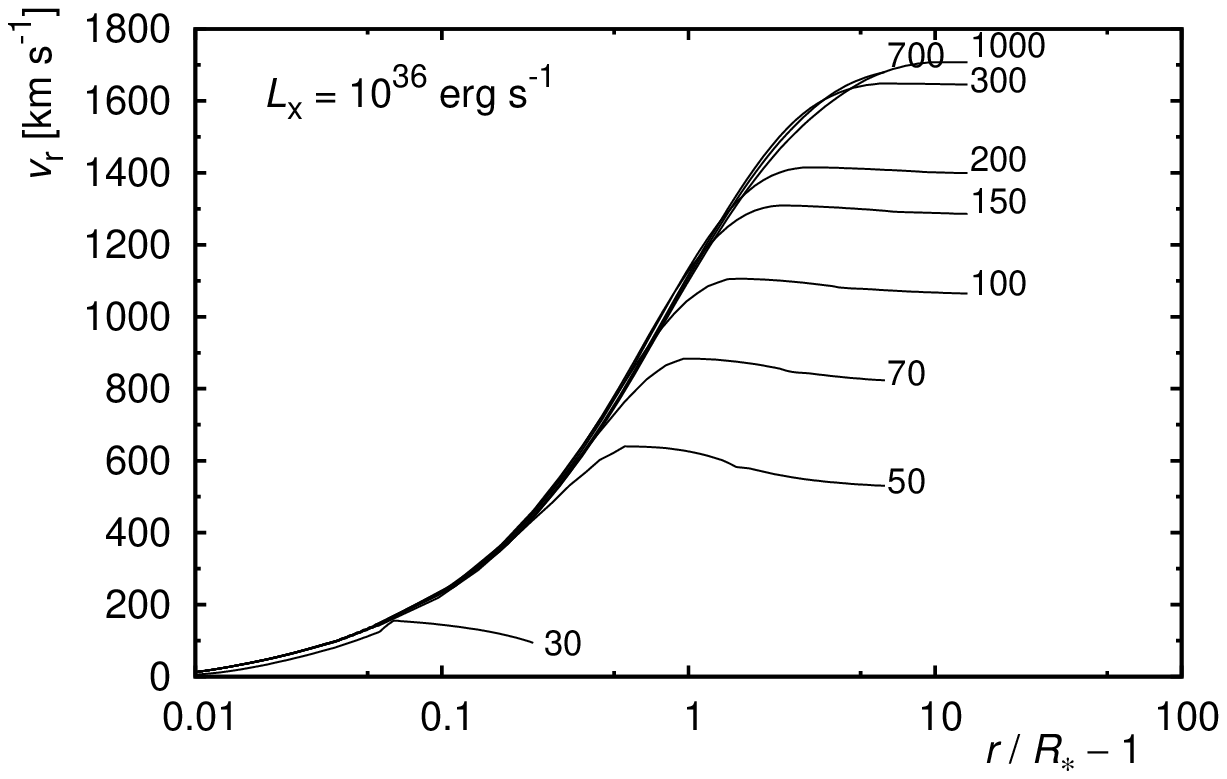}}
\resizebox{0.33\hsize}{!}{\includegraphics{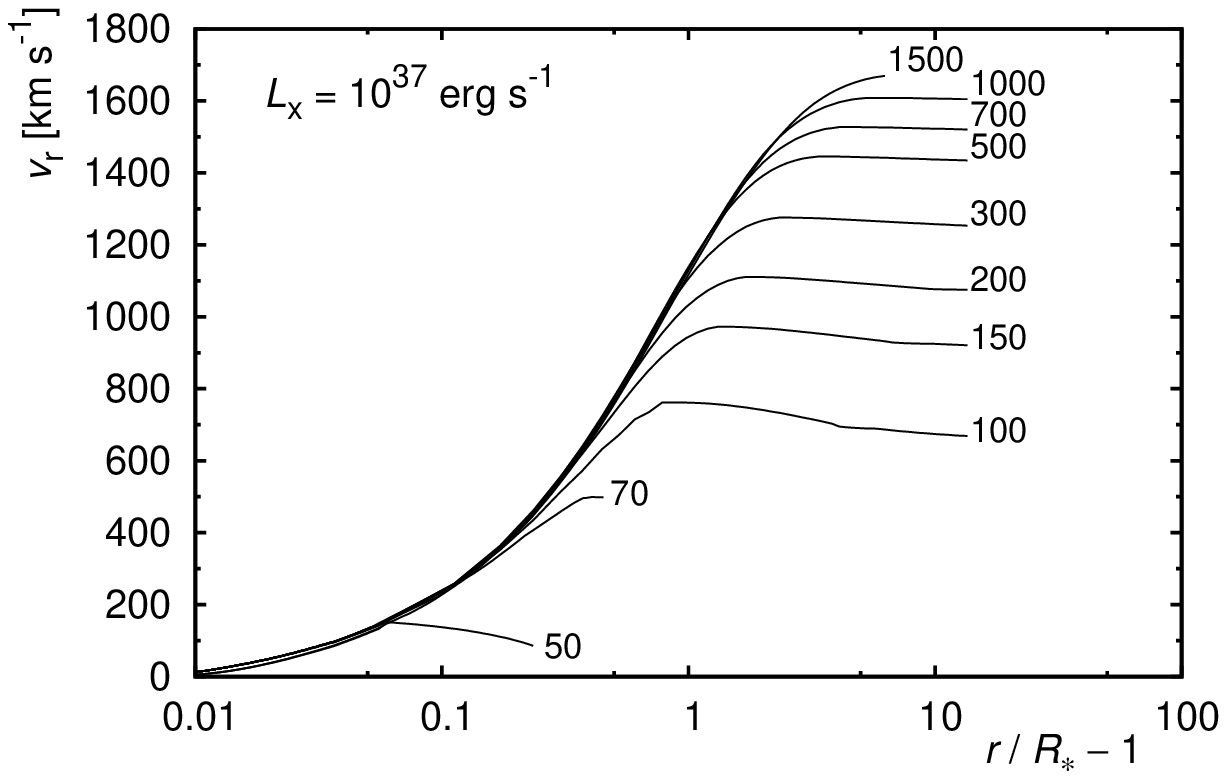}}
\resizebox{0.33\hsize}{!}{\includegraphics{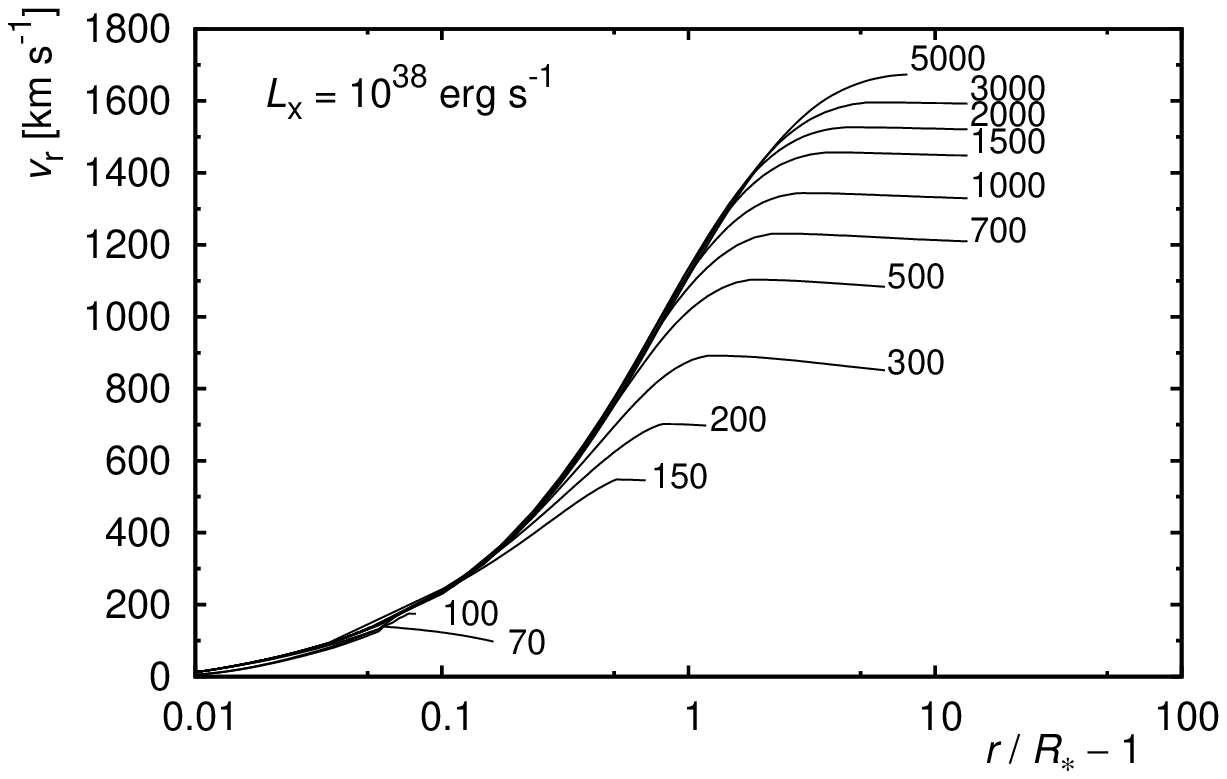}}
\caption{Same as Fig.~\ref{300-1vr}, except for a model star 375-1.}
\label{375-1vr}
\end{figure}

\begin{figure}[ht]
\centering
\resizebox{0.33\hsize}{!}{\includegraphics{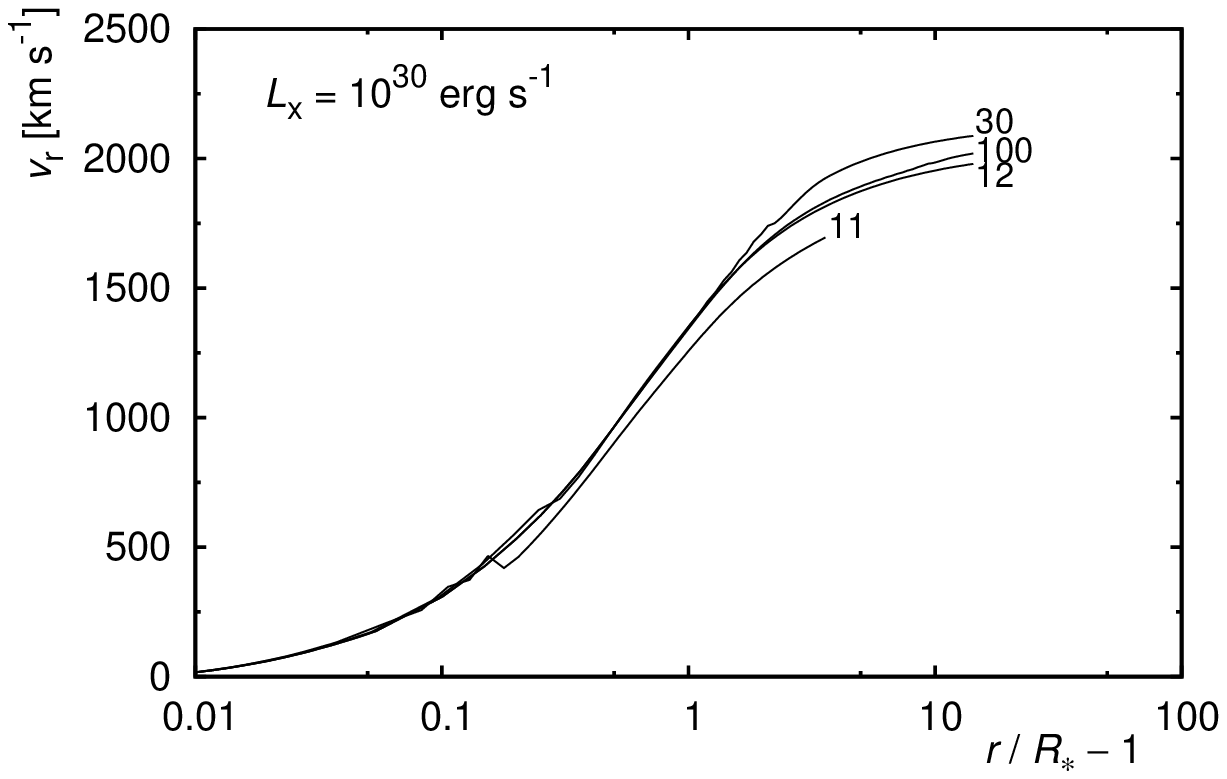}}
\resizebox{0.33\hsize}{!}{\includegraphics{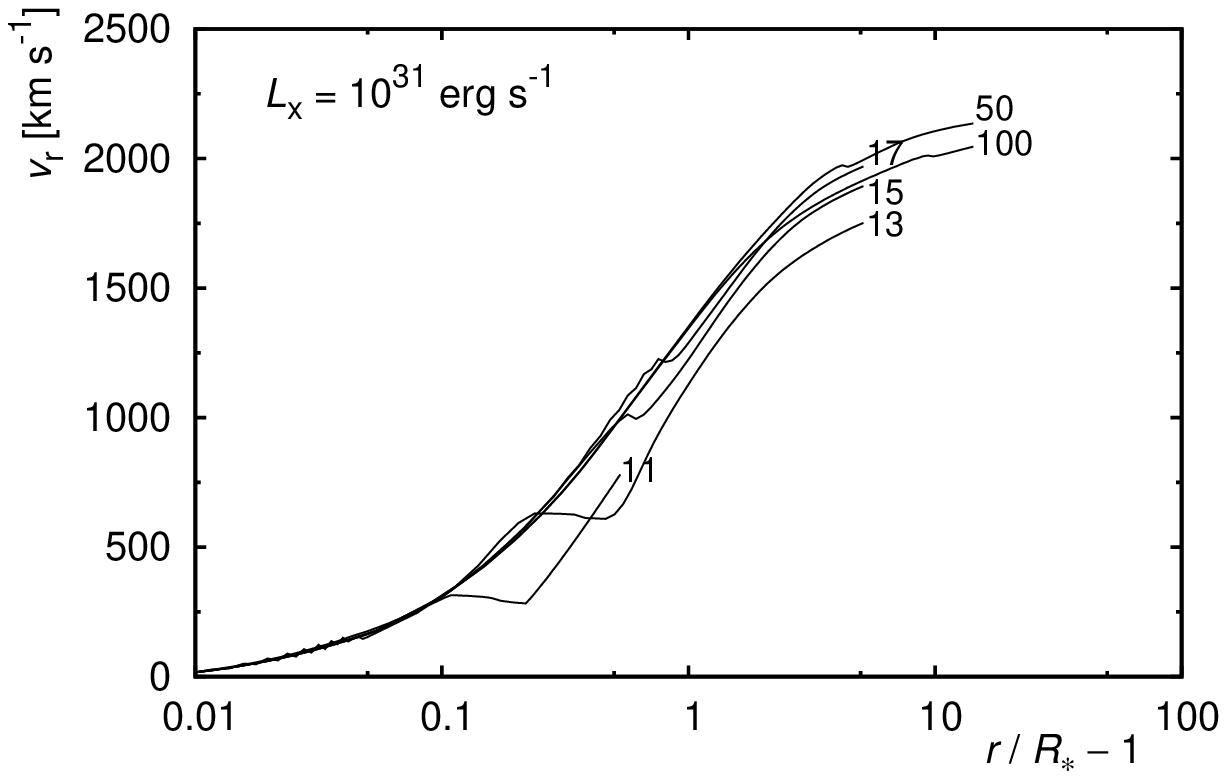}}
\resizebox{0.33\hsize}{!}{\includegraphics{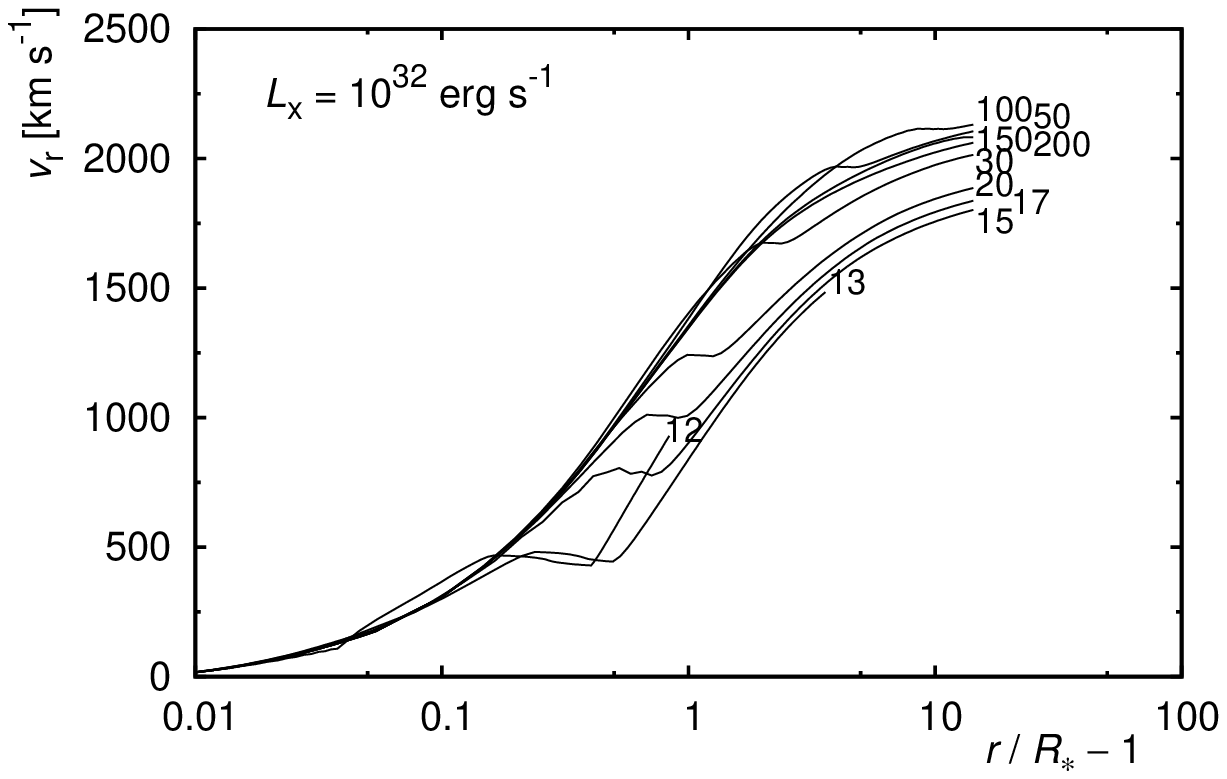}}
\resizebox{0.33\hsize}{!}{\includegraphics{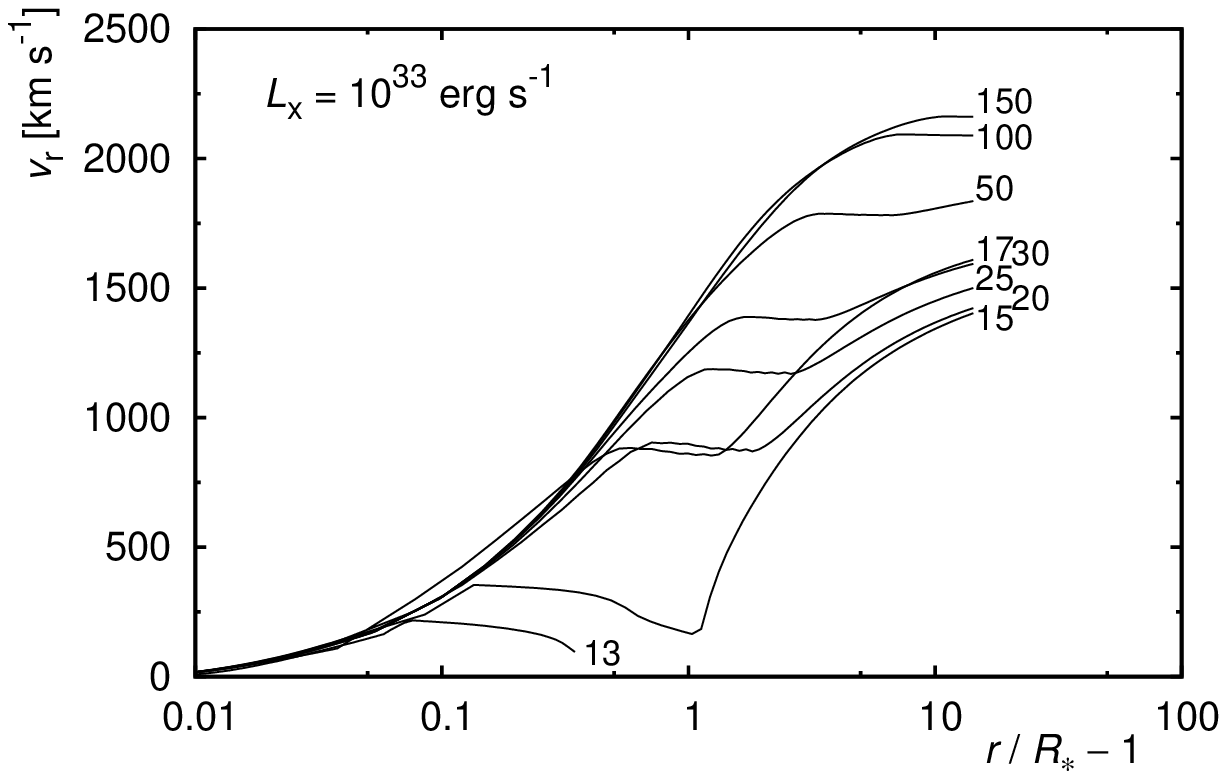}}
\resizebox{0.33\hsize}{!}{\includegraphics{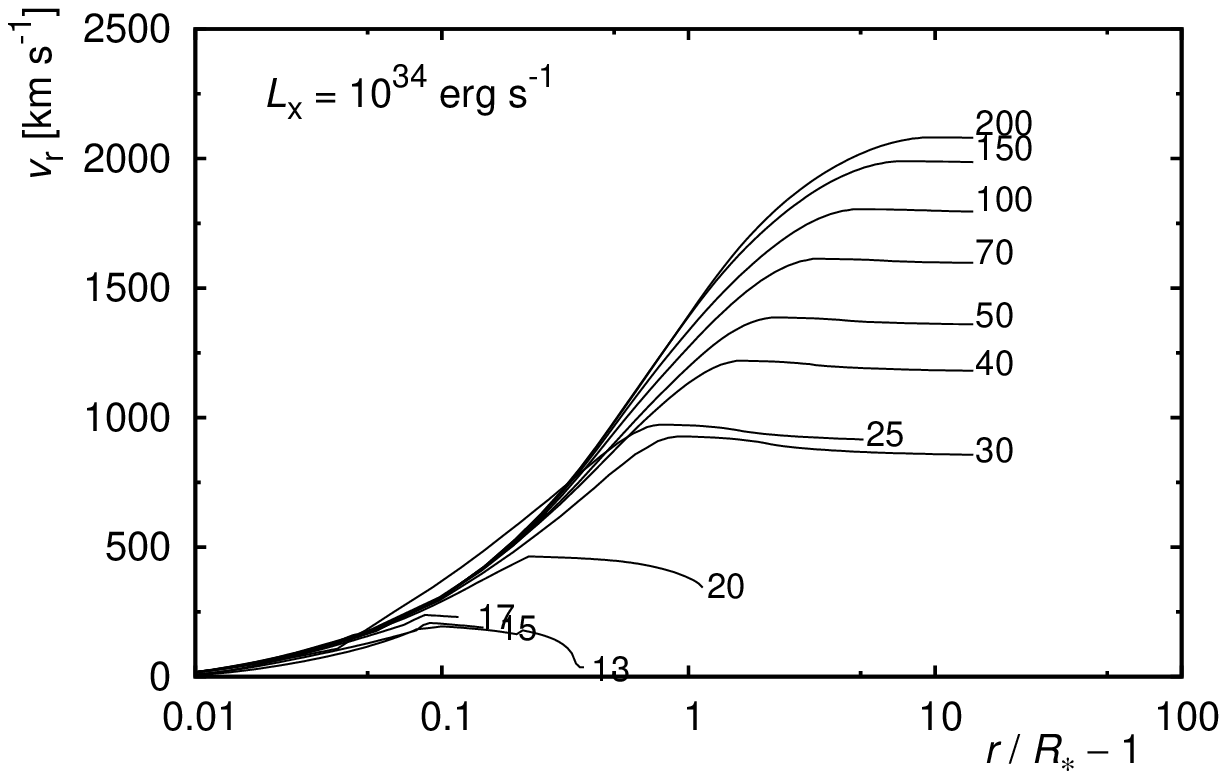}}
\resizebox{0.33\hsize}{!}{\includegraphics{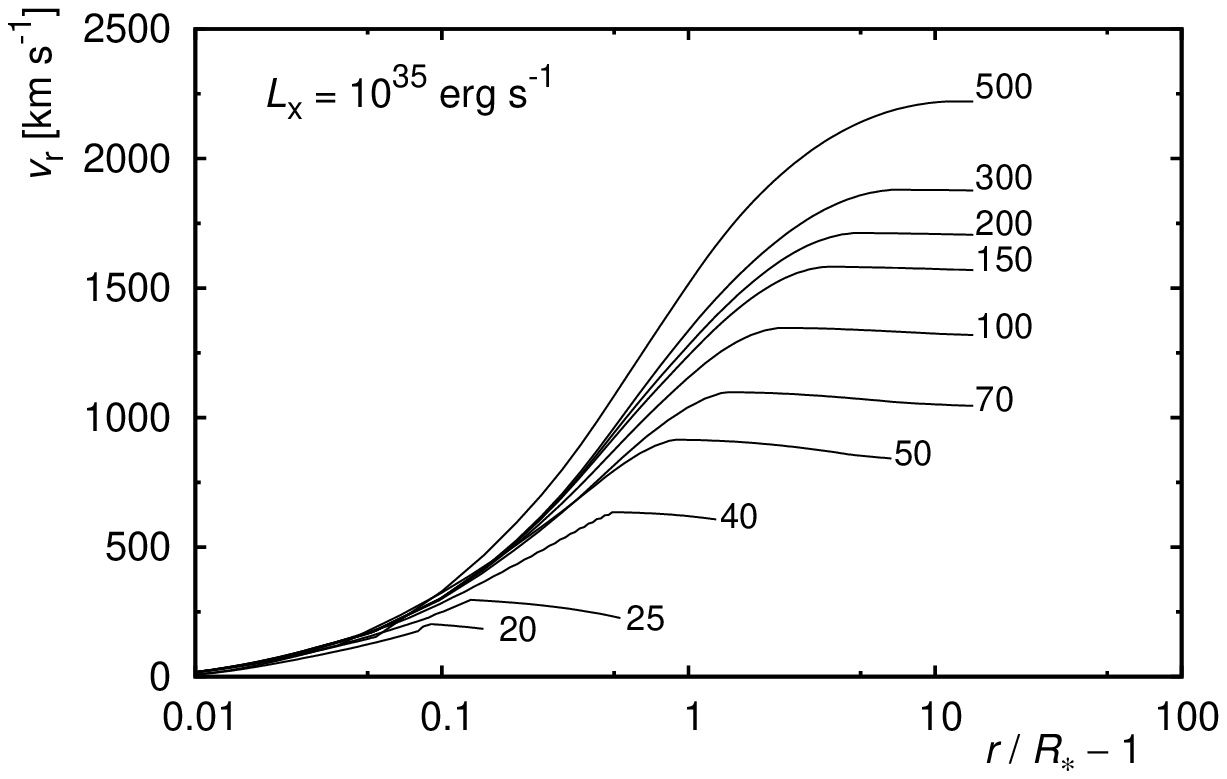}}
\resizebox{0.33\hsize}{!}{\includegraphics{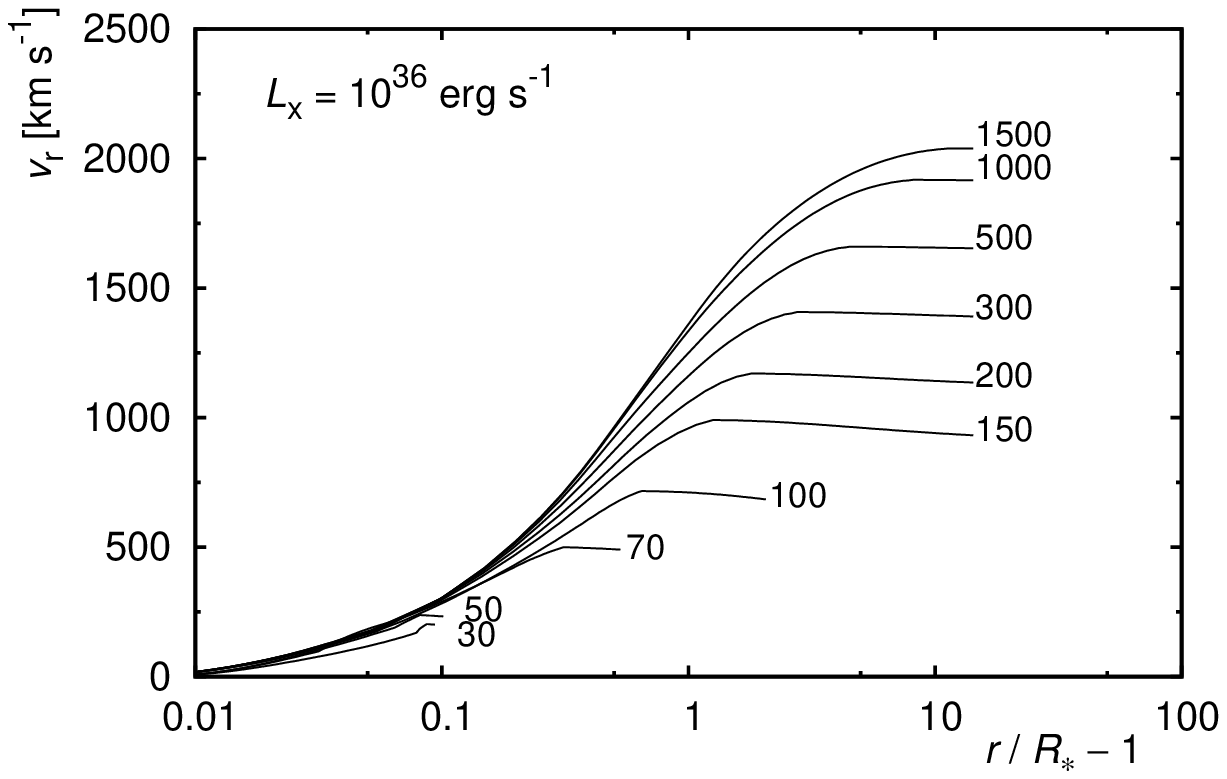}}
\resizebox{0.33\hsize}{!}{\includegraphics{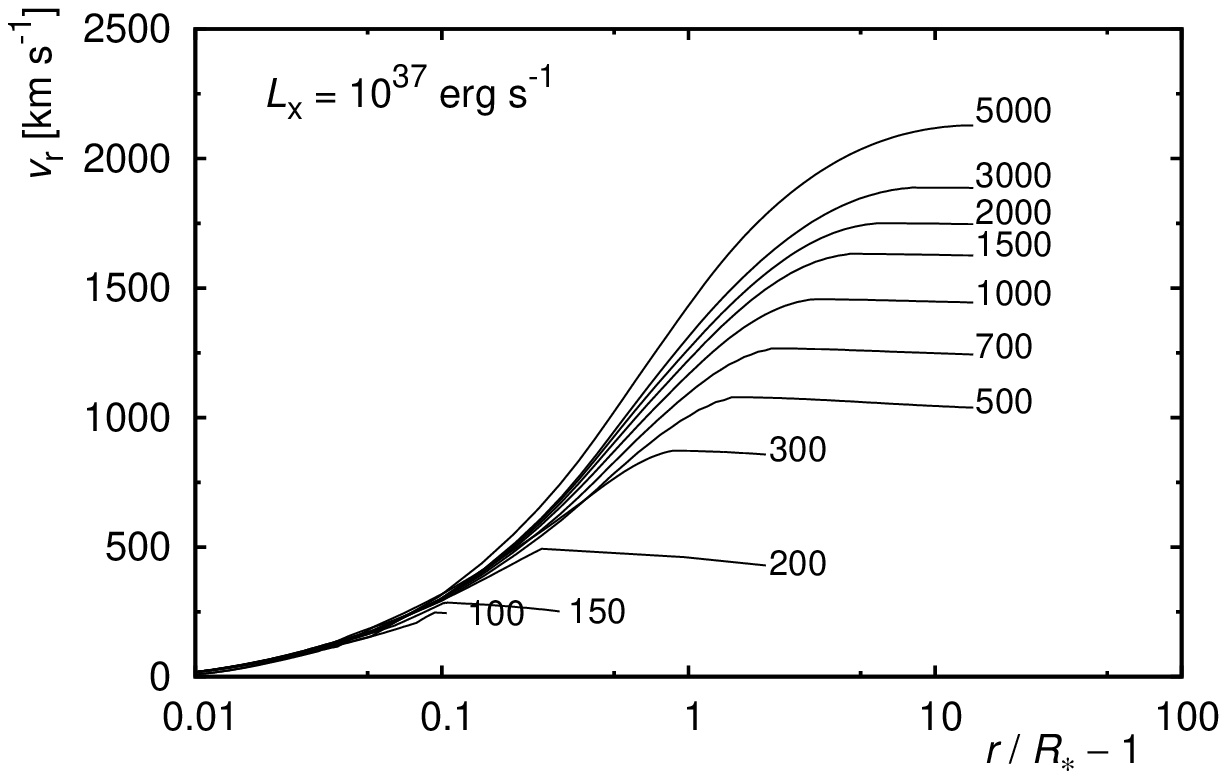}}
\resizebox{0.33\hsize}{!}{\includegraphics{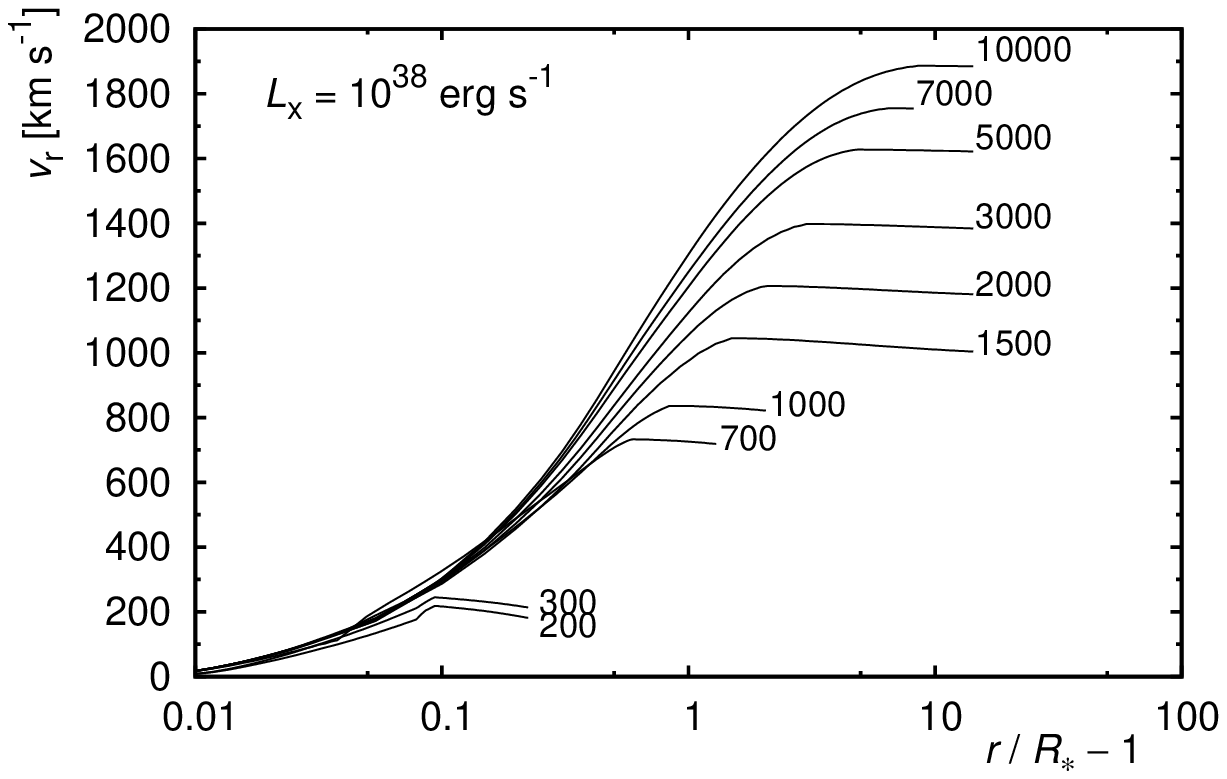}}
\caption{Same as Fig.~\ref{300-1vr}, except for a model star 375-5.}
\label{375-5vr}
\end{figure}

\begin{figure}[ht]
\centering
\resizebox{0.33\hsize}{!}{\includegraphics{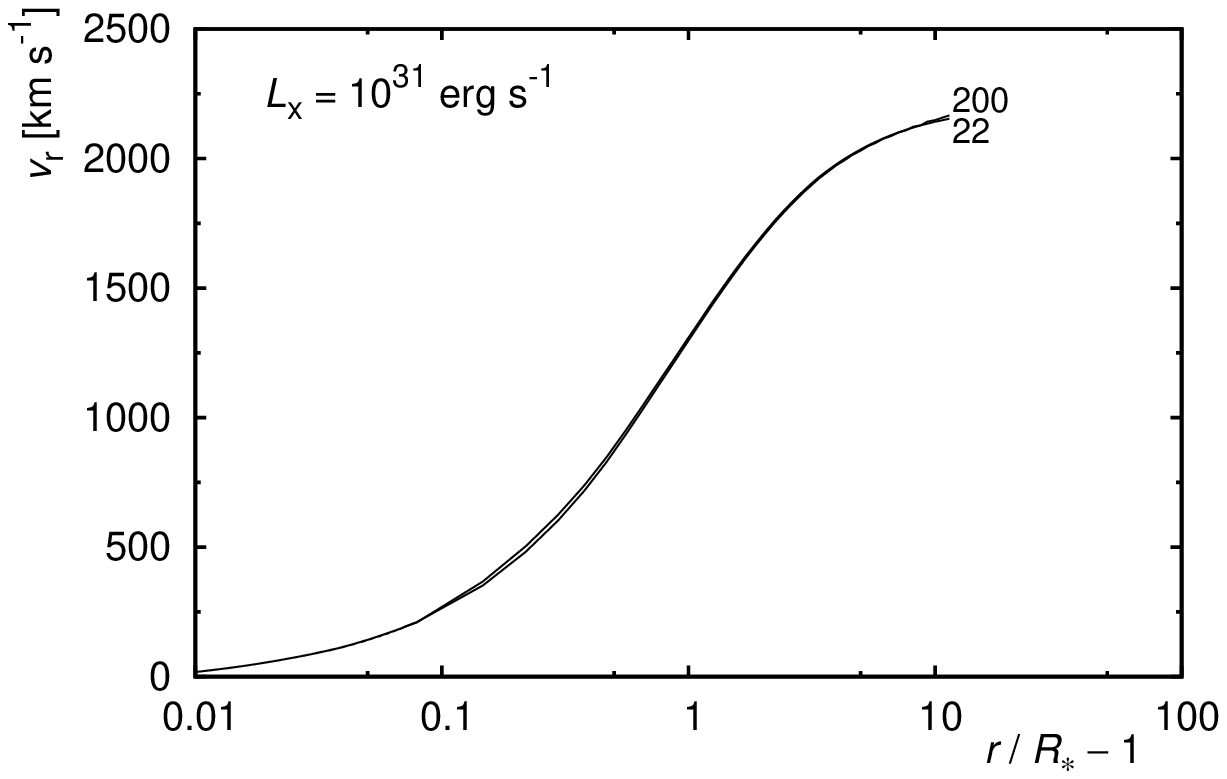}}
\resizebox{0.33\hsize}{!}{\includegraphics{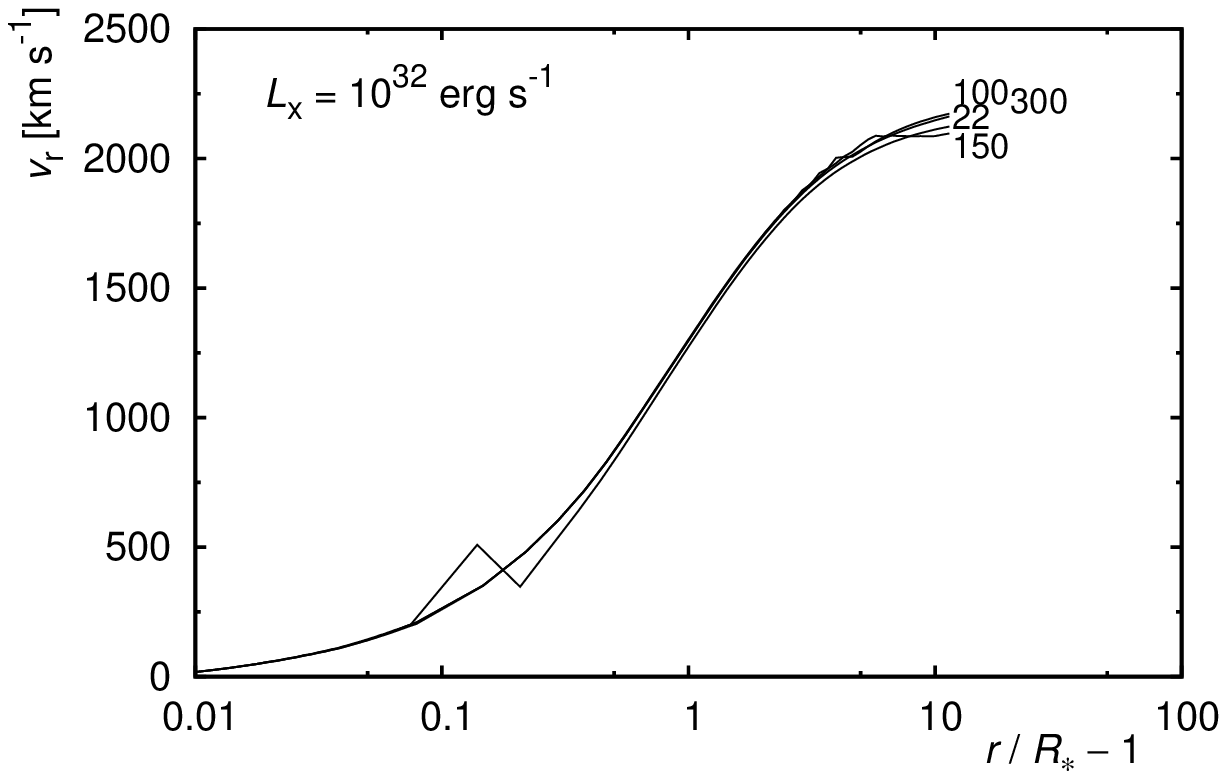}}
\resizebox{0.33\hsize}{!}{\includegraphics{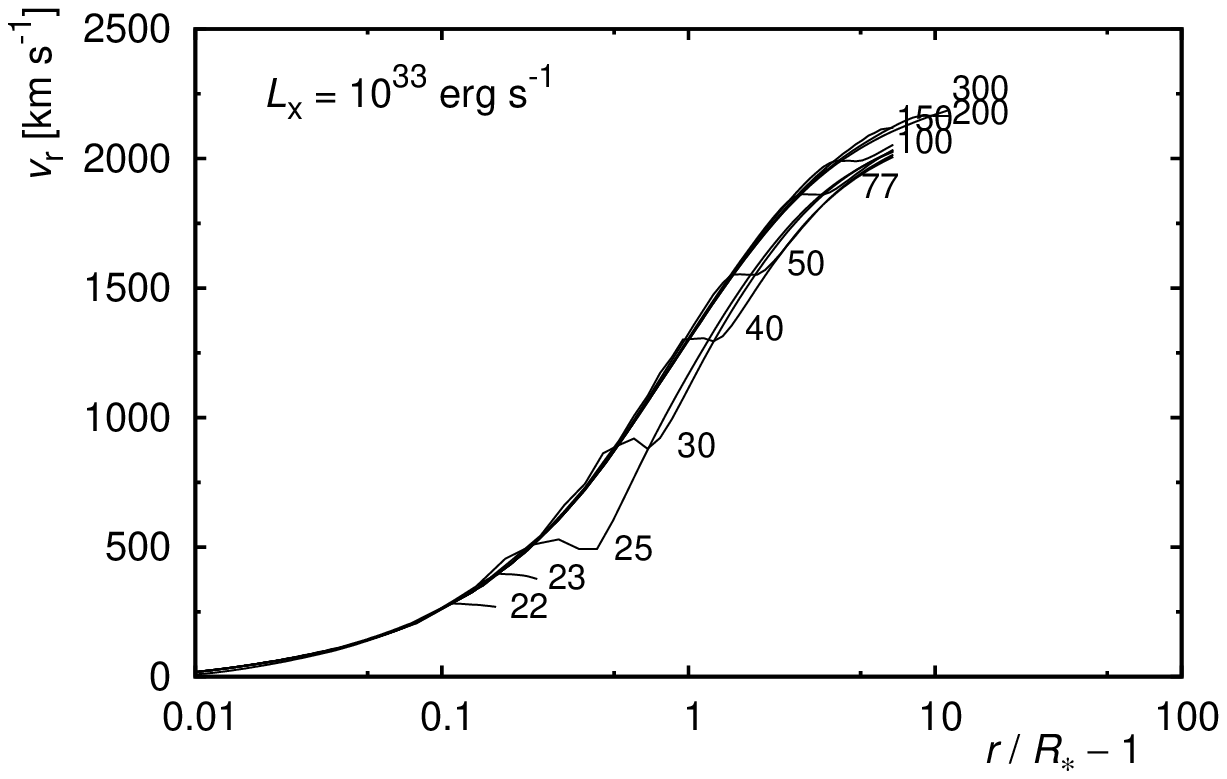}}
\resizebox{0.33\hsize}{!}{\includegraphics{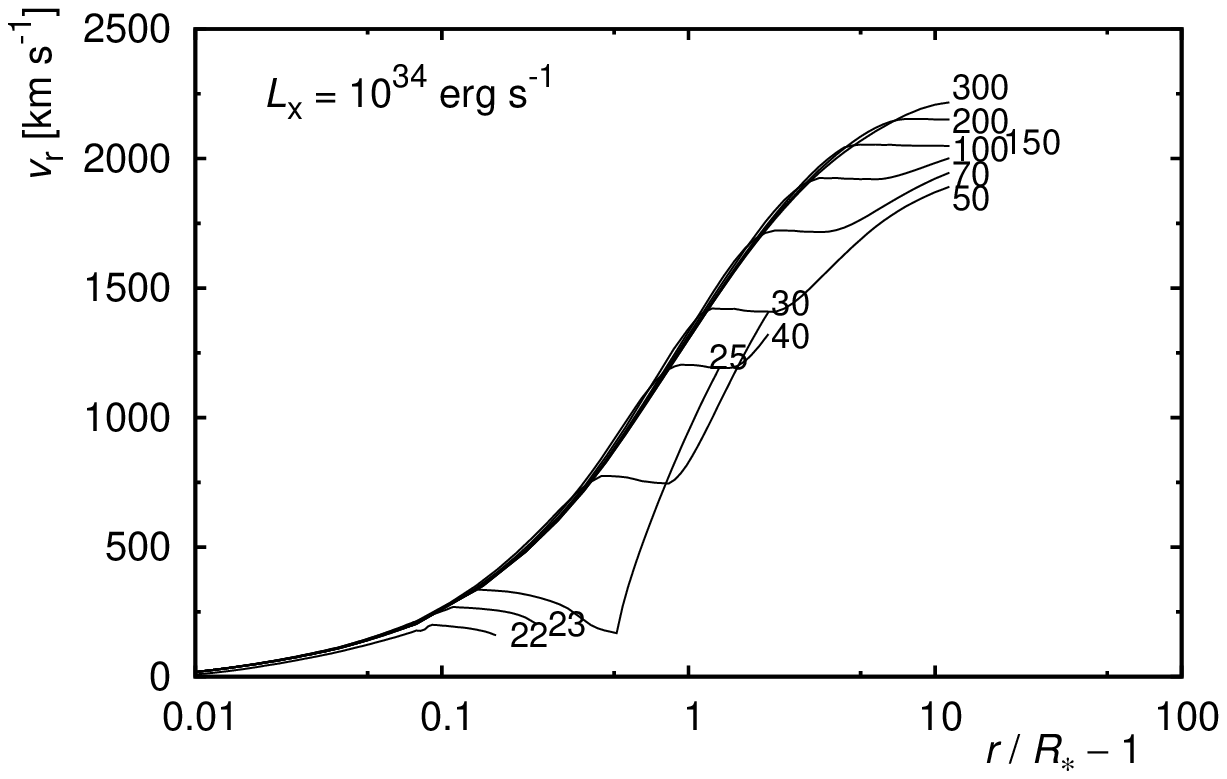}}
\resizebox{0.33\hsize}{!}{\includegraphics{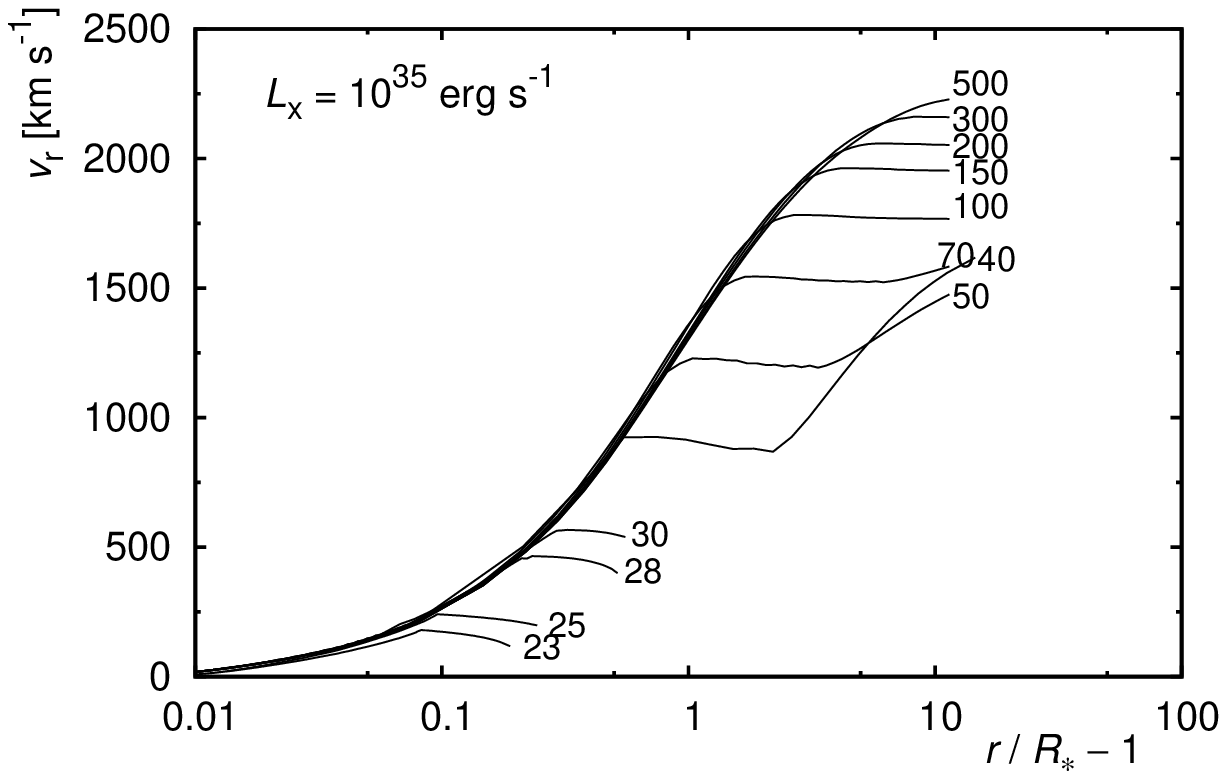}}
\resizebox{0.33\hsize}{!}{\includegraphics{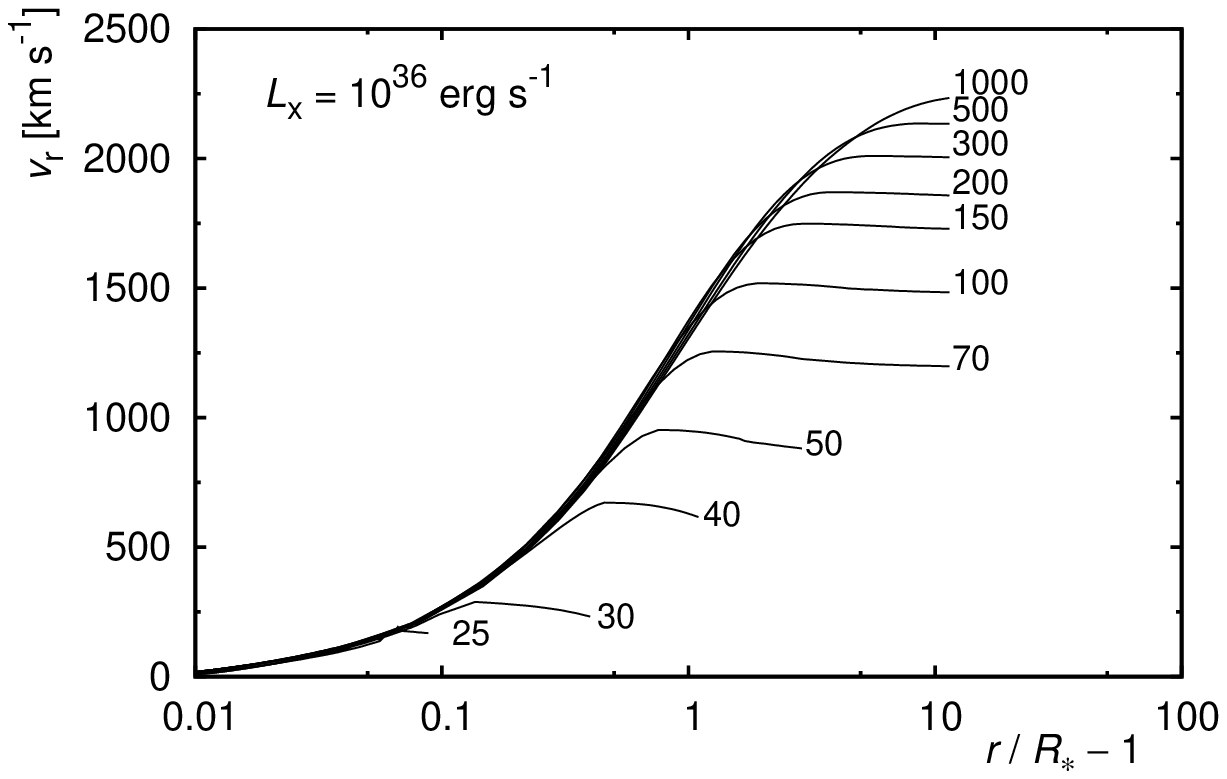}}
\resizebox{0.33\hsize}{!}{\includegraphics{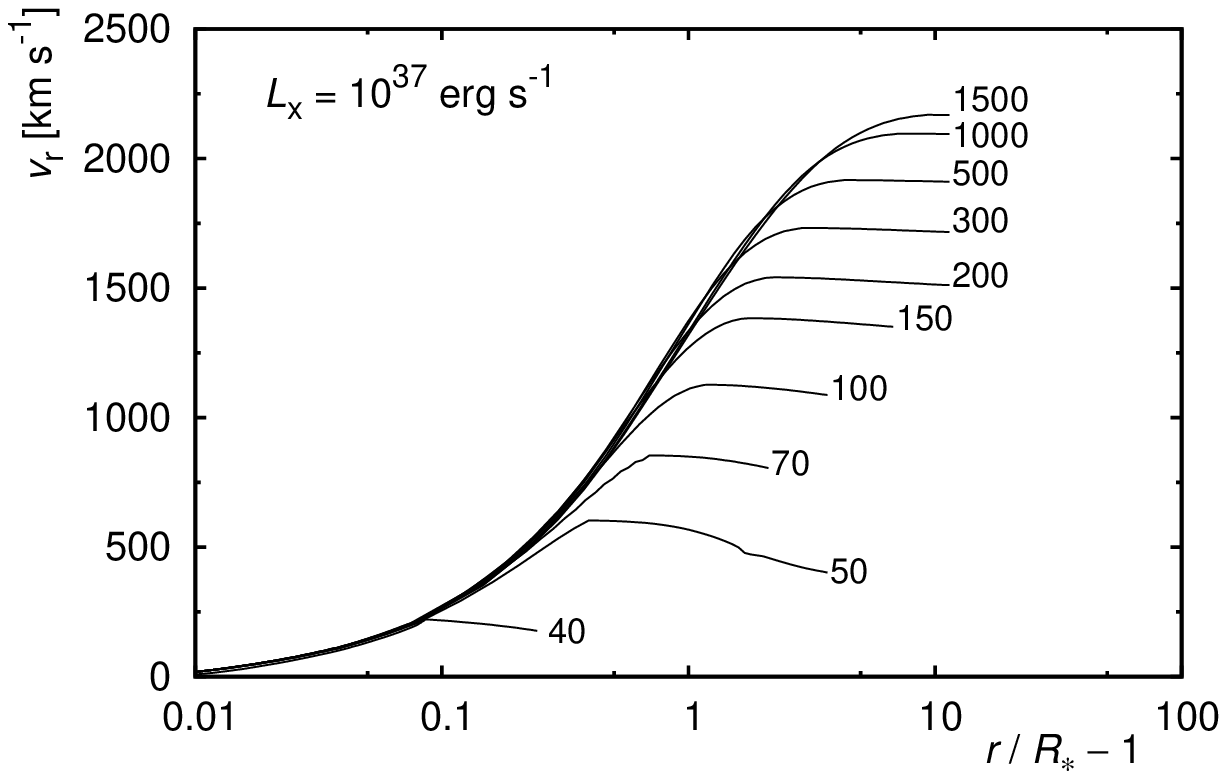}}
\resizebox{0.33\hsize}{!}{\includegraphics{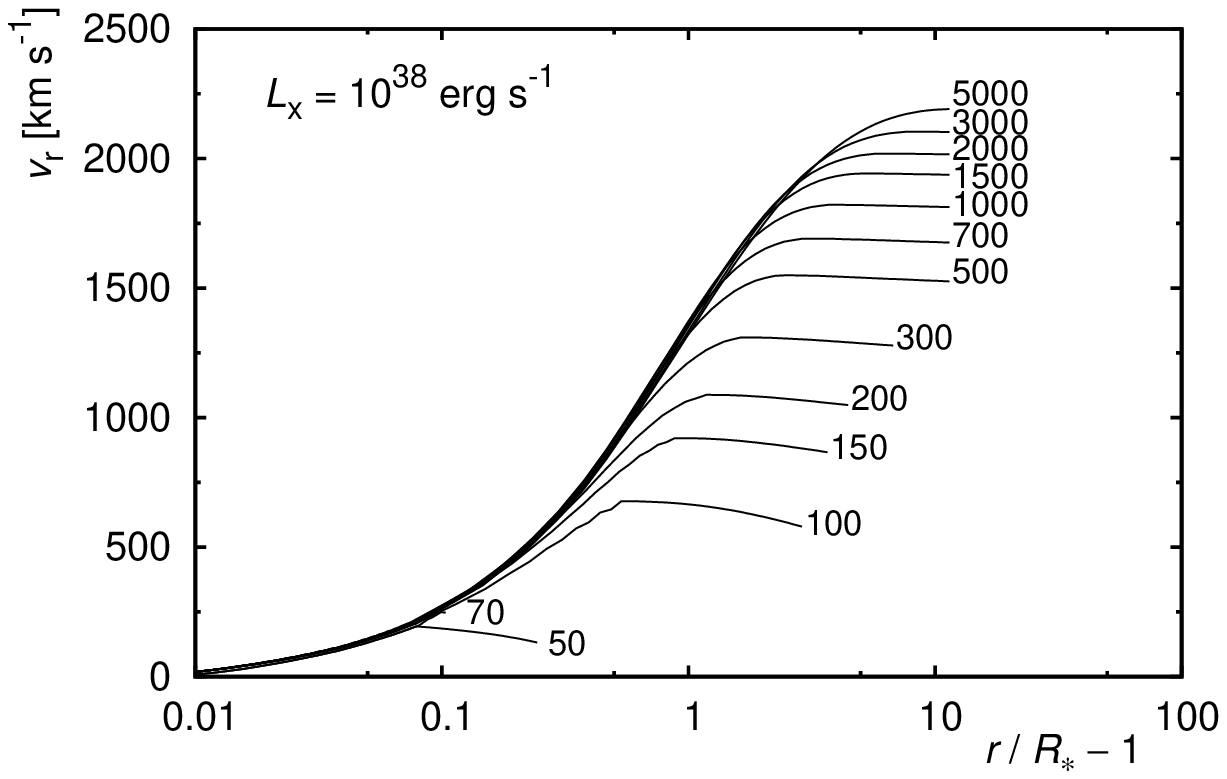}}
\caption{Same as Fig.~\ref{300-1vr}, except for a model star 425-1.}
\label{425-1vr}
\end{figure}

\begin{figure}[ht]
\centering
\resizebox{0.33\hsize}{!}{\includegraphics{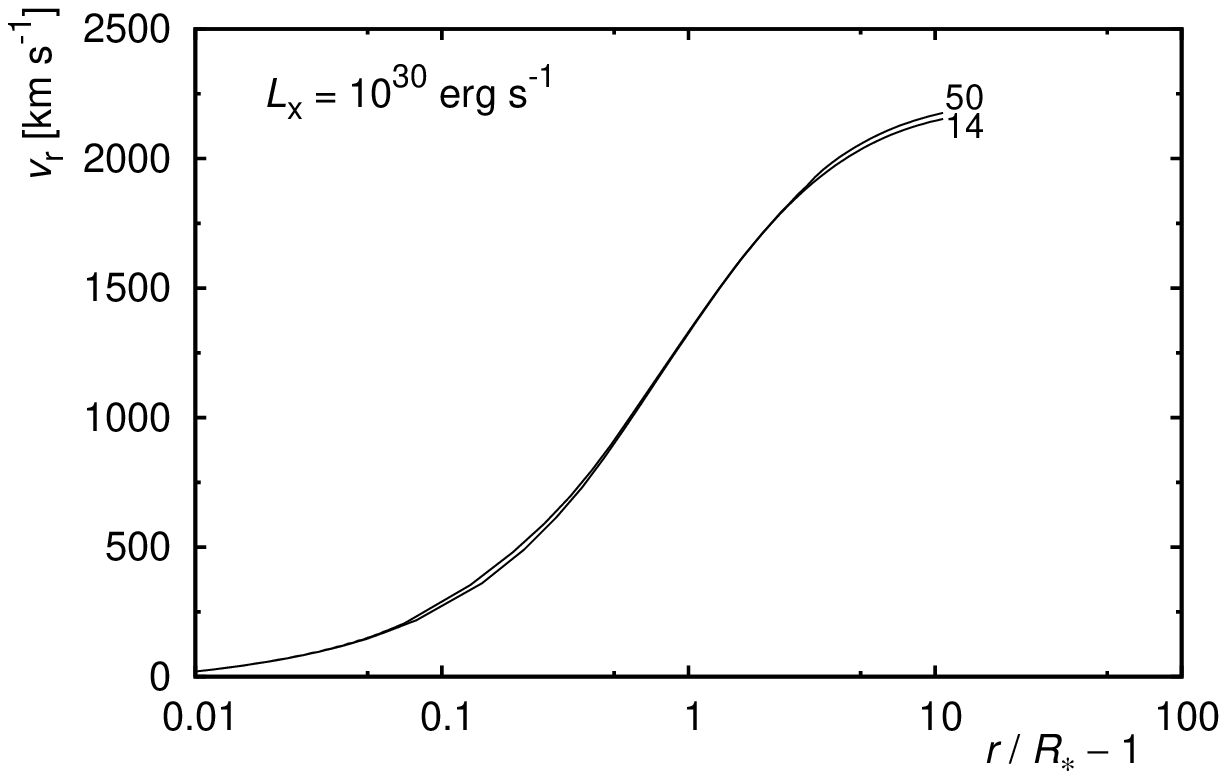}}
\resizebox{0.33\hsize}{!}{\includegraphics{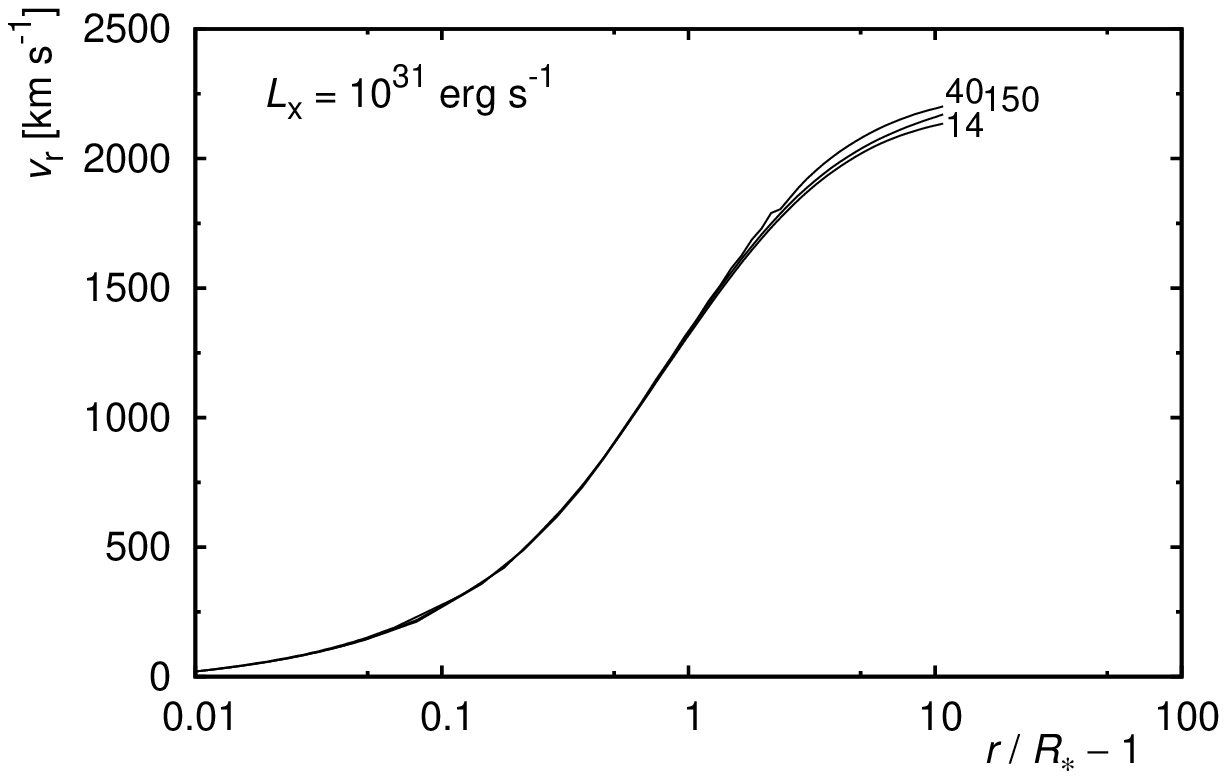}}
\resizebox{0.33\hsize}{!}{\includegraphics{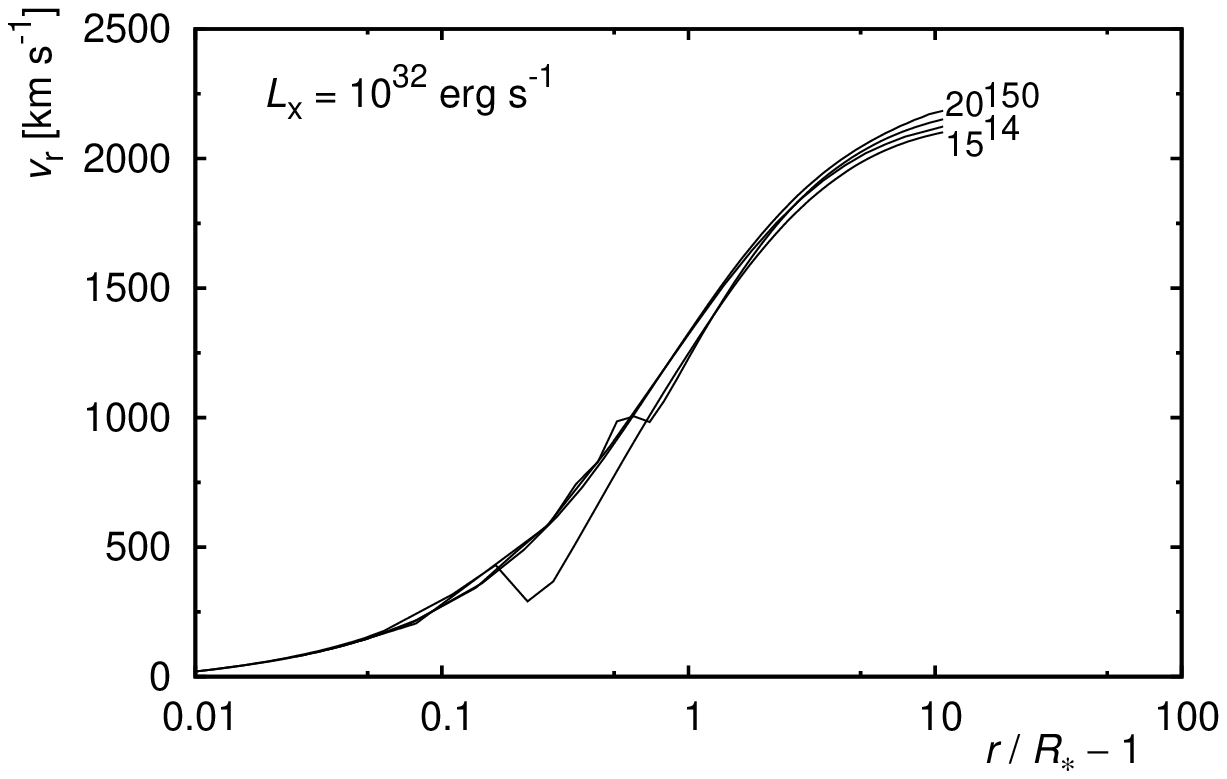}}
\resizebox{0.33\hsize}{!}{\includegraphics{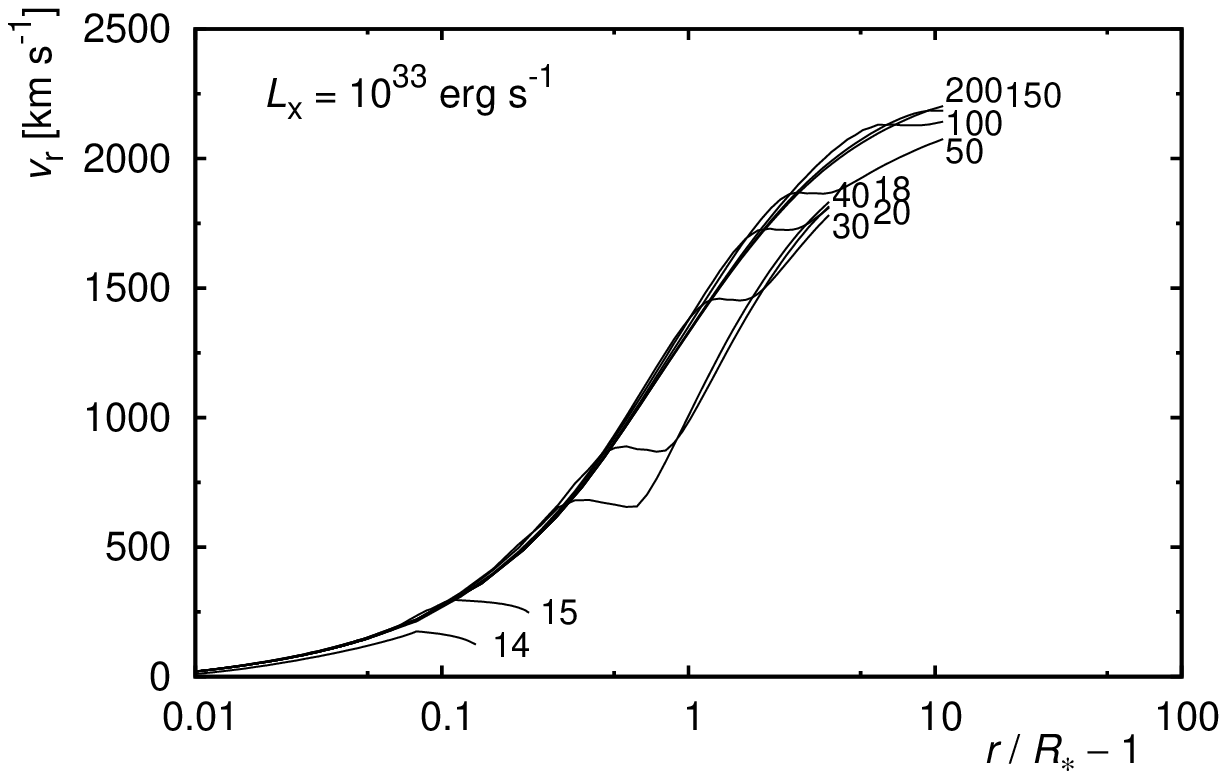}}
\resizebox{0.33\hsize}{!}{\includegraphics{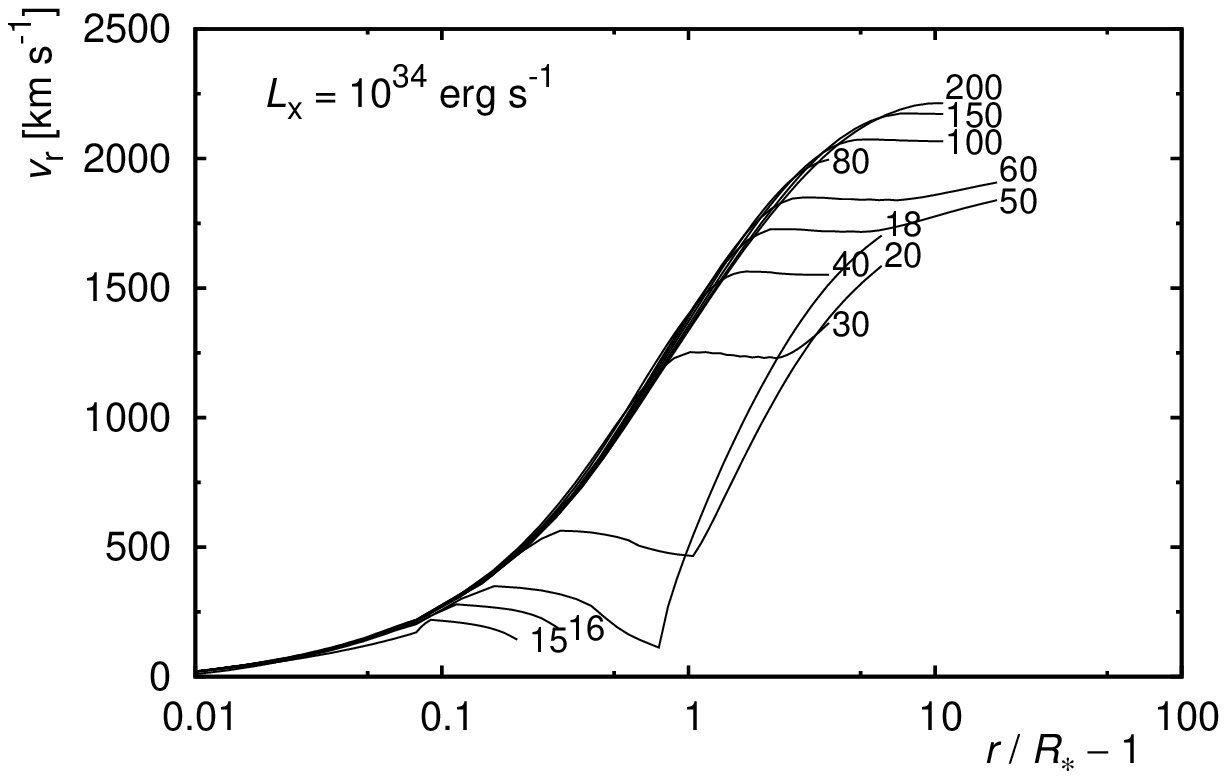}}
\resizebox{0.33\hsize}{!}{\includegraphics{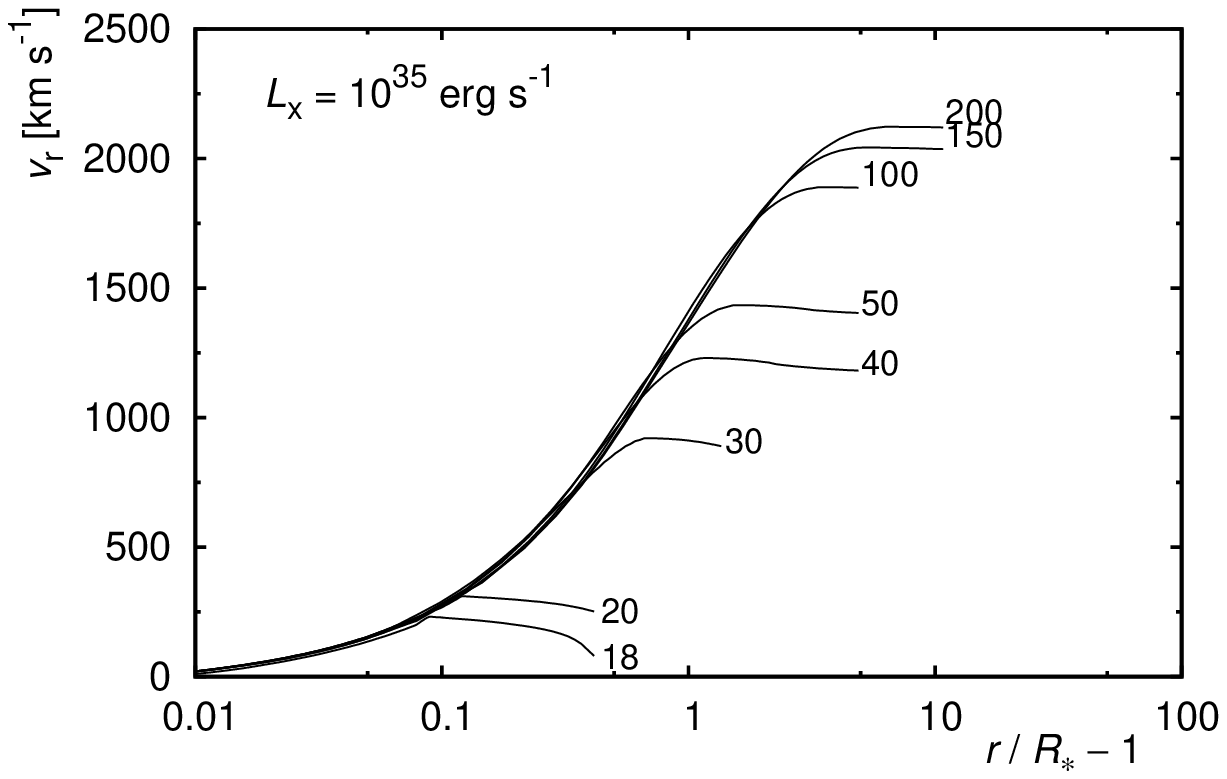}}
\resizebox{0.33\hsize}{!}{\includegraphics{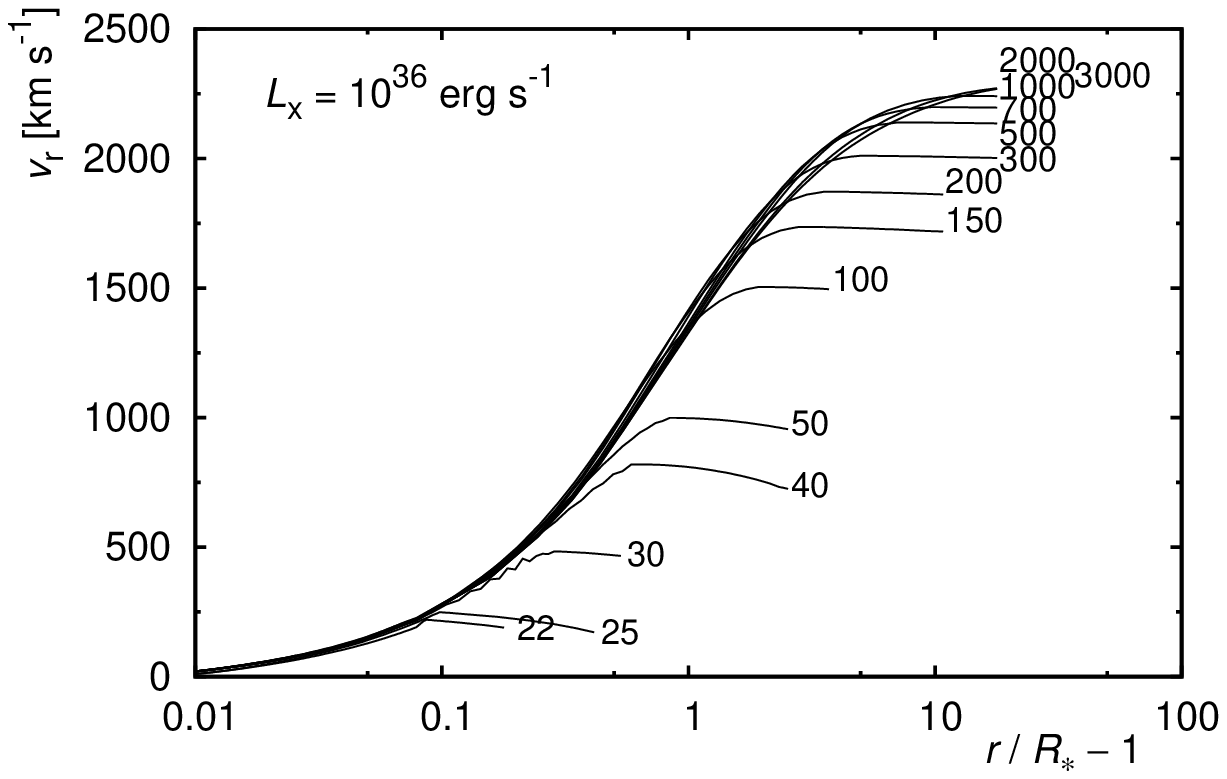}}
\resizebox{0.33\hsize}{!}{\includegraphics{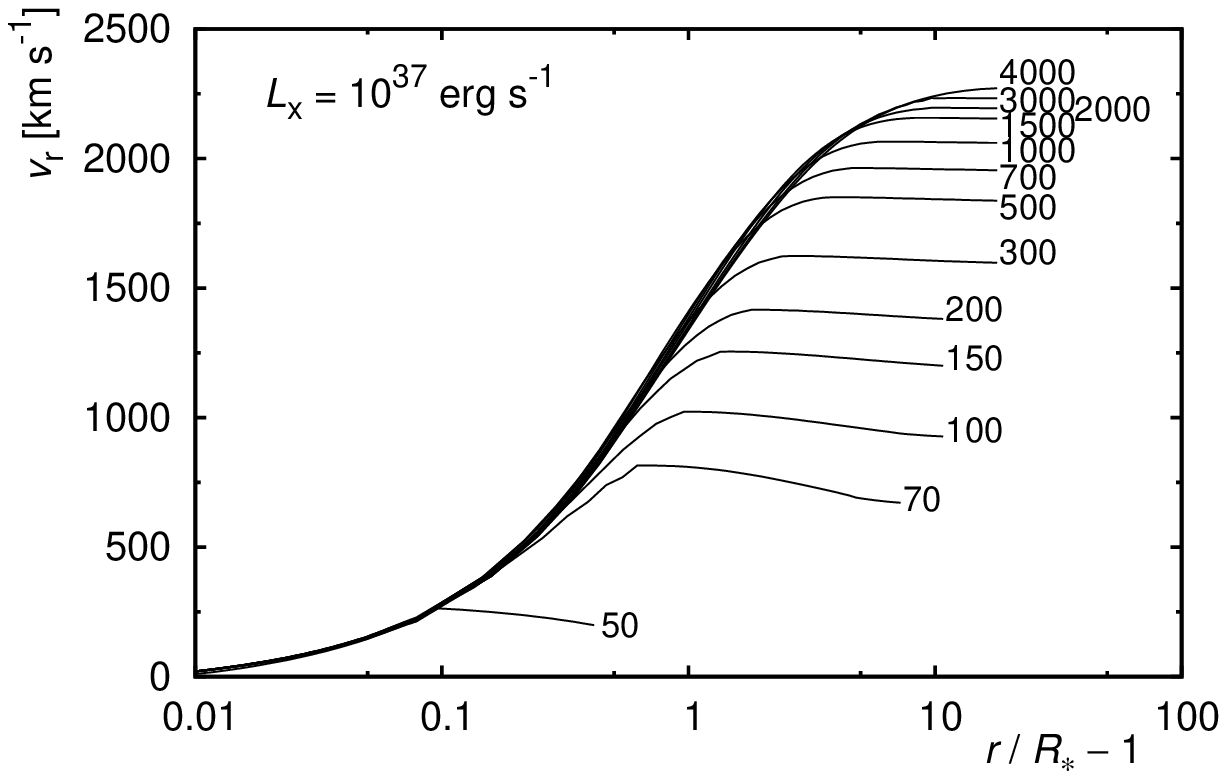}}
\resizebox{0.33\hsize}{!}{\includegraphics{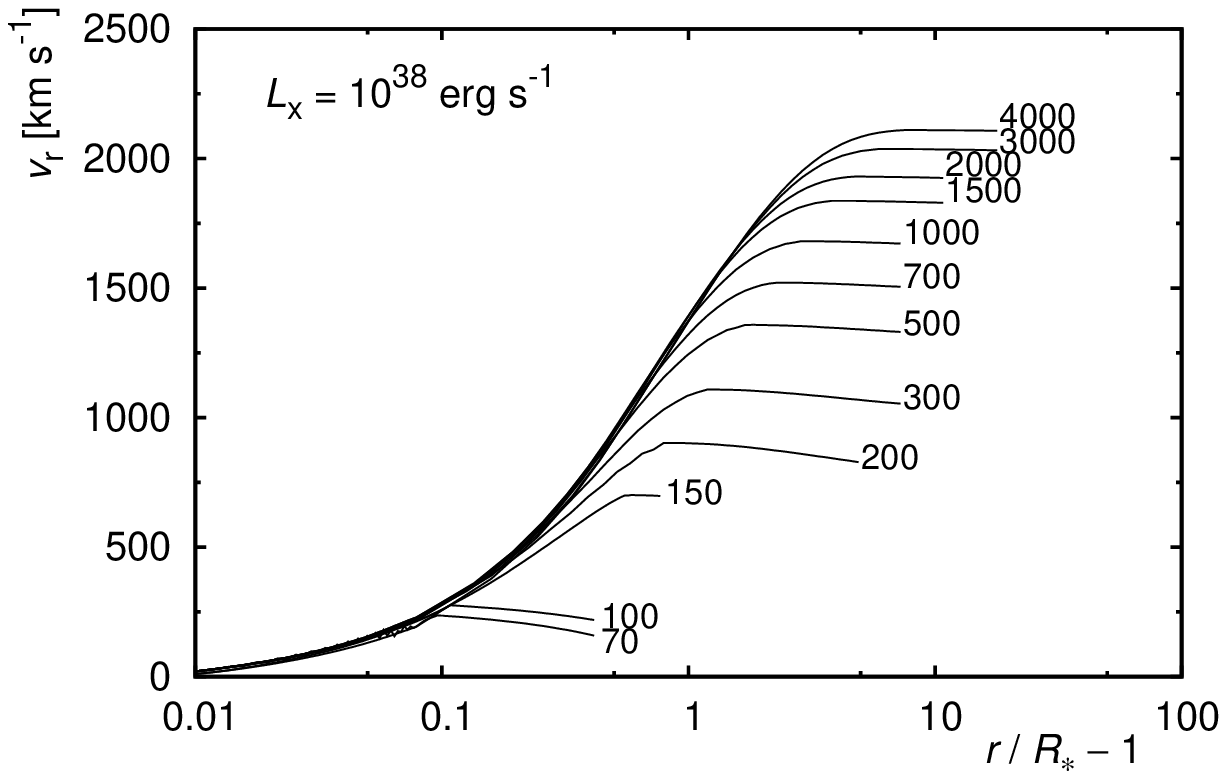}}
\caption{Same as Fig.~\ref{300-1vr}, except for a model star 425-5.}
\label{425-5vr}
\end{figure}

The resulting wind models are given in Figs.~\ref{300-1vr}--\ref{425-5vr}. Here
we plot the dependence of the radial wind velocity on radius in the wind models
for individual studied model stars from Table~\ref{ohvezpar} for different X-ray
luminosities and X-ray source distances.

For a low X-ray luminosity \lx, or a large X-ray source distance $D$, the
influence of the external X-ray source on the wind model is only marginal. In
this case the X-ray ionization influences only trace ionization states
\citep[via direct and Auger ionization,][]{macown,lepsipasam}. This modifies the
emergent spectra because other trace ions with a high degree of ionization
appear \citep{snihmo,zimnila}, but the wind structure remains unaffected. Wind
models for supergiant stars 375-1 and 425-1 with
$\lx=10^{31}\,\text{erg}\,\text{s}^{-1}$ (see Figs.~\ref{375-1vr} and
\ref{425-1vr}) are examples of models with negligible influence of X-ray
irradiation. Because new ionization states appear that may drive the wind, the
wind terminal velocity may be slightly higher \citep{nlteiii}. For example, this
explains why 300-1 model with $\lx=10^{37}\,\text{erg}\,\text{s}^{-1}$ and
$D=5000\,R_\odot$ has higher terminal velocity than the same model, but with
$D=7000\,R_\odot$ (see Fig.~\ref{300-1vr}).

\newrgbcolor{modra}{0.27 0.27 0.88}
\newrgbcolor{seda}{0.27 0.27 0.27}
\newrgbcolor{cervena}{0.98 0.17 0.17}

\begin{figure*}[t]
\centering
\resizebox{0.8\hsize}{!}{\includegraphics{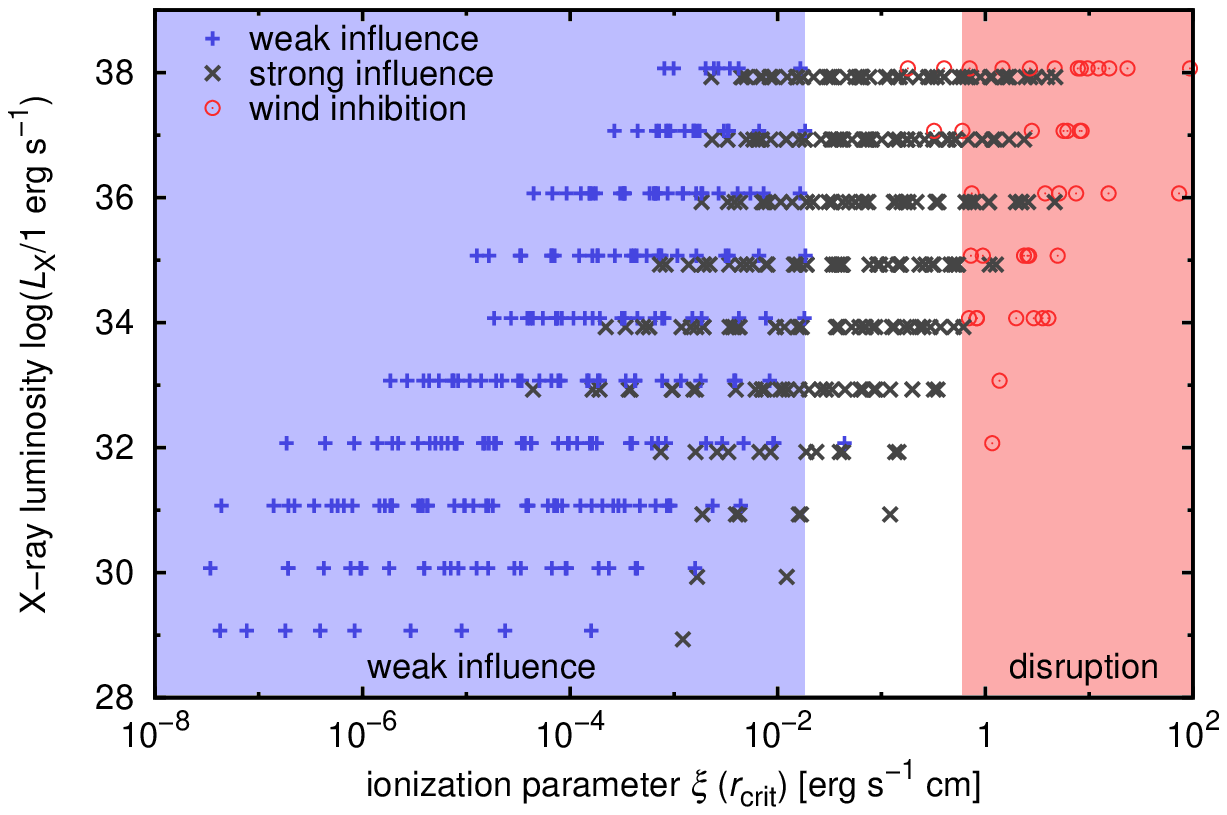}}
\caption{Parameters of all models from Fig.~\ref{txobr} plotted in the diagram
of X-ray luminosity vs.~the ionization parameter. The ionization parameter
Eq.~\eqref{xi} was evaluated at the wind critical point with radius
$r_\text{crit}$. The influence of X-rays is denoted using different symbols.
Blue plus signs {\modra $\boldsymbol+$ } denote the weak influence of X-rays
(possibly just on the ionization equilibrium), grey crosses {\seda
$\boldsymbol\times$ } denote the strong influence of X-rays leading to the decrease
of the terminal velocity, and empty red circles {\cervena $\boldsymbol\odot$ }
denote models with wind inhibition. The individual symbols are slightly
vertically shifted for a better readability. The extension of individual regions
is also denoted using coloured areas.}
\label{tau}
\end{figure*}

For higher X-ray luminosities \lx\ (or a lower X-ray source distance $D$), the
effect of X-ray ionization becomes stronger. The ionization states that drive a
wind in a case without X-ray irradiation become less populated close to the
X-ray source, which results in a decrease of the radiative force and the wind
velocity in the outer parts of the wind (e.g. 300-1 wind models with
$\lx=10^{38}\,\text{erg}\,\text{s}^{-1}$, Fig.~\ref{300-1vr}). The decrease of
the radiative force may be so strong that the accelerating wind solution is not
possible any more, and the wind velocity switches to decelerating overloaded
solutions \citep{feslop,fero} with a typical kink in the velocity profile. This
leads to a non-monotonic velocity law and decrease in the terminal velocity,
which were deduced from observations of HMXBs by \cite{kapka} and calculated for
Vela X-1 by \cite{velax1}. The decelerating regions of the kink may be followed
by accelerating regions if the kink is weak (e.g. 300-1 wind models with
$\lx=10^{32}\,\text{erg}\,\text{s}^{-1}$ in Fig.~\ref{300-1vr} or 425-1 models
with $\lx=10^{34}\,\text{erg}\,\text{s}^{-1}$ and $D\geq25\,R_\odot$ in
Fig.~\ref{425-1vr}). For stronger kinks, the switch back to the accelerating
solution is missing (e.g. 375-5 models with
$\lx=10^{35}\,\text{erg}\,\text{s}^{-1}$ in Fig.~\ref{375-5vr}). However,
our calculations simplify the treatment of the radiative force in the kink
because we do not account for the complex line resonances that occur in this
case \citep[e.g.][]{fero}.

With decreasing source distance $D$ (or increasing X-ray luminosity \lx) the
kink position moves towards the stellar surface (see model 300-1 with
$\lx=10^{34}\,\text{erg}\,\text{s}^{-1}$ in Fig.~\ref{300-1vr}). If the kink
occurs for velocities lower than the escape speed, the wind never leaves the
star and the wind material falls back to the star (for example, this happens for
the 300-1 model with $\lx=10^{37}\,\text{erg}\,\text{s}^{-1}$ and
$D=200\,R_\odot$). Our models only describe an accelerating part of the solution
and the fall back has to be studied using time-dependent models \citep[see][for
a similar situation]{obalka}.

In the case when the kink reaches the critical point, the X-rays start to
significantly influence the wind mass-loss rate. The wind may be completely
inhibited by the X-rays. In some cases, it is still possible to find a stationary
solution (with the kink now being a critical point, not shown here), however,
with a mass-loss rate typically lower by one or two orders of magnitude.
Time-dependent models are needed to study this problem.

Derived results are summarized in the diagrams of $\x L$ vs.~$\x t$ in Fig.~\ref{txobr}.
Each diagram indicates results for one of the model stars listed in
Table~\ref{ohvezpar}. For a given star (with fixed $\dot M$ and $\varv_\infty$),
the optical depth parameter $\x t$ in Eq.~\eqref{tx} only depends on the X-ray
source distance $D$, and $\x t$ is an increasing function of $D$. Wind models
with a negligible influence of X-rays appear in the right bottom part of
diagrams of $\x
L$ vs.~$\x t$ (high $\x t$ and low \lx). Wind models in which the X-ray source
causes a decrease in terminal velocity (denoted by the $\times$ symbol)
appear in a diagonal strip of the diagram of $\x L$ vs.~$\x t$. A forbidden
area, where the
X-rays lead to wind inhibition, is located in the upper left corner of the
diagrams of $\x L$ vs.~$\x t$ diagrams.

The influence of the X-rays may be simply described by the ionization
parameter $\xi$ introduced in Eq.~\eqref{xi}. This is shown in Fig.~\ref{tau},
where we plot the positions of all studied models in the diagrams of $\x L $
vs.~$\xi$.
It follows that for low values of the ionization parameter, $\xi\lesssim
0.01\,\text{erg}\,\text{s}^{-1}\,\text{cm}$, the X-rays may influence the wind
ionization state, but their influence on wind structure is negligible. The
X-rays significantly influence the winds for
$\xi\gtrsim0.01\,\text{erg}\,\text{s}^{-1}\,\text{cm}$. Large ionization
parameters $\xi\gtrsim1-10\,\text{erg}\,\text{s}^{-1}\,\text{cm}$ lead to the
wind inhibition. These parameter values correspond to the forbidden area in the
diagrams of $\x L$ vs.~$\x t$.

The borderline between the ionization parameters that lead to wind inhibition
and those that do not is not the same for all model stars in Fig.~\ref{tau}. There are
different reasons for this, the most important is likely the influence of the
energy distribution of the irradiating X-rays. The X-rays become harder with
increasing optical depth between the source and a given point (because of
frequency-dependent opacity, see Eq.~\ref{kapafit}), leading to a smaller
influence of the X-ray irradiation. Therefore, a distant obscured source has a
smaller effect on wind ionization than a closer and weaker unobscured source
with the same $\xi$. The wind optical depth in model 300-5 is so low that
there is no significant obscuration even for the distant sources. Consequently,
the X-rays already influence the wind for
$\xi\approx1\,\text{erg}\,\text{s}^{-1}\,\text{cm}$.

\begin{table*}[t]
\caption{Parameters of HMXBs with neutron star or black hole companion}
\label{neutron}
\centering
\begin{tabular}{l@{\hspace{2.5mm}}c@{\hspace{2.5mm}}c@{\hspace{2.5mm}}l@{\hspace{2.5mm}}c@{\hspace{2.5mm}}c@{\hspace{2.5mm}}c@{\hspace{2.5mm}}c@{\hspace{2.5mm}}c@{\hspace{2mm}}c@{\hspace{2.5mm}}c}
\hline
Binary & Sp. Type & $\log(L/L_\odot)$ & $T_\text{eff}$ [K] & $R$ [$R_\odot$] &
$M$ [$M_\odot$] & $a$ [$R_\odot$] & $\x L$ [$\text{erg}\,\text{s}^{-1}$]&
$\dot M$ [$M_\odot\,\text{year}^{-1}$] & \x t & Reference\\
\hline
\input{neutron.tex}
\hline
\end{tabular}
\tablefoot{Stellar parameters are taken from literature except for the mass-loss
rate, for which we used fits of \citet{fosfor}. \tablefoottext{a}{Derived from the spectral type using the
expressions of \citet{okali}. \tablefoottext{b}{Periastron distance.}
\tablefoottext{c}{Possible disk accretion.}\tablefoottext{d}{Polar
values.}\tablefoottext{e}{In the direction to the companion.}}
\tablefoottext{f}{Some alternative designations: \input{alt_neutron.tex}}
}
\tablebib{\input{ref_neutron.tex}}
\end{table*}

\section{Comparison with observations}
\subsection{High-mass X-ray binaries}

The existence of the forbidden area in the diagram of $\x L$ vs.~$\x t$ can be tested
against the observed parameters of HMXBs. For this purpose we searched the
literature for available parameters of wind-powered Galactic HMXBs (these are
listed in Table~\ref{neutron}). In this table we restricted ourselves to HMXBs
with primary effective temperatures $T_\text{eff}\gtrsim25\,000\,\text{K}$ to
avoid the bistability jump \citep{papubista,bista}, which may complicate the
estimate of wind parameters for cooler stars and deserves special study. Our
list of HMXBs with known parameters is likely incomplete, but it is a
representative subset that characterizes the complete sample.

Stellar and binary parameters (spectral type, $T_\text{eff}$, $R$, $M$,
semimajor axis $a$, and \x L) in Table~\ref{neutron} were taken from the literature
(however, see the notes below the table). We typically used the highest \x L in
the case of variable X-ray sources. The mass-loss rate was derived using Eq.~(1)
of \citet{fosfor} from the stellar luminosity $L$ and the luminosity class.

We included the parameters of HMXBs in the $\x L-\x t$ plots in
Fig.~\ref{txobr}. We assumed that the X-ray source distance $D$ is equal to the
semimajor axis $a$. The positions of all HMXBs in the diagrams of $\x L$ vs.~$\x t$ lie
outside the forbidden area. Moreover, many HMXBs appear just at the border
between the forbidden area and the diagonal strip with a strong influence of
X-rays on the wind. This indicates that the winds of these stars are in a
self-regulated state. In the self-regulated state an increase in X-ray
luminosity causes wind inhibition and therefore a decrease in \x L
\citep{velax1}, whereas with a lower X-ray luminosity the influence of X-rays on
the wind becomes weaker and the wind is faster, which leads to a stronger X-ray
production. The combination of the two effects may keep the location of the star
in the diagram  of $\x L$ vs.~$\x t$ in the vicinity of the border line with forbidden
solutions.

Some objects (e.g. IGR~J16479-4514, IGR~J17252-3616, and CPD-63$\degr$2495)
show strong X-ray variability, which would appear as a vertical shift of their
position in the $\x L-\x t$ plots. A horizontal shift in $\x L-\x t$ plots is
introduced in eccentric systems as a result of the orbital motion (e.g.
IGR~J11215-5952, CPD-63$\degr$2495). This shows that ionization conditions
in the wind may change with time.

Parameters of most HMXBs lie in the area with a strong influence of X-rays
on the wind structure (see Fig.~\ref{txobr}), leading to the decrease in
wind terminal velocity with respect to non-irradiated winds. Within the
classical Bondi-Hoyle-Lyttleton picture \citep{holy,boho} this affects the
accretion rate and the X-ray luminosity. This effect can, for example, explain,
why the wind velocity required to explain the X-ray luminosity of \object{4U
2206+54} is significantly lower than the terminal velocity expected for this
type of star \citep{hmxb24}.

The existence of the forbidden area in the diagram of $\lx$ vs.~$\x t$ may be one of the
reasons, why the X-ray luminosity of IGR~J16479-4514 is by two orders of
magnitude lower than that estimated from the Bondi-Hoyle accretion theory
\citep{hmxb11}. This star with
$\lx=1.09\times10^{34}\,\text{erg}\,\text{s}^{-1}$ and $\x t=0.4$ lies very
close to the forbidden area. Moreover, during flares it reaches peak
luminosities of $\lx=6.4\times10^{34}\,\text{erg}\,\text{s}^{-1}$, which 
corresponds to the boundary of the forbidden area.

Stars exhibiting Roche-lobe overflow may appear in the forbidden area, since
forces other than the radiative force drives the flow. This is likely the case of
\object{Cen X-3}, which for parameters from \citet{semafor} and \citet{atoyan}
gives $\x t=0.3$; this is too low for its X-ray luminosity
$\lx=5.4\times10^{37}\,\text{erg}\,\text{s}^{-1}$ \citep{semafor}. There are
also observational indications of the Roche-lobe overflow in this system
\citep{semafor,preteka}.

We also tested additional HMXBs (not listed in Table~\ref{neutron}) with
$T_\text{eff}<25\,000\,\text{K}$. Their wind parameters may be influenced by
the presence of the bistability jump, and consequently, we applied the mass-loss rate
predictions of \citet{vikola} for these stars. For \object{IGR J00370+6122}, with
parameters from \citet{hmxbv1}, we derive $\x t=3.3$, which is for
$\lx=2.5\times10^{35}\,\text{erg}\,\text{s}^{-1}$ well below the forbidden
region in Fig.~\ref{txobr}. A similar result was derived for \object{Swift
J1700.8-4139}. With parameters from \citet{hmxbv2} and \citet{hmxbv3} we derive
the optical depth parameter $\x t=1.5$, which is for
$\lx=3\times10^{36}\,\text{erg}\,\text{s}^{-1}$ again well below the threshold
in Fig.~\ref{txobr}. The object \object{IGR J16318-4848}, with $\x
L=1.3\times10^{35}\,\text{erg}\,\text{s}^{-1}$ \citep{filch} and $\x t=0.5$
\citep[with parameters from][]{chatazarohem}, lies just at the border of the
forbidden area.

\subsection{Binaries with non-degenerate components}

The comparison with observations in the case of massive X-ray binaries with
non-degenerate components is complicated by the fact that the X-ray source
distance can not be uniquely determined from observations. To overcome this, we
assume that the wind collisional front is in equilibrium, which means that the
momenta deposited per unit of time and surface by the wind of both components
are equal \citep[e.g.][]{igor},
\begin{equation}
\rho_1 \varv_1^2=\rho_2 \varv_2^2,
\end{equation}
where the subscripts 1 and 2 denote the wind parameters of the primary and
secondary, respectively. Here we study the wind at the intersection of both
components. Denoting the radial distances from the individual star centres $D_1$
and $D_2$ and using the continuity equation, we find
\begin{equation}
\label{momrov}
\frac{D_1^2}{D_2^2}=\frac{\dot M_1\varv_{1,\infty}}{\dot M_2\varv_{2,\infty}},
\end{equation}
with $a=D_1+D_2$. Here we assumed that both winds reached their terminal
velocities.

Stellar and wind parameters of individual binaries are given in
Table~\ref{dvojhve}. The list is based on the known binaries included in the
\citet{dvojx1} sample supplemented by several binaries found in the literature.
Stellar and binary parameters are taken from the literature. The mass-loss rates
were derived using Eq.~(1) of \citet[corrected for actual radius]{metuje} for
main-sequence B stars and Eq.~(1) of \citet{fosfor} for other stars. For the
terminal velocity we assumed the relation between the escape velocity
$\varv_\text{esc}$ and the terminal velocity
$\varv_\infty=2.6\,\varv_\text{esc}$. This relation follows from both
observations and theory \citep{lsl,nltei}.

We included the parameters of X-ray binaries in the $\x L-\x t$ plots in
Fig.~\ref{txobr}. If the X-ray emission originates from the wind-wind collision,
then any of the winds should not be inhibited by X-ray emission. Therefore,
each binary appears in Fig.~\ref{txobr} and Table~\ref{dvojhve} twice, once for
the primary and once for the secondary.

Most of the binary member parameters studied lie in the allowed part of the $\x
L-\x t$ plots in Fig.~\ref{txobr}. This shows that the winds of both components
are not inhibited by X-ray emission and therefore X-ray emission in
these objects may originate in wind-wind collisions. Many parameters
correspond to the region with a strong influence of X-rays on the wind velocity
structure. These binaries may be in a self-regulated state, but of a different
type than we identified in the HMXBs. In binaries with non-degenerate
components,
an increase in X-ray luminosity \x L causes a decrease in the wind velocity
and therefore a decrease in \x L \citep{pars}. Similarly, the decrease in \x L
leads to increase in the terminal velocity, which subsequently causes an
increase in \x L. This may be one of the effects that contribute to the
observational conclusion that most binaries are not stronger X-ray sources than
corresponding single stars \citep[e.g.][]{sane,igorkar}.

Some of the secondaries' parameters lie in the forbidden region in
Fig.~\ref{txobr} (these are denoted by a superscript {\em b} in
Table~\ref{dvojhve}). This indicates that their wind may be inhibited by the
X-ray emission. The emission may originate in this case from the collision of the primary
wind with the secondary star. It is also possible that 1D models
are not applicable for modelling the wind in these systems, or that some
parameters are not correct (e.g. mass-loss rate). Another possibility is that
the basic stellar parameters (luminosities, masses, and radii) need revision.

For some secondaries, the X-ray source distance $D_2$ calculated from
Eq.~\eqref{momrov} is lower than their radius $R_2$ (see a superscript {\em a}
in Table~\ref{dvojhve}). For these stars, the primary wind may reach the
secondary and the secondary wind may be completely inhibited in the direction
toward
the primary. Parameters of these secondaries are not plotted in Fig.~\ref{txobr}.
Our models may also be oversimplified in this case. For example, we neglect the
radiative force from the secondary acting on the primary wind \citep{gocbrzda},
which may avert the secondary wind inhibition. Moreover, a significant part of
X-rays may also originate in shocks in supersonic winds similar to the case
of single stars \citep{luciebila,ocr,felpulpal,bamo,udopec}.

\section{Implications for other massive X-ray sources}

\subsection{Be/X-ray binaries}

The X-ray emission in Be/X-ray binaries originates from the accretion of the Be
star disk material on the compact source \citep[see][for a review]{reig}. The
origin of the disk is still a matter of debate. Some theories propose that
the disk is fed by the wind. We used the list of Be/X-ray binaries given in
\citet{mrikkk} to test whether the radiatively driven wind can survive the
strong X-rays source in Be/X-ray binaries. In many Be/X-ray binaries the compact
source is located at large distance from the Be star, and, consequently, it can
not disrupt its wind. However, in some Be/X-ray binaries the compact object is
located in such proximity to the Be star, that it disrupts its wind. This is
another argument that disfavours the wind origin of Be star disks.

\subsection{X-ray binaries in the Magellanic Clouds}

X-ray binaries in the Magellanic Clouds typically show large X-ray luminosities
of the order of $10^{37}-10^{38}\,\text{erg}\,\text{s}^{-1}$. Because of lower
metallicity of LMC and SMC compared
\linebreak
\begin{table*}[h!]
\caption{Parameters of binaries with non-degenerate components.}
\label{dvojhve}
\centering
\begin{tabular}{l@{\hspace{2.5mm}}c@{\hspace{2.5mm}}c@{\hspace{2.5mm}}l@{\hspace{2.5mm}}c@{\hspace{2.5mm}}c@{\hspace{2.5mm}}c@{\hspace{2.5mm}}c@{\hspace{2.5mm}}c@{\hspace{2.5mm}}c@{\hspace{2.5mm}}c@{\hspace{2.5mm}}c@{\hspace{2.5mm}}c}
\hline
Binary & Sp. Type & $\log(L/L_\odot)$ & \multicolumn{1}{c}{$T_\text{eff}$} & $R$ &
$M$  & $\dot M$ & $\varv_\infty$ & \x t & $a$  & $\x L$  & Reference\\
& &  & \multicolumn{1}{c}{[K]} & [$R_\odot$] &
[$M_\odot$] & [$M_\odot\,\text{year}^{-1}$]& 
[$\text{km}\,\text{s}^{-1}$] & & [$R_\odot$] &
[$\text{erg}\,\text{s}^{-1}$] & \\
\hline
\input{dvojhve.tex}
\hline
\end{tabular}
\tablefoot{Stellar parameters are taken from literature except for the mass-loss
rate, for which we used fits of \citet{fosfor} and \citet{metuje}. Alternative
designations: HR 8281 (HD 206267), HD 152219 (V1292 Sco), HD 47129 (V640 Mon, HR
2422, and Plaskett star), HR 6187 (HD 150136), HR 6736 (9 Sgr, HD 164794), V712
Car (WR 20a), HR 65 (AO Cas, HD 1337), HR 5664 ($\delta$ Cir, HD 135240), HR
8406 (14 Cep, LZ Cep, HD 209481), HD 93206 (QZ Car), V1007 Sco (HD 152248), HR
1931 ($\sigma$ Ori, HD 37468), HR 6535 (V1036 Sco, HD 159176), HR 1899 ($\iota$
Ori A, HD 37043), HR 1852 ($\delta$ Ori, HD 36486), V918 Sco (HR 6164, HD
149404), HR 7767 (HD 193322), and HD 93205 (V560 Car). \tablefoottext{a}{The
primary wind reaches the secondary surface ($D_2<R_2$).} \tablefoottext{b}{The
wind parameters lie in the forbidden area.} \tablefoottext{c}{Derived from the
effective temperature or spectral type using the expressions of \citet{har} for
main-sequence B stars and \citet{okali} for others.}}
\tablebib{\input{ref_dvojhve.tex}}
\end{table*}
\clearpage
\noindent
 to our Galaxy, the wind mass-loss rates are
expected to be also correspondingly lower \citep[e.g.][]{bourak}. We
tested three LMC and SMC binaries, \object{LMC X-1}, \object{LMC X-4}, and
\object{SMC X-1}, with parameters taken from literature \citep{mm1,mm2,mm3,mm4}.
In all these cases, the corresponding stellar parameters lie in a forbidden zone,
indicating that the wind of these stars is inhibited in the direction toward the
companion and the X-rays most likely originate in the Roche lobe overflow. There
is an observational support for this indicating that many of these binaries
indeed likely fill their Roche lobes \citep{mm5,mm6,mm1}.

\subsection{X-rays from wind shocks}

The X-ray emission of single non-magnetic hot stars presumably originates in the
shocks caused by the instability connected with line driving
\citep{ocr,felpulpal}. There may be a feedback effect of the shock generated
X-rays on the ionization structure of the wind. The typical X-ray luminosity
scales with the stellar luminosity, $\x L\approx 10^{-7}\,L$
\citep{sane,igorkar}, however, simulations predict stochastic X-ray variability
on some level. Taking as an example the predicted \x L\ variations of the O9.7Ib
star $\zeta$~Ori~A (in the order of
$10^{33}-10^{34}\,\text{erg}\,\text{s}^{-1}$) from the numerical simulations of
\citet{felpulpal} and assuming the distance of X-ray source from the star
$1.5\,R_*-5\,R_*$, the derived value of the \x t parameter is $\x
t\approx0.5-1.5$. This lies in the region of a strong influence of X-rays on the
wind ionization state (see Fig.~\ref{txobr} for a model 300-1). Consequently,
shock generated X-rays may cause kinks in the velocity profile, which in turn
may lead to moving absorption features in the P~Cygni line profiles
\citep[e.g.][]{cro} or increasing wind inhomogeneities.

\begin{figure}[ht]
\centering
\resizebox{0.33\hsize}{!}{\includegraphics{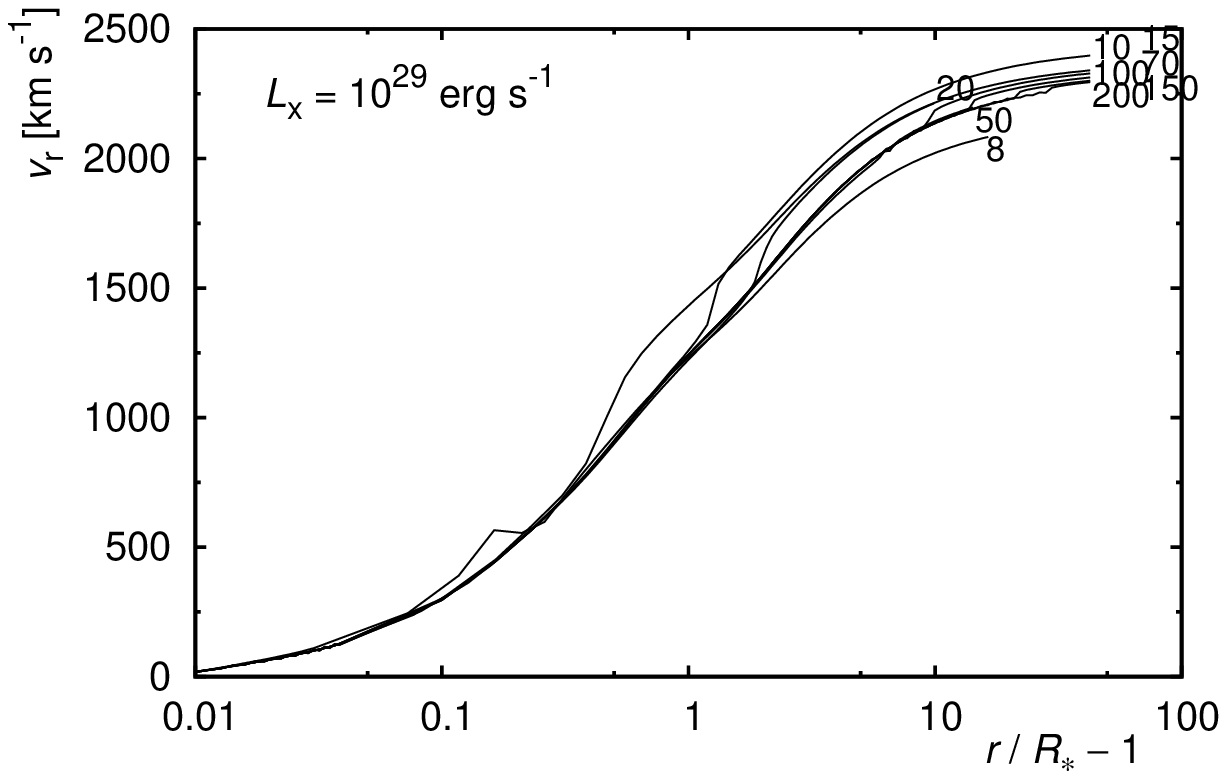}}
\resizebox{0.33\hsize}{!}{\includegraphics{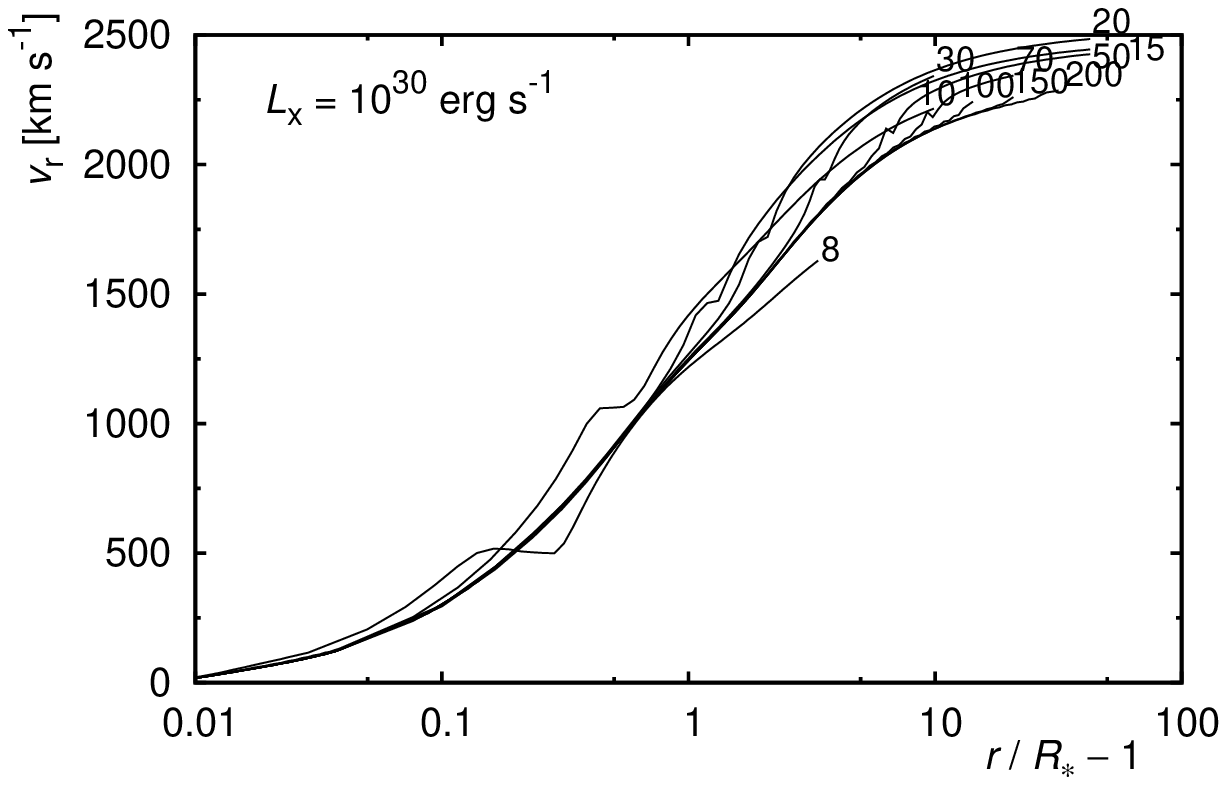}}
\resizebox{0.33\hsize}{!}{\includegraphics{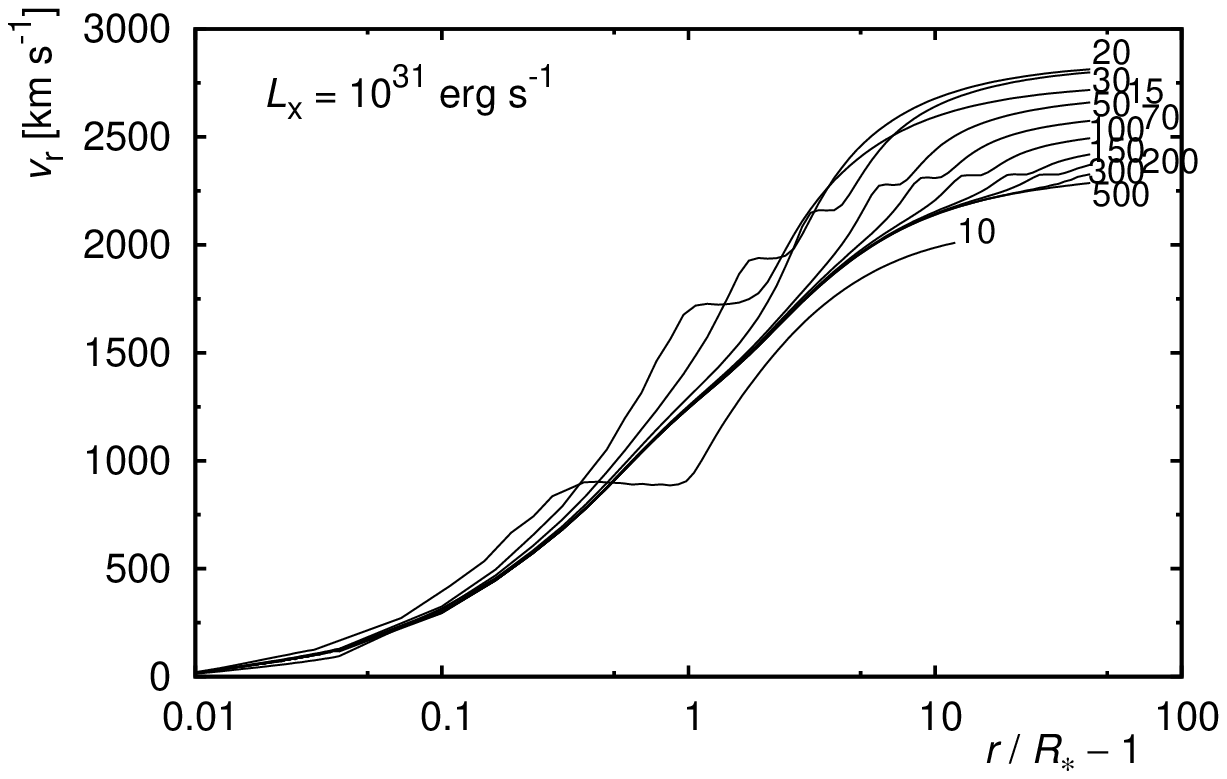}}
\resizebox{0.33\hsize}{!}{\includegraphics{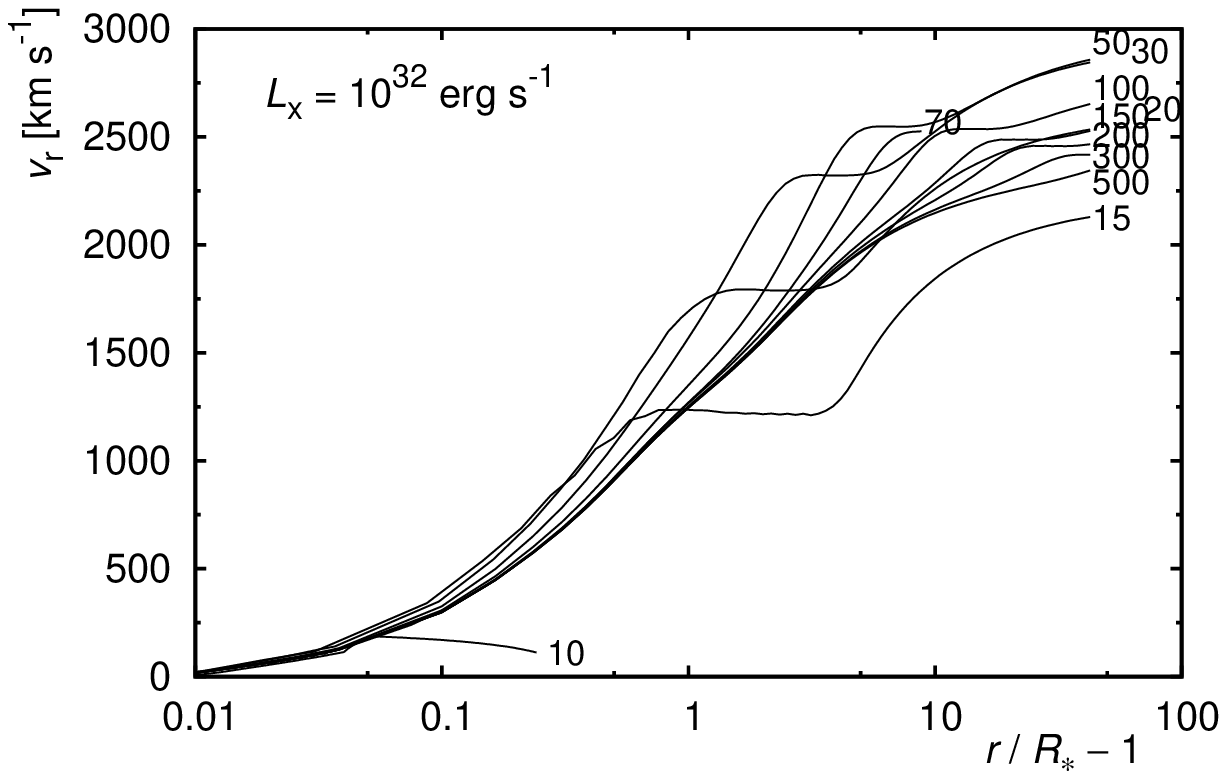}}
\resizebox{0.33\hsize}{!}{\includegraphics{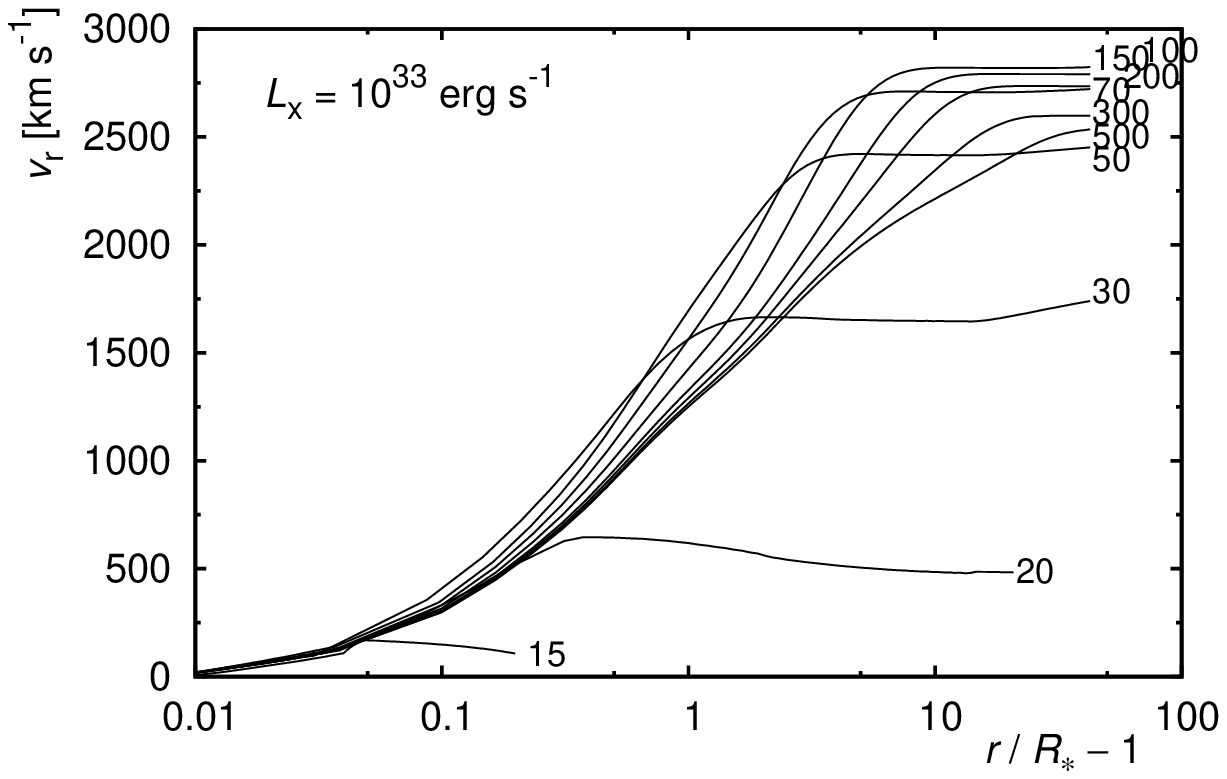}}
\resizebox{0.33\hsize}{!}{\includegraphics{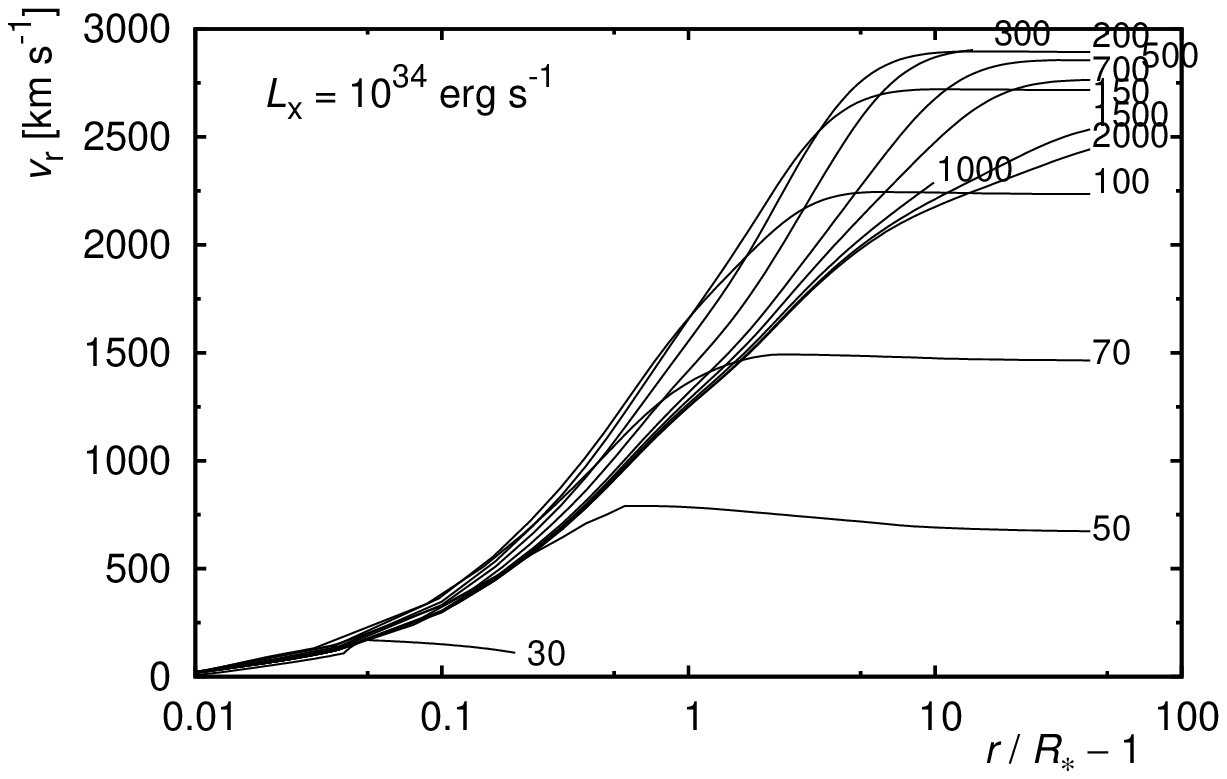}}
\resizebox{0.33\hsize}{!}{\includegraphics{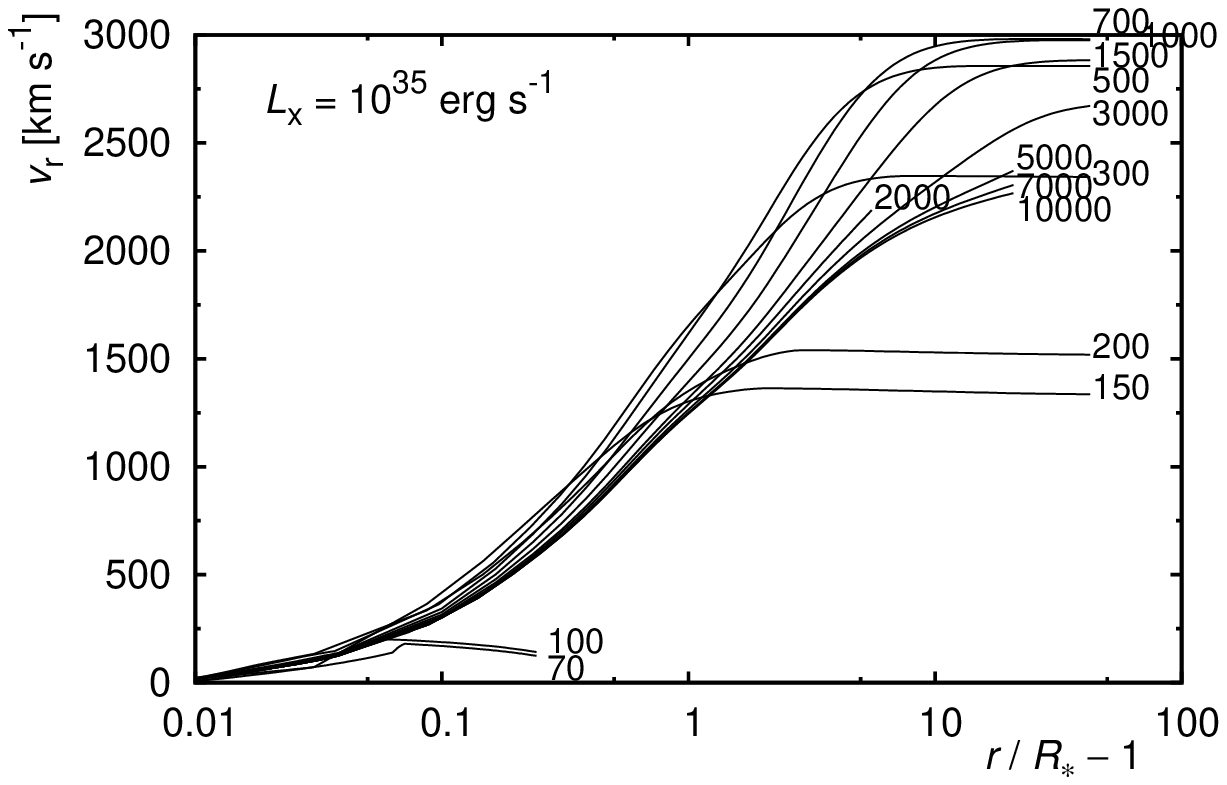}}
\resizebox{0.33\hsize}{!}{\includegraphics{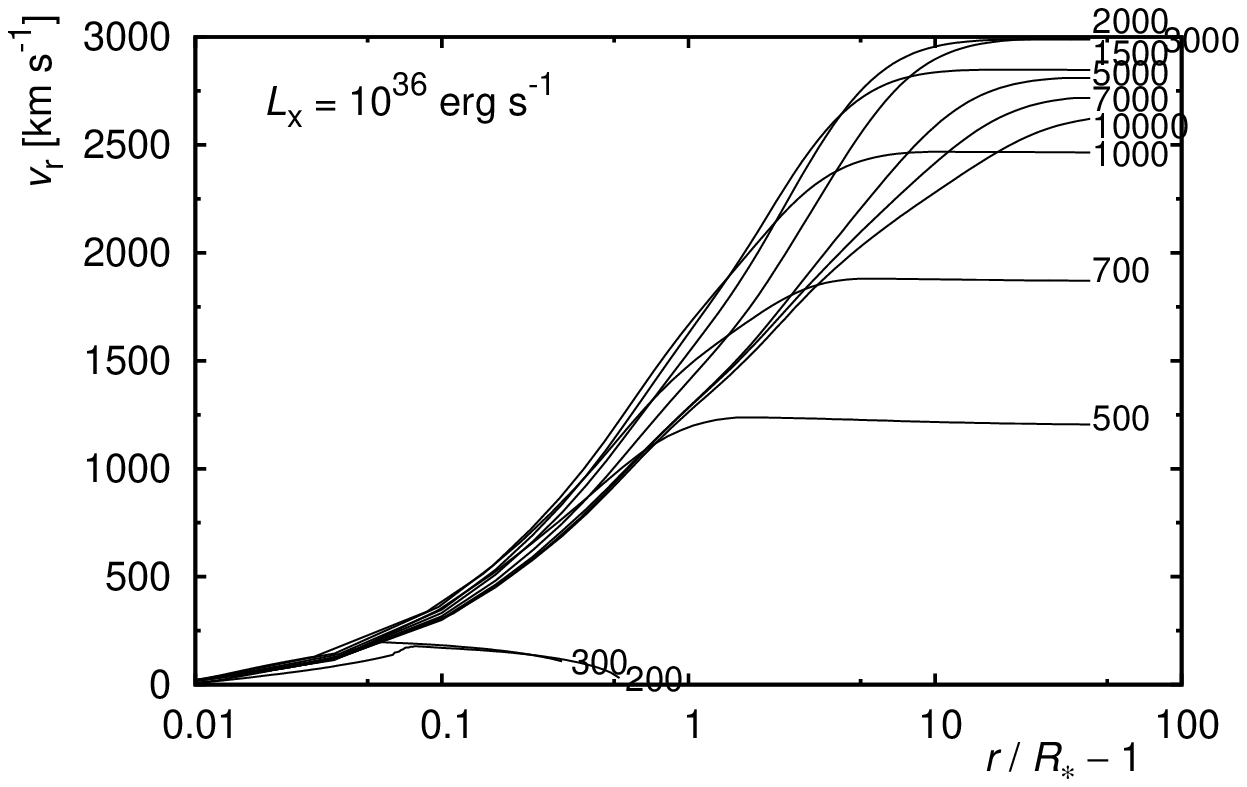}}
\resizebox{0.33\hsize}{!}{\includegraphics{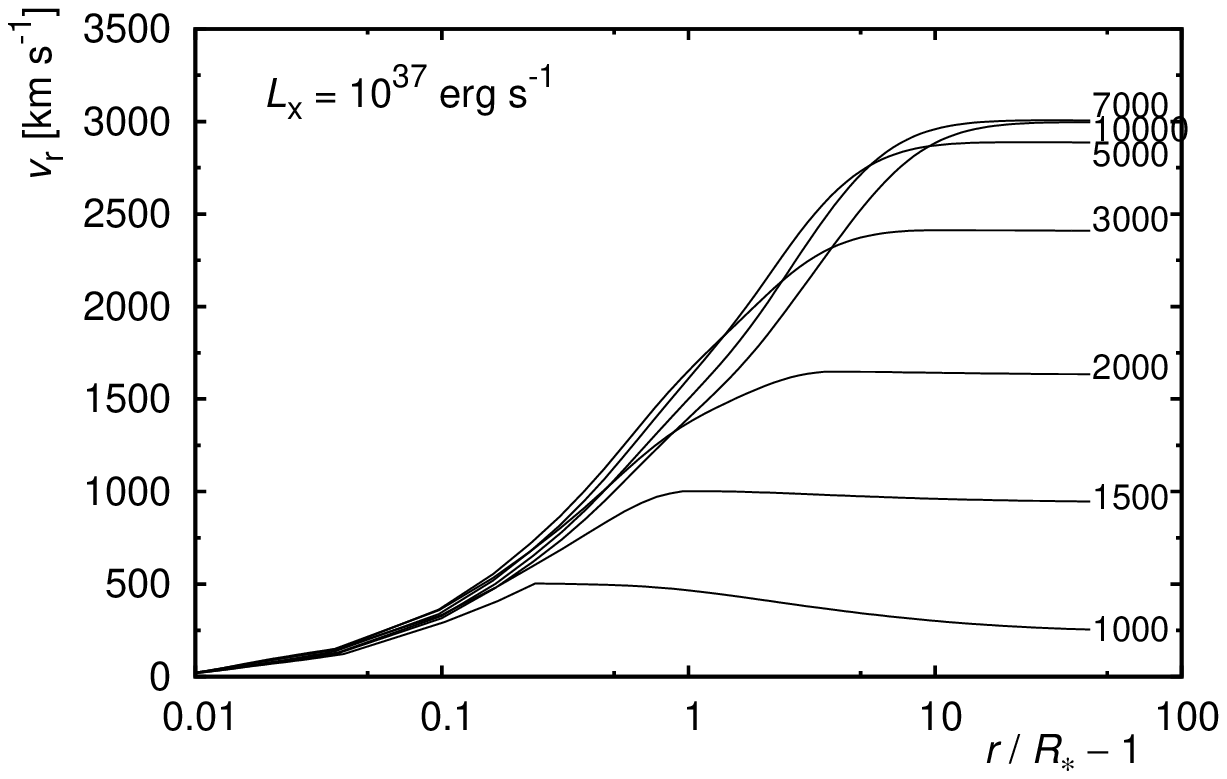}}
\resizebox{0.33\hsize}{!}{\includegraphics{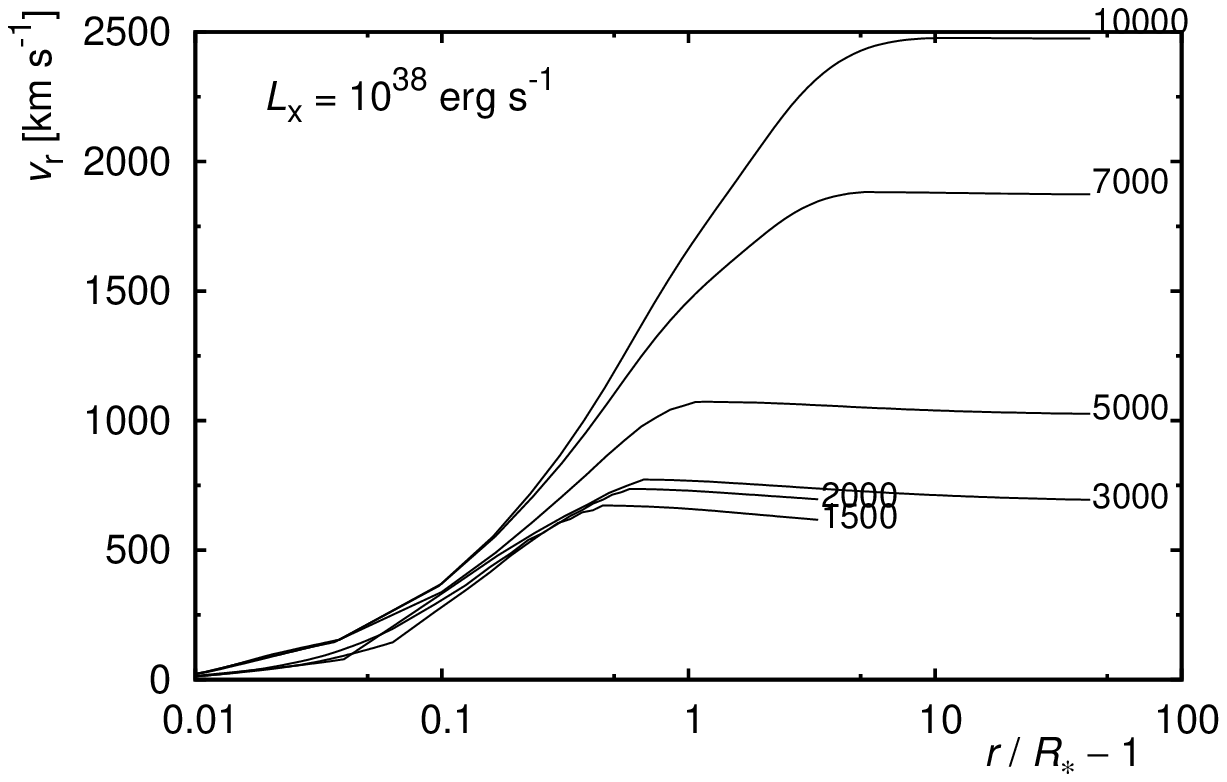}}
\caption{Same as Fig.~\ref{300-1vr}, except for a model star 300-5.}
\label{300-5vr}
\end{figure}

For some low-luminosity O stars the predicted mass-loss rates are significantly
higher than those inferred from observations \citep[e.g.][]{bourak,martclump}.
This disagreement is termed as a "weak wind problem". 
This problem could possibly be explained as a consequence of X-ray influence on the mass-loss rate
\citep{nemajpravdu}. For a typical luminosity of stars showing the weak wind
problem, which have $\log(L/L_\odot)=5$, the X-ray luminosity can be inferred
using the relation $\x L\approx 10^{-7}\,L$ \citep{sane,igorkar} as
$\log\zav{\lx/1\,\text{erg}\,\text{s}^{-1}} \approx31.6$. If an effect of X-rays
on the wind is to appear for these low X-ray luminosities, the X-ray source has
to be in a very close proximity to the star ($D\lesssim1.5\,R_*$, see
Fig.~\ref{300-5vr}). However, X-rays typically originate at larger distances
\citep{felpulpal}. Moreover, in our calculations we assumed one spatially
localized X-ray source, which is unlikely if the X-rays have to influence the
entire wind. For a spatially distributed X-ray source the effect of X-ray
ionization is expected to be lower. Consequently, our results disfavour this
explanation of "weak wind problem", and the weak wind problem seems to be more
likely caused by an excessively
large
cooling length in the post-shock region
\citep{luciebila,martclump,cobecru,nlteiii,lucyjakomy}.

\subsection{Winds in active galactic nuclei}

The X-ray overionization may also be problematic for line-driven winds in the
active galactic nuclei. Using typical parameters of these winds in simulations of
\citet{hb} ($\x L= 10^{44}\,\text{erg}\,\text{s}^{-1}$ and
$D=10^{16}\,\text{cm}$), we derive the ionization parameter $\xi$ of the order of
unity (neglecting absorption) even for the densest regions with
$\rho=10^{-12}\,\text{g}\,\text{cm}^{-3}$. Since typical wind densities are by
two to four orders of magnitude lower, this supports the conclusion that the
winds of active galactic nuclei are strongly inhibited, unless a strong
absorption source is present \citep{hb}.

\section{Discussion and conclusions}

We study the effect of external X-ray irradiation on hot star winds in massive
binaries. Hot star winds are driven by light absorption in the lines of heavier
elements, therefore the wind acceleration is sensitive to the ionization stage
of the wind. We used our own computer code to calculate NLTE wind models for
typical O star and X-ray source parameters found in massive binaries to estimate
the influence of external X-ray irradiation.

The influence of X-rays is determined by the X-ray luminosity, and the X-ray
optical depth between a given point in the wind and the X-ray source, which
follows from the distance to the X-ray source and X-ray opacity along the path.
Therefore, our results can be interpreted in the diagrams of X-ray luminosity
vs.~the optical
depth parameter (in this paper referred to as the diagrams $\x L$ vs.~$\x t$).
The effects of X-rays are negligible in binaries with low X-ray
luminosities or large distances between the binary component with wind and the
X-ray source. In this kind of a case the external irradiation leads just to a
modification of the wind ionization state, which has a relatively low influence
on the radiative driving.

With increasing X-ray luminosity, the influence of X-rays on the wind structure
becomes stronger. There appears to be a typical kink in the velocity profile at the
point where the X-ray overionization is so strong that it lowers the radiative
force so much that is not able to accelerate the wind any more. With decreasing
distance between the star and the X-ray source the kink moves towards the star.
If the kink approaches the critical point, where the mass-loss rates is
determined, the wind becomes inhibited by X-rays. Consequently, there is a
forbidden area in the diagram of $\x L$ vs.~$\x t$ with high X-ray luminosities and low
optical depth parameters, where the X-ray ionization leads to wind inhibition.

We compared the positions of known wind-accreting, high-mass X-ray binaries in
the diagrams of X-ray luminosity vs.~the optical depth parameter with respect to the
location of the forbidden area. All primaries of binaries with compact
components lie outside the forbidden area, which is consistent with our wind
models. Many of primaries lie close to the border of the forbidden area, which
means that the increase in X-ray luminosity would lead to wind inhibition.
This indicates that the X-ray luminosities of HMXBs are self-regulated. 

The X-rays may also strongly affect the ionization state of circumstellar matter in
colliding wind binaries. For many of them we predict a strong influence of
external irradiation on the wind ionization state, pointing to wind
self-regulation of another kind. In some cases the wind of secondary may be
inhibited by the X-ray irradiation. We also discussed the application of our
results for the X-rays generated by wind instabilities and for disk winds in
active galactic nuclei.

There are further observational tests of our models. The decrease in wind
terminal velocity may be detectable for individual stars as the shift of the
blue edge of the UV P~Cygni line profiles \citep{kapka} or as the shift of the
X-ray emission lines \citep{viteal}. Moreover, the kink in the velocity profile
found in the models with strong X-ray ionization may lead to a localized
stronger absorption in the blue part of the P~Cygni line profiles resembling the
discrete absorption components \citep[DACs,][]{huib,cro}.

Despite a good agreement between observations and theory, our models have
limitations. First, we assume a spherical symmetry and solve the wind equations
only in the direction towards the X-ray source. While this may be applicable to
estimate the overall influence of X-ray irradiation, we are missing many
important observational effects that are connected with departures of the wind
from spherical symmetry \citep{frc,blok,felan}. We also neglected the
time-dependent phenomena, which may be important especially in binaries with
eccentric orbits. Moreover, there is a mounting evidence that hot star winds
show small-scale structures (clumping). This may weaken the effect of X-ray
ionization because of increased recombination \citep{osfek}.

Regardless of these limiting assumptions, our models clearly demonstrate the
effect of X-rays on the ionization balance of stellar winds in massive binaries.
Depending on the strength of the X-ray source and its distance from the wind
losing-star, the X-rays are able, via their influence on the ionization
structure, to
change the wind driving force and, consequently, the properties of the wind. In
a limiting case of a strong X-ray source, this leads to wind inhibition. This
inhibition is supported by observed properties of binaries.

\begin{acknowledgements}
This research was supported GA\,\v{C}R  13-10589S.
Access to computing and storage facilities owned by parties and projects
contributing to the National Grid Infrastructure MetaCentrum, provided under the
programme "Projects of Large Infrastructure for Research, Development, and
Innovations" (LM2010005), is greatly appreciated.
\end{acknowledgements}

\clearpage
\onecolumn

\Online

\end{document}

%% file: neutron.tex
\object{2S 0114+650\tablefootmark{f}} & B1Iae & 5.61 & 24000 & 37 & 16 & 54 & $ 1.1\times10^{36 }$ &$ 6.8\times10^{-7  }$ & 0.9 & 1, 2, 37 \\
\object{Vela X-1\tablefootmark{f}} & B0Ia & 5.63 & 27000 & 30 & 23.5 & 53.4 & $ 3.5\times10^{36 }$ &$ 7.5\times10^{-7  }$ & 1.0 & 3, 4, 5 \\
\object{4U 1700-377\tablefootmark{f}} & O6.5Iaf & 6.02 & 40200 & 21.2 & 58 & 37 & $ 2.2\times10^{36 }$ &$ 4.6\times10^{-6  }$ & 4.6 & 6, 7, 8 \\
\object{IGR J16418-4532} & O8.5I & 5.69 & 32800 & 21.7 & 30 & 31.6 & $ 2\times10^{36 }$ &$ 9.8\times10^{-7  }$ & 1.0 & 9, 38 \\
\object{IGR J18029-2016} & B1I & 5.59 & 32500 & 19.8 & 20.2 & 33.1 & $ 2\times10^{36 }$ &$ 6.3\times10^{-7  }$ & 1.1 & 10, 29 \\
\object{IGR J16479-4514} & O8.5Iab & 5.52 & 31000\tablefootmark{a} & 20 & 35 & 31.6 & $ 1.09\times10^{34 }$ &$ 4.5\times10^{-7  }$ & 0.4 & 11 \\
\object{IGR J17252-3616} & B0Ia & 5.79 & 30000 & 29 & 15 & 51 & $ 1.6\times10^{37 }$ &$ 1.5\times10^{-6  }$ & 11.3 & 12, 13, 40 \\
\object{IGR J18483-0311} & B0.5Ia & 5.57 & 24600 & 33.8 & 33 & 95.7 & $ 3.7\times10^{35 }$ &$ 5.7\times10^{-7  }$ & 0.8 & 14, 15, 42 \\
\object{IGR J18450-0435} & O9Ia & 5.58 & 30000 & 23 & 30 & 72 & $ 7\times10^{35 }$ &$ 6.0\times10^{-7  }$ & 1.2 & 16, 17, 43 \\
\object{X Per\tablefootmark{f}} & B0Ve & 4.69 & 32000\tablefootmark{a} & 7.2\tablefootmark{a} & 15.5 & 420 & $ 2.7\times10^{33 }$ &$ 2.2\times10^{-8  }$ & 0.1 & 18, 30 \\
\object{IGR J11215-5952} & B0.5Ia & 5.73 & 24700 & 40 & 29 & 80\tablefootmark{b} & $ 3\times10^{36 }$ &$ 1.2\times10^{-6  }$ & 1.5 & 19, 20 \\
\object{PSR B1259-63\tablefootmark{f}\tablefootmark{c}} & O9.5Ve & 4.9 & 34000\tablefootmark{d} & 8.1\tablefootmark{d} & 10 & 1100 & $ 3.5\times10^{34 }$ &$ 5.5\times10^{-8  }$ & 0.4 & 21, 22 \\
\object{IGR J19140+0951} & B0.5I & 5.47 & 28000\tablefootmark{a} & 23.2\tablefootmark{a} & 25.4\tablefootmark{a} & 62.5 & $ 3\times10^{35 }$ &$ 3.6\times10^{-7  }$ & 0.7 & 23 \\
\object{4U 2206+54} & O9.5Ve & 4.59 & 30000 & 7.3 & 16 & 53 & $ 1.8\times10^{35 }$ &$ 1.4\times10^{-8  }$ & 0.1 & 24 \\
\object{4U 1907+09} & O8.5Iab & 5.68 & 29760 & 26.2 & 26.0 & 54 & $ 2\times10^{36 }$ &$ 9.6\times10^{-7  }$ & 1.6 & 25 \\
\object{LS 5039\tablefootmark{f}} & O6.5V & 5.19 & 37500 & 9.3 & 22.9 & 34.5 & $ 6\times10^{34 }$ &$ 2.0\times10^{-7  }$ & 0.7 & 26, 39, 44, 45 \\
\object{Cyg X-1\tablefootmark{f}} & O9.7Iab & 5.57 & 32000 & 19.9\tablefootmark{e} & 24.0 & 42.4 & $ 1.4\times10^{37 }$ &$ 5.7\times10^{-7  }$ & 1.1 & 27, 28, 41 \\
\object{IGR J16465-4507} & O9.5Ia & 5.69 & 30000\tablefootmark{a} & 26 & 27.8 & 124 & $ 6.8\times10^{36 }$ &$ 9.9\times10^{-7  }$ & 2.4 & 31, 32 \\
\object{IGR J17544-2619} & O9Ib & 5.53 & 31000 & 20.3 & 26.5 & 36.3 & $ 1.7\times10^{35 }$ &$ 4.7\times10^{-7  }$ & 0.7 & 33, 34 \\
\object{IGR J16207-5129} & B1Ia & 5.4 & 29000\tablefootmark{a} & 20 & 20 & 40 & $ 2\times10^{34 }$ &$ 2.6\times10^{-7  }$ & 0.5 & 35, 36 \\
\object{XTE X1739-302} & O8.5Iab & 5.71 & 33000\tablefootmark{a} & 21.9\tablefootmark{a} & 30.8\tablefootmark{a} & 173 & $ 4.8\times10^{33 }$ &$ 1.1\times10^{-6  }$ & 3.0 & 46, 47 \\
\object{IGR J17354-3255} & O9.5Iab & 5.56 & 30000\tablefootmark{a} & 22.4\tablefootmark{a} & 28.8\tablefootmark{a} & 54 & $ 2.6\times10^{36 }$ &$ 5.4\times10^{-7  }$ & 0.9 & 48, 49 \\

%% file: alt_neutron.tex
2S 0114+650 (V662 Cas),
Vela X-1 (GP Vel, HD 77581),
4U 1700-377 (V884 Sco, HD 153919),
X Per (HR 1209),
PSR B1259-63 (CPD-63$\degr$2495),
LS 5039 (V479 Sct),
and  Cyg X-1 (V1357 Cyg. HD 226868).

%% file: ref_neutron.tex
(1)~\citet{hmxb1};
(2)~\citet{hmxb2};
(3)~\citet{hmxb3};
(4)~\citet{hmxb4};
(5)~\citet{viteal};
(6)~\citet{hmxb6};
(7)~\citet{hmxb7};
(8)~\citet{hmxb8};
(9)~\citet{hmxb9};
(10)~\citet{hmxb10};
(11)~\citet{hmxb11};
(12)~\citet{hmxb12};
(13)~\citet{hmxb13};
(14)~\citet{hmxb14};
(15)~\citet{hmxb15};
(16)~\citet{hmxb16};
(17)~\citet{hmxb17};
(18)~\citet{hmxb18};
(19)~\citet{hmxb19};
(20)~\citet{hmxb20};
(21)~\citet{hmxb21};
(22)~\citet{hmxb22};
(23)~\citet{hmxb23};
(24)~\citet{hmxb24};
(25)~\citet{hmxb25};
(26)~\citet{hmxb26};
(27)~\citet{hmxb27};
(28)~\citet{hadrvitr};
(29)~\citet{hmxb29};
(30)~\citet{dvojx1};
(31)~\citet{hmxb31};
(32)~\citet{hmxb32};
(33)~\citet{hmxb33};
(34)~\citet{hmxb34};
(35)~\citet{hmxb35};
(36)~\citet{hmxb36};
(37)~\citet{hmxb37};
(38)~\citet{hmxb38};
(39)~\citet{hmxb39};
(40)~\citet{hmxb40};
(41)~\citet{hmxb41};
(42)~\citet{hmxb42};
(43)~\citet{hmxb43};
(44)~\citet{hmxb44};
(45)~\citet{hmxb45};
(46)~\citet{hmxb46};
(47)~\citet{hmxb47};
(48)~\citet{hmxb48};
(49)~\citet{hmxb49}.

%% file: dvojhve.tex
\object{HR 8281} & O6.5V & 5.7 & 41200 & 13.9 & 40.2  & $ 2.0\times10^{-6} $ &  2290 & 3.6 & 40  & $ 2.3\times10^{32}  $ & 1, 2, 3 \\
\object{ } & O9V\tablefootmark{a} & 5.02 & 34900 & 8.9 & 22.4  & $ 9.7\times10^{-8} $ &  2400 &   &   & $   $ &  \\
\object{HD 152219} & O9.5III & 5.07 & 31900 & 11.2 & 26.2  & $ 8.0\times10^{-8} $ &  2320 & 0.2 & 37.1  & $ 8\times10^{30}  $ & 4 \\
\object{ } & B1V\tablefootmark{a} & 3.74 & 21800 & 5.2 & 10.3  & $ 3.4\times10^{-10} $ &  2250 &   &   & $   $ &  \\
\object{HD 47129} & O8I & 5.35 & 33500 & 14.1 & 54.  & $ 2.0\times10^{-7} $ &  2980 & 0.4 & 128.5  & $ 8.34\times10^{32}  $ & 5, 30 \\
\object{ } & O7.5III & 5.09 & 33000 & 10.8 & 56.  & $ 9.0\times10^{-8} $ &  3560 & 0.2 &   & $   $ &  \\
\object{HR 6187} & O3V & 5.78 & 44500 & 13.14 & 63  & $ 2.9\times10^{-6} $ &  3090 & 3.9 & 38.0  & $ 3.2\times10^{33}  $ & 1, 6, 7, 8 \\
\object{ } & O5.5V\tablefootmark{a} & 5.28 & 39000 & 9.54 & 40.  & $ 3.0\times10^{-7} $ &  3100 &   &   & $   $ &  \\
\object{HR 6736} & O3.5V & 5.75 & 43850 & 13.1\tablefootmark{c} & 55  & $ 2.6\times10^{-6} $ &  2860 & 6.9 & 4100  & $ 1.2\times10^{33}  $ & 1, 9 \\
\object{ } & O5V & 5.5 & 40850 & 11.2\tablefootmark{c} & 36  & $ 8.1\times10^{-7} $ &  2560 & 2.8 &   & $   $ &  \\
\object{V712 Car} & O3I & 6.06 & 43000 & 19.3 & 82.7  & $ 5.4\times10^{-6} $ &  2720 & 2.8 & 53  & $ 7.5\times10^{33}  $ & 10, 11 \\
\object{ } & O3I & 6.06 & 43000 & 19.3 & 81.9  & $ 5.4\times10^{-6} $ &  2700 & 2.8 &   & $   $ &  \\
\object{HR 65 } & O9III & 5.2 & 32000\tablefootmark{c} & 13 & 15  & $ 1.5\times10^{-7} $ &  1490 & 0.3 & 32  & $ 7.2\times10^{32}  $ & 12, 13 \\
\object{ } & O9III & 4.83 & 32000\tablefootmark{c} & 8.5 & 21  & $ 2.7\times10^{-8} $ &  2430 & 0.0 &   & $   $ &  \\
\object{HR 5664} & O7V & 5.27 & 37500 & 10.2 & 21.6  & $ 2.9\times10^{-7} $ &  2090 & 0.9 & 33.8  & $ 6.9\times10^{31}  $ & 1, 14 \\
\object{ } & O9.5V\tablefootmark{b} & 4.64 & 33000 & 6.4 & 12.4  & $ 1.7\times10^{-8} $ &  2140 & 0.0 &   & $   $ &  \\
\object{HR 8406} & O8.5III & 5.11 & 32000 & 11.7 & 16.9  & $ 9.7\times10^{-8} $ &  1750 & 0.2 & 25.5  & $ 7.1\times10^{31}  $ & 1, 15 \\
\object{ } & O9.5V\tablefootmark{a} & 4.69 & 28000 & 9.4 & 6.7  & $ 2.2\times10^{-8} $ &  1230 &   &   & $   $ &  \\
\object{HD 93206A} & O9.7I & 5.7 & 32000 & 23 & 40  & $ 1.0\times10^{-6} $ &  1780 & 2.0 & 116  & $ 3.4\times10^{32}  $ & 16, 29 \\
\object{ } & B2V\tablefootmark{a} & 3.7 & 20000 & 5.9 & 10  & $ 2.3\times10^{-10} $ &  2080 &   &   & $   $ &  \\
\object{HD 93206B} & O8III & 5.3 & 32600 & 14 & 14  & $ 2.3\times10^{-7} $ &  1310 & 0.6 & 49  & $ 1.2\times10^{32}  $ & 16, 29 \\
\object{ } & O9V & 4.9 & 32500 & 8.9 & 28  & $ 5.6\times10^{-8} $ &  2750 & 0.1 &   & $   $ &  \\
\object{V1007 Sco} & O7.5III & 5.4 & 34000 & 14.47 & 27.24  & $ 3.7\times10^{-7} $ &  1950 & 0.5 & 51.7  & $ 4.4\times10^{32}  $ & 17, 18 \\
\object{ } & O7III & 5.5 & 34350 & 15.99 & 27.88  & $ 5.9\times10^{-7} $ &  1810 & 0.9 &   & $   $ &  \\
\object{HR 1931} & O9.5V & 4.64 & 32600 & 6.6\tablefootmark{c} & 19  & $ 1.8\times10^{-8} $ &  2650 & 0.1 & 375  & $ 2.1\times10^{32}  $ & 1, 19, 31 \\
\object{ } & B0V\tablefootmark{b} & 4.45 & 30600\tablefootmark{c} & 6.0\tablefootmark{c} & 15.4  & $ 2.6\times10^{-9} $ &  2520 & 0.0 &   & $   $ &  \\
\object{HR 6535} & O6V & 5.26 & 39000\tablefootmark{c} & 9.4 & 32  & $ 2.9\times10^{-7} $ &  2750 & 0.8 & 56.4  & $ 5.2\times10^{32}  $ & 1, 20, 21, 22 \\
\object{ } & O7V & 5.17 & 37000\tablefootmark{c} & 9.4 & 32  & $ 1.9\times10^{-7} $ &  2790 & 0.5 &   & $   $ &  \\
\object{HR 1899} & O9III & 4.81 & 32000\tablefootmark{c} & 8.3 & 23  & $ 2.5\times10^{-8} $ &  2580 & 0.1 & 80  & $ 7.4\times10^{32}  $ & 1, 21, 23, 24 \\
\object{ } & B0.5V\tablefootmark{b} & 4.19 & 28000\tablefootmark{c} & 5.3 & 13  & $ 1.8\times10^{-9} $ &  2480 & 0.0 &   & $   $ &  \\
\object{HR 1852} & O9.5II & 5.15 & 31000\tablefootmark{c} & 13 & 11.2  & $ 7.8\times10^{-8} $ &  1250 & 0.3 & 44  & $ 6.5\times10^{32}  $ & 1, 21, 25, 26 \\
\object{ } & B0.5III\tablefootmark{b} & 4.2 & 29000\tablefootmark{c} & 5 & 5.6  & $ 1.5\times10^{-9} $ &  1640 & 0.0 &   & $   $ &  \\
\object{V918 Sco} & O7.5I & 5.9 & 35100 & 24.3 & 57  & $ 2.7\times10^{-6} $ &  2000 & 3.1 & 91  & $ 5.2\times10^{32}  $ & 1, 27 \\
\object{ } & O9.7I & 5.79 & 30500 & 28.1 & 37  & $ 1.5\times10^{-6} $ &  1430 & 0.7 &   & $   $ &  \\
\object{HR 7767} & O8.5III & 5.25 & 32500 & 13.4\tablefootmark{c} & 23  & $ 1.9\times10^{-7} $ &  1900 & 0.7 & 620  & $ 8.9\times10^{31}  $ & 1, 28 \\
\object{ } & B2.5V\tablefootmark{b} & 3.29 & 20000 & 3.7\tablefootmark{c} & 9  & $ 3.5\times10^{-11} $ &  2500 & 0.0 &   & $   $ &  \\
\object{DH Cep} & O5.5V & 5.37 & 44000 & 8.31 & 29.4  & $ 4.5\times10^{-7} $ &  2720 & 0.9 & 26.4  & $ 5.4\times10^{33}  $ & 12, 32 \\
\object{ } & O6.5V\tablefootmark{b} & 5.27 & 43000 & 7.76 & 25.0  & $ 2.9\times10^{-7} $ &  2620 & 0.5 &   & $   $ &  \\
\object{V1034 Sco} & O9V & 4.78 & 33200 & 7.45 & 16.8  & $ 3.3\times10^{-8} $ &  2310 & 0.1 & 22.65  & $ 3.2\times10^{31}  $ & 33, 34 \\
\object{ } & B1V\tablefootmark{a} & 3.87 & 26200 & 4.18 & 9.4  & $ 5.8\times10^{-10} $ &  2390 &   &   & $   $ &  \\
\object{V1294 Sco} & O9V & 5.1 & 34000 & 10.3 & 24.5  & $ 1.4\times10^{-7} $ &  2320 & 0.4 & 42.8  & $ 6.9\times10^{31}  $ & 35 \\
\object{ } & O9.7V & 4.71 & 31200 & 7.8 & 18.2  & $ 2.4\times10^{-8} $ &  2370 & 0.0 &   & $   $ &  \\
\object{HD 93205} & O3V & 6.18 & 49000 & 17.2\tablefootmark{c} & 45  & $ 1.8\times10^{-5} $ &  1130 & 60.8 & 56.3  & $ 3.4\times10^{33}  $ & 36 \\
\object{ } & O8V\tablefootmark{a} & 5.11 & 36500 & 9.0\tablefootmark{c} & 20  & $ 1.4\times10^{-7} $ &  2200 &   &   & $   $ &  \\

%% file: ref_dvojhve.tex
(1)~\citet{dvojx1};
(2)~\citet{dvojx2};
(3)~\citet{dvojx3};
(4)~\citet{dvojx4};
(5)~\citet{dvojx5};
(6)~\citet{dvojx6};
(7)~\citet{dvojx7};
(8)~\citet{dvojx8};
(9)~\citet{dvojx9};
(10)~\citet{dvojx10};
(11)~\citet{dvojx11};
(12)~\citet{dvojx12};
(13)~\citet{dvojx13};
(14)~\citet{dvojx14};
(15)~\citet{dvojx15};
(16)~\citet{dvojx16};
(17)~\citet{dvojx17};
(18)~\citet{dvojx18};
(19)~\citet{dvojx19};
(20)~\citet{dvojx20};
(21)~\citet{dvojx21};
(22)~\citet{dvojx22};
(23)~\citet{dvojx23};
(24)~\citet{dvojx24};
(25)~\citet{dvojx25};
(26)~\citet{dvojx26};
(27)~\citet{dvojx27};
(28)~\citet{dvojx28};
(29)~\citet{dvojx29};
(30)~\citet{dvojx30};
(31)~\citet{dvojx31};
(32)~\citet{dvojx32};
(33)~\citet{dvojx33};
(34)~\citet{dvojx34};
(35)~\citet{dvojx35};
(36)~\citet{dvojx36}.